\documentclass[prd, preprint,nofootinbib,eqsecnum,superscriptaddress,12pt]{revtex4-1}
\pdfoutput=1

\newcount\showkey \showkey=0  

\usepackage{color}
\usepackage{tabularx}

\usepackage{hyperref}

\ifnum\showkey=1
\usepackage{seqsplit}
\usepackage[color]{showkeys}
\definecolor{refkey}{rgb}{0.9, 0.43, 0.63}
\definecolor{labelkey}{rgb}{0.59, 0.43, 0.63}
\usepackage{xstring}
\renewcommand*\showkeyslabelformat[1]{%
\noexpandarg%
\StrSubstitute{\(\{\)#1\(\}\)}{ }{\textvisiblespace}[\TEMP]%
\parbox[t]{\marginparwidth}{\raggedright\normalfont\small\ttfamily\expandafter\seqsplit\expandafter{\TEMP}}}
\fi

\usepackage{amsmath}
\usepackage{amssymb}
\usepackage{graphicx,multirow,subfigure}
\usepackage{float} 
\usepackage[T1]{fontenc}

\usepackage{rotating} 

\usepackage{enumitem}
\setlist[enumerate,2]{leftmargin=0.45em}

\usepackage{comment}

\usepackage[dvipsnames]{xcolor}

\usepackage[capitalize]{cleveref}    

\newcommand{\nn}{\nonumber}
\newcommand{\A}{\mathcal{A}}

\renewcommand{\bar}{\overline}

\usepackage{physics}
\usepackage{bbold}
\usepackage{mathtools}

\renewcommand \ket[1]{
        \left| #1 \right>
}

\renewcommand \bra[1]{
        \left< #1 \right|
}

\newcommand{\beq}{\begin{equation}}
\newcommand{\eeq}{\end{equation}}
\newcommand{\beqa}{\begin{eqnarray}}
\newcommand{\eeqa}{\end{eqnarray}}
\newcommand{\bea}{\begin{eqnarray}}
\newcommand{\eea}{\end{eqnarray}}
\newcommand{\bi}{\begin{itemize}}
\newcommand{\ei}{\end{itemize}}
\newcommand{\ben}{\begin{enumerate}}
\newcommand{\een}{\end{enumerate}}
\newcounter{mycount}
\newcommand{\pauseen}{\setcounter{mycount}{\value{enumi}}\end{enumerate}}
\newcommand{\resumeen}{\begin{enumerate}\setcounter{enumi}{\value{mycount}}}

\newcounter{defcount}
\newcounter{myenum}
\newcommand{\myben}{\setcounter{myenum}{1}}
\newcommand{\myitt}{(\roman{myenum})~\addtocounter{myenum}{1}}

\begin{document}

\title{The Mathematical Structure of $U$-Spin Amplitude Sum Rules}

\author{Margarita Gavrilova}
\email{mg2333@cornell.edu}
\affiliation{Department of Physics, LEPP, Cornell University, Ithaca, NY 14853, USA}

\author{Yuval Grossman}
\email{yg73@cornell.edu}
\affiliation{Department of Physics, LEPP, Cornell University, Ithaca, NY 14853, USA}

\author{Stefan Schacht}
\email{stefan.schacht@manchester.ac.uk}
\affiliation{Department of Physics and Astronomy, University of Manchester, Manchester M13 9PL, United Kingdom}

\begin{abstract}
We perform a systematic study of $SU(2)$ flavor amplitude sum rules with particular emphasis on $U$-spin. This study reveals a rich mathematical structure underlying the sum rules that allows us to formulate an algorithm for deriving all $U$-spin amplitude sum rules to any order of the symmetry breaking. This novel approach to deriving the sum rules does not require one to explicitly compute the Clebsch-Gordan tables, and allows for simple diagrammatic interpretation. Several examples that demonstrate the application of our novel method to systems that can be probed experimentally are provided.
\end{abstract}

\maketitle

\tableofcontents

\section{Introduction}

The main challenge in probing the weak interaction using hadrons is
the presence of non-perturbative QCD dynamics. $U$-spin symmetry can be utilized to probe short distance physics when we do not have the ability to calculate the effect of the strong interactions directly.

$U$-spin is an approximate $SU(2)$ symmetry of the QCD Lagrangian under the unitary rotation of down and strange quarks. 
Using this approximate symmetry between down and strange quarks, which form doublets under $U$-spin,
\begin{equation} \label{eq:sd-uspin}
\begin{bmatrix}
\,d\,\, \\ s
\end{bmatrix} =
\begin{bmatrix}
\ket{\frac{1}{2}, +\frac{1}{2}}\\
\ket{\frac{1}{2}, -\frac{1}{2}}
\end{bmatrix}, \hspace{25pt}
\begin{bmatrix}
\,\bar{s}\,\, \\
-\bar{d}
\end{bmatrix} =
\begin{bmatrix}
\ket{\frac{1}{2}, +\frac{1}{2}}\\
\ket{\frac{1}{2}, -\frac{1}{2}}
\end{bmatrix}\,,
\end{equation}
we are able to derive relations between amplitudes of various processes that involve $d$ and $s$ quarks. Such relations are called
\emph{$U$-spin amplitude sum rules}.
Using amplitude sum rules we can reduce the number of unknown hadronic parameters. Then, in some cases, that is what is needed in order to 
transform a system of measurements that we cannot solve, into one which we can use to extract fundamental parameters.

$U$-spin symmetry is broken by a small parameter of order $(m_s-m_d)/\Lambda_{QCD} \sim 0.3$.
We can systematically
expand in this small parameter and obtain $U$-spin sum rules that
hold beyond the symmetry limit.

Approximate flavor symmetries for non-leptonic decays have been extensively discussed in the literature~\cite{Kingsley:1975fe, Einhorn:1975fw, Altarelli:1974sc, Abbott:1979fw, Golden:1989qx, Quigg:1979ic, Voloshin:1975yx, Savage:1991wu, Chau:1991gx, Falk:2001hx, Pirtskhalava:2011va, Grossman:2006jg, Pirtskhalava:2011va, Hiller:2012xm, Grossman:2012ry, Grossman:2013lya, Muller:2015rna, Adolph:2020ema, deBoer:2018zhz, Brod:2012ud, Grinstein:2014aza, Bhattacharya:2012ah, Franco:2012ck, Hiller:2012xm, Grossman:2018ptn, Buccella:1994nf, Cheng:2012xb, Feldmann:2012js, Atwood:2012ac, Buccella:2019kpn, Muller:2015lua, Pirtskhalava:2011va, Chau:1993ec, Zeppenfeld:1980ex, Jung:2014jfa, Buras:2004ub, Gronau:2000zy, Fleischer:1999pa, Gronau:2000md,Jung:2009pb, Grossman:2003qp, Ligeti:2015yma}. 
They are also especially important in the context of the theoretical interpretation of the recent first observation of charm CP
violation \cite{LHCb:2019hro, Grossman:2019xcj, Khodjamirian:2017zdu, Chala:2019fdb, Li:2019hho, Soni:2019xko, Dery:2021mll, Schacht:2021jaz}.
In particular, sum rules that are valid up to second order had been pointed out in the past, for example
in Refs.~\cite{Kingsley:1975fe, Voloshin:1975yx, Barger:1979fu, Brod:2012ud, Grossman:2012ry}.  
Some results on general sum rules were also given in Ref.~\cite{Hassan:2022ucn}.

Note that we discuss linear sum rules,~i.e. sum rules linear in the decay amplitudes. The expansion parameter is relevant only when we talk about such relations. Clearly, one can get an arbitrary precision by using non-linear relations. Some examples of non-linear relations can be found in Refs.~\cite{Gronau:2013xba, Gronau:2015rda}.

In this work we focus on exploring the mathematical structure of $U$-spin amplitude sum
rules. 
The procedure of generating relations between amplitudes is then
needed to be transformed into physical observables like decay rates and CP
asymmetries. This step is simple in a few cases, but in general it is
not. 
Our primary objective in this paper is to analyze the underlying mathematical structure of higher order amplitude sum rules, and not the practicality of these results for phenomenological analyses. 
The latter is left for future work, which also has to consider phase space effects and the possible effects from the differing resonance structure of different decay channels.

Our analysis reveals a rich mathematical structure underlying amplitude
sum rules. This structure enables us to derive all the
sum rules to any order of $U$-spin breaking without performing any calculation. 
In particular, we develop an algorithm to derive the complete set of sum rules for an arbitrary order of 
$U$-spin breaking, by mapping the amplitudes onto a  
multi-dimensional lattice, from which the sum rules can be directly read.

The standard method of deriving sum rules obscures the underlying structure. It requires one to compute a table of Clebsch-Gordan coefficients and then read the sum rules from it.  As a result one could obtain sum rules of many different forms depending on the basis choice and the specific method used to read off the sum rules. The novel method that we present below, on the contrary, is very transparent. It utilizes the symmetry of the problem and allows for a systematic derivation of the sum rules. An additional advantage of the algorithm that we propose is that it is straightforward and simple to execute.

Even though in this paper we focus on $U$-spin, all the results are also applicable to any $SU(2)$ flavor symmetry. While the result is also valid for isospin, as we explain, its applicability to observables may be limited.

The rest of this paper is organized as follows. In Sec.~\ref{sec:definitions} we present our definitions, assumptions and notations and introduce basic concepts that are used throughout the paper. 
In Sec.~\ref{sec:n_doublet_system} we discuss the systematics of amplitude sum rules at arbitrary order in the $U$-spin breaking for systems of $U$-spin doublets. Furthermore, we present a method for deriving the sum rules in a purely geometric way.
We generalize our results in Sec.~\ref{sec:gen-arbitrary-irreps} to the case of arbitrary irreducible representations (irreps) and provide several examples of the application of our algorithm in Sec.~\ref{sec:gen_algo}.
We conclude in Sec.~\ref{sec:conclusions}. All formal derivations and technical details are provided in appendices. 

\section{Definitions, assumptions, and $U$-spin sum rules \label{sec:definitions}}

There are two main ideas that allow one to
write sum rules for a physical system. First, the basis rotation between
the physical basis and the $U$-spin basis, that is, the basis of
definite values of $U$-spin.   
Second, the application of the
Wigner-Eckart theorem that is used to reduce the number of basis
elements that are used to describe the amplitudes. Then, we can have
a situation where the number of different basis elements in the $U$-spin basis
becomes less than the number of amplitudes in the physical system thus
yielding linear relations between amplitudes. These relations are
called sum rules.

In this section we start by describing the system under consideration in subsections~\ref{sec:Uspin} and~\ref{sec:comments}. Then we review the standard approach to deriving amplitude sum rules in subsections~\ref{sec:expansion-in-b}--\ref{sec:sum-rules}. Finally, in subsection~\ref{sec:universality} we discuss the universality of sum rules and motivate our novel approach to amplitude sum rules.

\subsection{$U$-spin systems}\label{sec:Uspin}

We consider a general set of processes where all the initial and final state particles have definite properties under $U$-spin. We denote a set of amplitudes that correspond to physical decay processes that include
all the processes that are related by $U$-spin as {\em a $U$-spin set of
processes} or simply as {\em a $U$-spin set} or {\em a $U$-spin system}. To describe a $U$-spin set  it is necessary to describe the $U$-spin properties of the initial state, final state and the Hamiltonian. We use $n_A$ to denote the
number of amplitudes in the $U$-spin set, and $\A_j$ to denote these amplitudes.

The dynamics of the processes is encoded in the effective Hamiltonian
$\mathcal{H}_{\mathrm{eff}}$. In the $U$-spin limit the most general Hamiltonian is as follows
\begin{equation}\label{eq:H_LO}
    \mathcal{H}_{\text{eff}}^{(0)} = \sum_{u,m,\Gamma} f_{u,m} H^{u}_m(\Gamma)\,,
\end{equation}
where the superscript $(0)$ indicates the $U$-spin limit,
$H_m^u(\Gamma)$ are 
different $U$-spin operators 
with total $U$-spin $u$, third component of $U$-spin $m$, and Dirac structure $\Gamma$. The
factors $f_{u,m}$ encode the weak interaction factors like CKM matrix
elements, the Fermi constant $G_F$ and loop factors. Note that $f_{u,m}$ depends on $\Gamma$ but we keep it implicit.

We define $H^u$ without subscript index and without the $\Gamma$ dependence to refer to a set of
Hamilton operators with a common~$u$ and a common Dirac structure. In this work we only consider the cases where the effective
Hamiltonian in Eq.~\eqref{eq:H_LO} is dominated by one specific
$H^u$. In
this limit, there is no sum over $u$ and $\Gamma$ in Eq.~(\ref{eq:H_LO}), and the effective Hamiltonian takes the following form
\begin{equation}\label{eq:H_LO_u}
        \mathcal{H}_{\text{eff}}^{(0)} = \sum_{m} f_{u,m} H^{u}_m\,.
\end{equation}
Note that in what follows, unless explicitly mentioned otherwise, when we say ``Hamiltonian'' we refer to the zeroth order expression given in Eq.~\eqref{eq:H_LO_u}.

\subsection{Comments about the assumptions}\label{sec:comments}

In the above section we make two working assumptions about the $U$-spin properties of the states and the Hamiltonian:
\begin{itemize}
\item[${(i)}$]
All initial and final state particles are arranged into pure $U$-spin multiplets. 
\item[${(ii)}$]
The Hamiltonian contains only operators with one fixed value of $U$-spin and one type of Dirac structure.
\end{itemize}
These two assumptions are similar in nature. As we discuss in detail in Section~\ref{sec:universality}, from the $U$-spin point of view it does not matter if a multiplet belongs to a state or the Hamiltonian. Thus, these two assumptions are simply stating that the description of the $U$-spin set is given in terms of pure multiplets.

To put things into context, consider, for example, the $U$-spin limit Hamiltonian for charm decays
\begin{equation}\label{eq:Heff-charm}
    \mathcal{H}^{(0)}_{\text{eff, charm}} = f_{0,0}H^0_0 + \sum_{m = -1}^{m = 1} f_{1,m} H^1_m.
\end{equation}
In Eq.~\eqref{eq:Heff-charm}, $H^{0}_0$ is an operator for Singly-Cabibbo suppressed (SCS) decays
    \begin{equation} \label{H0-charm}
	    H^0_0 =  \frac{(\bar{u} s) (\bar{s} c)+(\bar{u} d) (\bar d c)}{\sqrt{2}}. 
    \end{equation}
Here, and in what follows, the Dirac structure is implicit. In terms of $U$-spin, $H^0_0$ is given by an antisymmetrized combination of two doublets:
    \begin{equation}\label{eq:0-u-spin-form}
        H_0^0 = \frac{\ket{+-} - \ket{-+}}{\sqrt{2}}.
\end{equation}
Similarly, the three operators that form the triplet in charm decay, $H^1$, 
are given by
    \begin{equation} \label{H1-charm}
        H^1_{1} =  (\bar{u} s) (\bar d c),\qquad
H^1_{-1}= -(\bar{u} d) (\bar s c), \qquad
H^1_0 =  {(\bar{u} s) (\bar s c)-(\bar{u} d) ( \bar d c)\over \sqrt{2}},
    \end{equation}
which can be written in terms of $U$-spin doublets as
    \begin{equation}\label{eq:1-u-spin-form}
        H^1_{1} = \ket{++},\qquad
H^1_{-1}= \ket{--}, \qquad
H^1_0 =  \frac{\ket{+-} + \ket{-+}}{\sqrt{2}}.
    \end{equation}
Here, $H^1_{1}$ is the Hamiltonian for doubly-Cabibbo suppressed (DCS) charm decays, and $H^1_{-1}$ is the one for Cabibbo-favored (CF) charm decays. $H_0^1$ is the CKM-leading part of the Hamiltonian for SCS charm decays, and $H^0_0$ is the corresponding CKM-suppressed part. 

The CKM-factors in Eq.~\eqref{eq:Heff-charm} are given by 
\begin{equation}\label{eq:charmCKM-f00}
f_{0,0} = \frac{V_{cs}^* V_{us} + V_{cd}^* V_{ud}}{2}\approx 0,
\end{equation}
\begin{equation}\label{eq:charmCKM-f1m}
f_{1,1} = V_{cd}^* V_{us}, \qquad 
f_{1,-1} = -V_{cs}^* V_{ud}, \qquad 
f_{1,0} = \frac{V_{cs}^* V_{us} - V_{cd}^* V_{ud}}{\sqrt{2}}\approx \sqrt{2}\,\left(V_{cs}^* V_{us}\right).
\end{equation}
The approximations used for $f_{0,0}$ and $f_{1,0}$ hold up to $O(\lambda^4)$, where $\lambda \approx 0.22$ is the Wolfenstein parameter.

In this example, due to the large ratio of CKM factors, we can neglect the singlet operator $H^0$ compared to the triplet $H^1$, resulting in the Hamiltonian of the form as in Eq.~\eqref{eq:H_LO_u}, that is
\begin{equation}
    \mathcal{H}^{(0)}_{\text{eff, charm}} \simeq \sum_{m = -1}^{m = 1} f_{1,m} H^1_m.
\end{equation}
Thus the Hamiltonian for the charm decays when the SCS part is neglected satisfies our assumptions. This example also explicitly demonstrates how arbitrary $U$-spin representations can be build from doublets, which is important for the discussion that follows.

The two assumptions (\emph{i}) and (\emph{ii}) could have different degrees of validity in different systems. In principle, when one aims at writing higher order sum rules for a $U$-spin system, the degree to which these assumptions are satisfied needs also to be compared with the breaking parameter.

In this work we focus on group-theoretical properties of $U$-spin systems and leave the questions related to the applicability of the results to physical systems for future study, giving only for illustration several amplitude level examples in Sec.~\ref{sec:gen_algo}. Thus, in what follows, we do not discuss much the validity of our assumptions.

\subsection{Expansion in $U$-spin breaking}\label{sec:expansion-in-b}

Our main focus in this work is the systematic expansion in $U$-spin breaking. We define $\varepsilon$ to be the breaking parameter. The concrete numerical value of $\varepsilon$ can vary depending on the process.
On the fundamental level, the breaking arises from the mass difference between the $s$ and $d$ quarks. The relevant terms in the Lagrangian  are $m_i q_i \bar q_i$ for $q_i=d,s$. Since the quarks form a doublet under $U$-spin, $q_i \bar q_i$ transforms as a sum of a singlet, $u=0,m=0$, and a triplet, $u=1,m=0$. The singlet respects the symmetry. It is the triplet that corresponds to $U$-spin breaking. Consequently, the breaking can be described by a spurion that transforms under
$U$-spin as an operator with  $u=1$ and $m=0$. We denote this operator as
$H_{\varepsilon}$. We stress that, being a spurion, $H_{\varepsilon}$
has a definite $m=0$, and while we treat it as a triplet, only
its $m=0$ component is present.

Terms of order $b$ in $U$-spin breaking are expected to be suppressed by $\varepsilon^b$. This terms are obtained from the leading order Hamiltonian, Eq.~\eqref{eq:H_LO_u}, by taking a tensor product with $b$ spurions leading to the full Hamiltonian at all orders of $U$-spin breaking
\begin{equation}\label{eq:Heff}
   \mathcal{H}_{\text{eff}} = \sum_{m,b} f_{u,m} \left(H^u_m \otimes H_\varepsilon^{\otimes b}\right)\,,
\end{equation}
where 
\begin{align}
H_{\varepsilon}^{\otimes b}&\equiv \underbrace{H_{\varepsilon} \otimes \dots \otimes H_{\varepsilon}}_{\text{$b$}}
\end{align}
is a tensor product of $b$ copies of $H_{\varepsilon}$. 
The resulting direct sum decomposition of the tensor product above has the following property
\begin{equation} \label{eq:b-parity}
    \left(1, 0\right)^{\otimes b} = \mathop{\oplus}_{j = 0}^{[b/2]} (b - 2j, 0)\,.
\end{equation}
That is, only the terms with total values of $U$-spin that have the same parity as the 
parity of $b$ appear in the decomposition.
For example, for $b=3$ the
decomposition of the tensor product into direct sum  contains only
$u=1$ and $u=3$ terms, but not $u=0$ nor $u=2$.

Moreover, the $u=1$ term from the $b=3$ case can be absorbed into the $u=1$
term from the $b=1$ case. That is, from the group theory point of view we
cannot tell if a breaking term with $u=1$ comes from $b=1$ or
$b=3$. This implies that the relevant new Hamiltonian structure at each order $b$ is given by the highest representation only. Therefore, when performing the sum rule analysis it is enough to only consider the $u=b,m=0$ term out of $H_{\varepsilon}^{\otimes b}$. This property is discussed formally and in more detail in 
Appendix~\ref{app:SR_counting_doublets}. 

We close this subsection with two remarks.
\begin{enumerate}
\item 
The fact that the breaking comes only from a $u=1$, $m=0$ spurion is important to the results that follow. 
\item
Note that while we concentrate on $U$-spin, also for isospin the symmetry-breaking operator is a $(1,0)$ spurion.
\end{enumerate}

\subsection{Decomposition in terms of reduced matrix elements} \label{sec:basis-transformation}

We define the \emph{physical basis of amplitudes} as a basis in which each particle in the initial and final state is represented by a component of a multiplet with definite value of $U$-spin, and the operators in the effective Hamiltonian are written as tensor products of operators from the $U$-spin limit Hamiltonian and possibly several insertions of the $U$-spin breaking spurion, see Eq.~\eqref{eq:Heff}.
We use $\A_j$ to denote amplitudes in the physical basis.

The physical basis of amplitudes $\A_j$ is to be contrasted with the \textit{$U$-spin basis of amplitudes}. The $U$-spin basis is defined as the basis in which the initial state, final state and all the terms in the Hamiltonian have definite values of total $U$-spin. For both bases, it is also natural to talk about the physical and $U$-spin basis for states and operators.

For formal details and definitions of physical and $U$-spin bases see Appendix~\ref{app:physical_vs_Uspin_basis}. Here we continue with a more schematic discussion.

The $U$-spin set is defined by listing the $U$-spin representations of the particles in the initial state, final state, and the Hamiltonian. Each amplitude from the $U$-spin set is defined by its specific set of $m$ QNs each corresponding to a component of a representation in the initial or final state. We use the index $j$ to enumerate the amplitudes of the $U$-spin set and, since all the amplitudes are defined by the sets of $m$ QNs, the index $j$ implicitly contains information about $m$. We assume that $\mathcal{H}_\text{eff}$ is known and can be written up to an arbitrary order of $U$-spin breaking using Eq.~\eqref{eq:Heff}. Each decay amplitude is then given by
\begin{equation} \label{eq:Aj-phys}
\A_j = {\mel{\text{out}}{\mathcal{H}_{\text{eff}}}{\text{in}}}_j.
\end{equation}

The amplitude in Eq.~\eqref{eq:Aj-phys} is an amplitude written in the physical basis. The rotation from the physical basis to the $U$-spin basis is
performed by the decomposition of the tensor products in $\ket{\text{in}}$,
$\ket{\text{out}}$, and $\mathcal{H}_\text{eff}$ into direct sums of irreducible representations such that each term in the resulting decomposition of each $\A_j$ has a definite value of $U$-spin in its initial state, final state, and the Hamiltonian.

The application of the Wigner-Eckart theorem to this decomposition allows to rewrite the amplitudes $\A_j$ in terms of the so called \emph{reduced matrix elements} (RME) $X_{\alpha}$:
\begin{equation}\label{eq:A_decomp}
    \A_j = \sum_{\alpha} c_{j \alpha} X_\alpha\,.
\end{equation}
where $\alpha$ is a multi-index defined in Eq.~\eqref{eq:def-alpha}. For formal derivation of Eq.~\eqref{eq:A_decomp} see Appendix~\ref{app:RMEdecomposition}. We emphasize that the RMEs do not depend on the $m$-QNs associated with the states nor on the
$m$-QNs of the operators in the Hamiltonian. 
The information on $m$ is contained in the coefficients $c_{j\alpha}$ only. 

In our convention, the $c_{j\alpha}$ are products of Clebsch-Gordan (CG) coefficients $C_{j\alpha}$ and weak interaction parameters,
\begin{equation}
c_{j\alpha} = C_{j\alpha} \, f_{u,m}\,,
\end{equation}
where $m$ is fully defined by the initial and final states of the amplitude $\mathcal{A}_j$. The value of $m$ in the Hamiltonian is given as the difference between the $m$ QNs of the final and the initial states. Only operators with this specific QN $m$ contribute to a given amplitude. 

We stress that in our notation, the RMEs, $X_\alpha$ are complex. Yet,
they contain
strong phases only. They do not include CKM matrix elements and thus
do not carry a weak phase. 
The factorization formula that follows from the Wigner-Eckart theorem is then given as
\beq
\A_j = \sum_\alpha C_{j\alpha}  f_{u,m} X_\alpha\,. \label{eq:factorization-gen}
\eeq
Note that, since $m$ is fixed by the initial and final state, it is not contained in the multi-index $\alpha$. Moreover, since in this work we consider only cases where only one $u$ is present in the
leading order Hamiltonian, we can factor out the CKM
factor $f_{u,m}$ and thus Eq.~\eqref{eq:factorization-gen} can be rewritten as
\beq
\A_j =  f_{u,m} \sum_\alpha C_{j\alpha}  X_\alpha\,. \label{eq:factorization}
\eeq
We learn that in the case under consideration the CKM dependence can be factored out.
That is, we can factorize the weak and strong physics. We emphasize that this factorization takes place due to our assumption that there is only one $U$-spin and one Dirac structure  in the Hamiltonian. The fact that in this case the CKM dependence factors out is important for the results that follow.
 
\subsection{Amplitude sum rules}\label{sec:sum-rules}

We introduce the notion of \emph{CKM-free amplitudes} which we denote
by $A_j$ and define as 
\begin{equation} \label{eq:def-ckm-free-amp}
    A_j = \frac{\mathcal{A}_j}{f_{u,m}}\,.
\end{equation}
Note the following regarding Eq.~\eqref{eq:def-ckm-free-amp}:
\begin{enumerate}
\item
In general, we can define CKM-free amplitudes only for 
cases where in the $U$-spin limit the Hamiltonian has only one specific
$u$ and one Dirac structure.
\item
We can define CKM-free amplitudes only for 
cases where the CKM factor, $f_{u,m}$, is not zero.
\end{enumerate}
It is point number two above that makes our result of limited use for isospin. In many cases charge conjugation 
implies $f_{u,m}=0$ for the case of isospin.

In terms of the CKM-free amplitudes the decomposition 
in Eq.~\eqref{eq:factorization} takes the form
\begin{equation}\label{eq:CKMfree_decomposition}
    A_j = \sum C_{j \alpha} X_\alpha\,.
\end{equation}
From this point on we are working with CKM-free amplitudes.

It is possible that some of the RMEs, $X_\alpha$, enter the
decompositions of the amplitudes from the $U$-spin set as fixed linear
combinations. Thus, we define $n_X^{(b)}$ to be the number of linearly
independent combinations of RMEs in the decompositions of the
amplitudes to an order of breaking $b$. 
Note the following:
\begin{itemize}
\item
The number of such linearly independent combinations 
is equal to the rank of the matrix of Clebsch-Gordan coefficients,
that is,  $n_X^{(b)} = \text{rank } [ C_{j\alpha} ]$.
(Recall that the multi-index $\alpha$ includes $b$.) Here, the notation \lq\lq{}$[\dots]$\rq\rq{} is used to represent a matrix that corresponds to an object with given indices.
\item
The number of linearly independent combinations of RMEs can only increase when we go to higher
order in the breaking, that is, $n_X^{(b)} \le n_X^{(b+1)}$.
\item
The maximum value of $n_X^{(b)}$ is the same as the number of amplitudes in the $U$-spin system, that is, $n_X^{(b)} \le n_A$.
\end{itemize}
In particular, we are interested in the case when
\begin{equation}
     n_X^{(b)}< n_A\,.
\end{equation}
In cases like this, there are sum rules between the amplitudes. That
is, there are algebraic relations between amplitudes of the form 
\begin{equation}
\sum w_j A_j = 0, \label{eq:sum-rule-form}
\end{equation}
where $w_j$ are numerical constants.

We define  $n_{\text{SR}}^{(b)}$ to be
the number of sum rules that are valid up to order $b$. 
It is given by
\begin{equation}
    n_{\text{SR}}^{(b)} = n_A -  n_X^{(b)} = n_A- \text{rank } [C_{j \alpha}]\,.
\end{equation}
The sum rules can be found as the null space of the matrix
$[C_{j\alpha}]^{T}$~\cite{Grossman:2013lya}. This treatment thus suggests that in order to find
amplitude sum rules for a given system one needs first to explicitly
find the matrix $[C_{j \alpha}]$ and then, if its rank is less than the
number of amplitudes, $n_A$, the null space of this matrix gives the desired sum rules.

\subsection{Universality of sum rules}\label{sec:universality}

Consider as an example three systems of amplitudes whose $U$-spin structure can be described as follows:
\begin{itemize}
    \item System I: A $U$-spin singlet in the initial state, two $U$-spin doublets in the final state, and the
      Hamiltonian is a triplet, i.e. 
    $\ket{i}\sim \ket{0}$,
    $\mathcal{H}\sim\ket{1}$,
    $\ket{f}\sim\ket{1/2}\otimes \ket{1/2}$.
    \item System II: A $U$-spin doublet in the initial state, a
      $U$-spin triplet in the final state, the Hamiltonian is a doublet, i.e. 
      $\ket{i}\sim\ket{1/2}$, 
      $\mathcal{H}\sim \ket{1/2}$,
      $\ket{f}\sim\ket{1}$.
    \item System III: $U$-spin singlets in the initial state and the Hamiltonian, four $U$-spin doublets in the final state, i.e. $\ket{i}\sim \ket{0}$, 
    $\mathcal{H}\sim\ket{0}$\,,
    $\ket{f}\sim \ket{1/2}\otimes \ket{1/2}\otimes \ket{1/2}\otimes \ket{1/2}$\,.
\end{itemize}
Let us first discuss systems I and II. In the standard approach systems I and II look different 
and require one to find and study the matrix $C_{j \alpha}$ for
each of them separately. 
However, from the point of view of $U$-spin symmetry, these systems are
identical: the sum rules for the two systems are the same. In Appendix~\ref{app:signs} we provide a formal proof of this statement and explain what it means that sum rules for two such different systems are identical. Basically, by identical we mean that 
there is  a one-to-one mapping between the amplitudes of the two systems and that the sum rules for the two are the same up to relative signs between amplitudes.

We define a \emph{universality class} as a collection of all the $U$-spin sets that are described by the same representations independently  if these representations belong to the initial state, final state, or the Hamiltonian. According to this definition, systems I and II belong to the same universality class. As we show in Appendix~\ref{app:signs}, it is always enough to study only one $U$-spin set from any universality class and then the results for the rest of the systems in the class can be obtained trivially. All universality classes contain a $U$-spin set such that the $U$-spin limit Hamiltonian and the initial state are singlets and all the representations belong to the final state. This is one of the cases that is the most straightforward to study and in our work we focus on it. That is, we first obtain the sum rules for this case and then generalize our results to any other system in the same universality class.

To get an intuitive  understanding of the universality of sum rules
we recall the concept of crossing symmetry. Let us focus on the $U$-spin structure of amplitudes. Crossing symmetry suggests that $U$-spin multiplets can be moved freely between initial states, final states, and the Hamiltonian and the relationships between amplitudes, that is, the sum rules, be preserved up to possibly relative signs between amplitudes. Here we put the Hamiltonian in one line with initial and final states since from the $U$-spin point of view the products $H^{u}_m \ket{u^\prime, m^\prime}$ and $ \ket{u,m} \otimes \ket{u^\prime, m^\prime}$ are identical. In more technical terms, when
we move multiplets between the initial state, final state, and the Hamiltonian, the only change that takes place is the flipping of a sign of the $m$-QNs of some of the multiplets. As a result, the two matrices $C_{j\alpha}$ for systems I and II might look different, but their
structure, that is the number of sum rules and their form (up to relative signs between amplitudes) is preserved. 

In addition to the relations between $U$-spin sets belonging to the same universality class, one can also establish relations between $U$-spin sets from different universality classes. For that recall that all higher $U$-spin representations can be obtained from the tensor product of $U$-spin doublets with proper symmetrization taken into account. 
This fact implies that all the sum rules for any system with arbitrary $U$-spin representations, can be obtained from the sum rules of a system that is made only from doublets. For example, systems~I and~II can be obtained from system~III via symmetrization of the tensor product of two out of four doublets. As a result the sum rules for systems~I and~II can be derived from the sum rules for system~III.

The above two observations allow us to view any $U$-spin system as a set of $U$-spin doublets where for some of them a symmetrization rule is specified. 
Thus, we can define the number $n$ of ``would-be'' doublets as the minimal number of doublets needed to describe a $U$-spin system. For  systems~I,~II, and III considered above, $n = 4$.

In our work we show that for an arbitrary $U$-spin set with the number of would-be doublets given by $n$, all the sum rules can be derived from the sum rules for a $U$-spin set described by $n$ doublets. Moreover, one never needs to explicitly find the matrix $C_{j \alpha}$ to write the sum rules and the $U$-spin symmetry ensures that the sum rules take a very simple form.

\section{Systems of $n$ doublets}\label{sec:n_doublet_system}

In the previous section we motivated the consideration of the $U$-spin set of processes with the following $U$-spin structure:
\begin{equation}\label{eq:only_doublets_process}
    0 \xrightarrow{\hspace{3pt} u = 0 \hspace{3pt}} \left(\frac{1}{2}\right)^{\otimes n},
\end{equation}
where the initial state is a $U$-spin singlet, the final state has $n$
doublets and the process is realized via a singlet operator in the
$U$-spin limit. The breaking of order $b$
is realized via the insertion of a tensor product of $b$ spurion
operators $H_\varepsilon^{\otimes b}$, as discussed in Section~\ref{sec:expansion-in-b}. 
Note that in order to have non-zero amplitudes $n$ must be even. At this point we focus on the $U$-spin structure of
processes, that is, we consider an abstract system of $U$-spin
doublets. We further assume that all the $U$-spin doublets are distinguishable. This assumption is motivated by the fact that the physical multiplets generally are distinguishable due to the additional momentum variables assigned to them, unless a specific kinematic region is studied.

\subsection{Amplitude $n$-tuples}\label{sec:An-tuples-doublets}
We consider a set of CKM-free amplitudes, $A_j$, that form a $U$-spin
set of the processes in Eq.~\eqref{eq:only_doublets_process}. We map any set of amplitudes $A_j$ onto a set of \emph{amplitude $n$-tuples} which we define as follows. We order
the $U$-spin doublets in an arbitrary (but defined) order and for each amplitude we represent the up components of the doublets as \lq\lq{}$+$\rq\rq{} and the down components as \lq\lq{}$-$\rq\rq{}.
The $n$-tuple representation of each process is then defined as a string of pluses and minuses for all the doublets according to the set order. 

Note the following: 
\begin{enumerate}
\item
The length of the amplitude $n$-tuple is $n$, which is even.
\item
The numbers of pluses \lq\lq{}$+$\rq\rq{} and minuses
\lq\lq{}$-$\rq\rq{} in the $n$-tuple are equal, that is there are $n/2$
of each.
\item
While we could write $n$-tuples where the numbers of \lq\lq{}$+$\rq\rq{} and \lq\lq{}$-$\rq\rq{} are not equal, the corresponding amplitudes vanish.
\item
In what follows we are using the terms $n$-tuple and amplitude interchangeably.
\end{enumerate}

As a bookkeeping device we assign numbers to amplitudes according to the binary code given by the $n$-tuple and use the assignment of 
\begin{itemize} 
\item \lq\lq{}$+$\rq\rq{} to \lq\lq{}one\rq\rq{}.
\item \lq\lq{}$-$\rq\rq{} to \lq\lq{}zero\rq\rq{}.
\end{itemize}
Everywhere in this paper, if not stated otherwise explicitly, we use the following index notation: 
\begin{itemize}
    \item $i$ takes values from $0, ..., 2^{n-1} - 1$.
    \item $\ell$ takes values from $2^{n-1}, ..., 2^n - 1$.
    \item $j$ and $k$ are used as generic indices.
\end{itemize}
Note that due to the constraint that the number of pluses and minuses in any $n$-tuples are equal, not all the values in the above ranges for $i$ and $\ell$ are used.

To demonstrate our notation we use the $n = 4$ case as an example. 
The non-vanishing $n$-tuples with corresponding indices $i$ and $\ell$
are given by 
\begin{equation}\label{eq:n4_example_notation}
\begin{gathered}
    A_3 = (-, -, +, +) \qquad A_{12} = (+, +, -, -)\\
    A_{5} = (-, +, -, +) \qquad A_{10} = (+, -, +, -)\\
    A_{6}=(-, +, +, -) \qquad A_{9} = (+, -, -, +)
\end{gathered}
\end{equation}
where the first column are the $A_i$ amplitudes with $i = 3, 5, 6$ and
the second column are the corresponding $A_\ell$ amplitudes with $\ell = 12, 10, 9$. The $U$-spin structure of any process is fully described by its
corresponding $n$-tuple.

\subsection{$U$-spin amplitude pairs}\label{sec:U-spin-amp-pairs}

We next define the notion of \emph{$U$-spin conjugation}. In physical
terms in our phase convention this operation is realized by simultaneously performing the following exchanges
\beq
s \leftrightarrow d, \qquad \bar s \leftrightarrow -\bar d.
\eeq
In the notation of $n$-tuples, $U$-spin conjugation
corresponds to a complete exchange between \lq\lq{}$+$\rq\rq{} and \lq\lq{}$-$\rq\rq{} for all the entries of the $n$-tuples. 
That is, the $U$-spin conjugation operator 
interchanges all the up and down components of all the $U$-spin
doublets. 

We call a pair of amplitudes that are $U$-spin
conjugate to each other a \emph{$U$-spin pair of amplitudes} or simply
\emph{$U$-spin pair}. In the example of
Eq.~\eqref{eq:n4_example_notation} the $U$-spin pairs are
\begin{equation}
    A_3 \text{ and } A_{12}, \qquad A_5 \text{ and } A_{10}, \qquad A_6 \text{ and } A_{9}.
\end{equation}
The relation between their indices is
\begin{equation} \label{eq:cong-ind}
    \ell = 2^n-1-i\,.
\end{equation}

Similarly to the case of amplitudes, we can use $n$-tuples to refer to
$U$-spin pairs. Since the number of $U$-spin pairs is half the number
of amplitudes
we adopt a convention of using the $n$-tuples that start with a minus sign to describe the $U$-spin pairs of amplitudes. In the example above we thus have the following correspondence between $U$-spin pairs and $n$-tuples:
\begin{equation}
\begin{gathered}
      A_3 \text{ and } A_{12}: \qquad \left(-, -, +, +\right),\\
      A_5 \text{ and } A_{10}: \qquad \left(-,+,-,+\right),\\
      A_6 \text{ and } A_{9}:\, \,  \qquad \left(-,+,+,-\right).
\end{gathered} \label{eq:notation-for-pairs}
\end{equation}

Our starting point in the analysis of this section is Eq.~\eqref{eq:CKMfree_decomposition} which we rewrite here for convenience for an amplitude $A_i$ 
\begin{equation}\label{eq:factorization-again}
    A_i = \sum_\alpha C_{i \alpha} X_\alpha \,.
\end{equation}
Recall that $C_{i \alpha}$ are products of CG coefficients. According to our result from Appendix~\ref{app:Upair_relation}, Eq.~\eqref{eq:Upair_decomp_theorem}, the decomposition of $A_\ell$ 
is given by 
\begin{equation}\label{eq:u-pair}
    A_\ell=   (-1)^{p} 
    \sum_\alpha (-1)^b C_{i \alpha} X_\alpha\,,
\end{equation}
where $C_{i\alpha}$ and $X_\alpha$ are the same sets that appear in
Eq.~\eqref{eq:factorization-again}, the relation between $\ell$ and $i$ is given in Eq.~\eqref{eq:cong-ind},
and
$p$ is an integer defined in Eq.~(\ref{eq:p_def}). 
Note that since we care only about the parity of $p$, adding multiples of two as well as multiplying $p$ by an overall sign does not matter: these operations leave $(-1)^p$ invariant.
For the 
case of $n$ doublets in the final state, and only singlets in the initial state and the Hamiltonian, we have
\begin{align}
p &= \frac{n}{2}\,. 
\end{align}
The order of the breaking of the RME $X_\alpha$ is denoted by
$b$. We emphasize that $b$ is included in the multi-index $\alpha$. 
We also recall that
$p$ is the same for all amplitudes from the $U$-spin set. 

We would also like to emphasize the following point regarding Eqs.~\eqref{eq:factorization-again} 
and~\eqref{eq:u-pair}. The relative sign between the terms in the decomposition of the $U$-spin pair amplitudes alternate with the order of the breaking $b$. For
example, for even $p$, there is no relative sign between terms at
$b=0$, there is a relative sign between the terms at $b=1$, no relative sign at $b=2$, and so on.

Before we go on and discuss the implications of these results we provide a simple intuition where the alternating minus signs come from. The reason is related to the fact that the $U$-spin breaking is realized by a spurion that transforms as $u=1$ and $m=0$. Writing the spurion in a matrix form using $a_{jk} = a_\alpha (\sigma_\alpha)_{jk}$, the spurion is proportional to $\sigma_3$, that is
\beq
    \begin{pmatrix} +1 & 0 \\ 0 & -1\end{pmatrix}\,. \label{eq:pauli-matrix}
    \eeq
We see that applying  the operation of $U$-spin conjugation, which in Eq.~(\ref{eq:pauli-matrix}) corresponds to $j \leftrightarrow k$, we gain a minus sign. Applying the spurion $b$ times, the resulting expression 
picks up a $(-1)^b$ under $U$-spin conjugation.
It is that property that makes the $U$-spin amplitude pairs simpler to work with.

\subsection{The $a$-type and $s$-type amplitudes}
We next define 
\beq\label{eq:as-comb-def}
a_i \equiv A_i- 
(-1)^p 
A_\ell , \qquad
s_i \equiv A_i + 
(-1)^p 
A_\ell,
\eeq
where $a_i$ are the \emph{a-type amplitudes} and are defined as the
anti-symmetric combinations of CKM-free amplitudes from a $U$-spin
pair, 
and $s_i$ are the \emph{s-type amplitude} and are defined as the symmetric
combinations of CKM-free amplitudes from a $U$-spin pair.
Note the following:
\begin{enumerate}
\item 
The $a$- and $s$-type amplitudes are eigenstates of $U$-spin conjugation. All $a$-type amplitudes of a given system have eigenvalue $(-1)^{p+1}$, while all $s$-type amplitudes have eigenvalue $(-1)^p$.
\item
The factor $(-1)^p$ in Eq.~(\ref{eq:as-comb-def}) cancels the same factor in Eq.~(\ref{eq:u-pair}). Thus, when writing the $a-$and $s-$type amplitudes in terms of the RMEs they do not depend on $p$
\beq \label{eq:a-s-rme}
a_i = \sum_\alpha \Big(1-(-1)^b\Big)C_{i \alpha} X_\alpha \,, \qquad
s_i = \sum_\alpha \Big(1+(-1)^b\Big)C_{i \alpha} X_\alpha \,. \eeq
\end{enumerate}

The definitions of the $a$-type and $s$-type amplitudes turn out to be
particularly helpful when constructing  sum rules. First, we note that
working with $a$- and $s$-type amplitudes instead of using $A_i$ and $A_\ell$ is merely a change of
basis. Thus, 
all the sum rules can be expressed in terms of $a_i$ and
$s_i$. In the following we present most of the results in the basis of the
$a$- and $s$-type amplitudes.

Using Eq.~\eqref{eq:a-s-rme}
we find the following properties: 
\begin{itemize}
\item
$a_i$ only contains terms that are odd in the  breaking $b$,
\item
$s_i$ only contains terms that are even in the breaking $b$.
\end{itemize}
One important direct result of the above is that the $a$- and $s$-type
amplitudes are fully decoupled from each other. Each of them contains
different sets of $X_\alpha$: the $a$-type amplitudes can be written as
a sum of
$X_\alpha$ with $b$ odd, while  the $s$-type amplitudes can be written as
a sum of
$X_\alpha$ with $b$ even. Thus, we can write sum rules for any system such that each sum rule involves only $a$-type or $s$-type amplitudes. We call these sum rules \emph{$a$-type sum rules} and \emph{$s$-type sum rules}, respectively.

Based on the above properties we also note the following: 
\begin{enumerate}
\item
To leading order, that is, for $b=0$,
\begin{equation}\label{eq:a-SR-LO}
    a_i = 0.
\end{equation}
This result provides us with $n/2$ sum rules and exhausts the set of
linearly independent $a$-type sum rules that are valid at the leading
order. We also call this type of sum rule \lq\lq{}trivial\rq\rq{}. 
These leading order $a$-type sum rules were pointed out before,
see Refs.~\cite{Gronau:2000zy, Fleischer:1999pa, Gronau:2000md,Jung:2009pb}. 
\item
For even $b$: Any $s$-type sum rule that is valid to order $b$ 
is also valid to order $b+1$.
\item
For odd $b$: Any $a$-type sum rule that is valid to order $b$ is also valid to order $b+1$.
\end{enumerate}
In particular, we learn that $s$-type sum rules that are valid at 
leading order also hold at the first order of breaking.

\subsection{Amplitude sum rules}\label{sec:amp-sum-rules}

In this subsection we show how to obtain all the sum rules for a given
system of $n$ $U$-spin doublets. 
The subsection is based on the results derived in Appendices~\ref{app:signs}
and~\ref{app:ThII_doublets}.
  
Employing the notation for $U$-spin pairs introduced in Eq.~(\ref{eq:notation-for-pairs}), we define $S_j^{(k)}$ to be a subset of all $U$-spin pairs that share $k$ minuses at the same position of their $n$-tuple representation. 
Given that there could be more than
one such subset, we also attach an index $j$ to them.
In the example of $n = 4$ the $U$-spin pairs are given explicitly in Eq.~(\ref{eq:notation-for-pairs}) and there are consequently four such subsets: three with $k=2$ and one with $k = 1$:
\begin{equation} \label{eq:S-defs-ex}
\begin{gathered}
    S_1^{(2)} = \{\left(-,-,+,+\right)\}, \hspace{25pt} S_2^{(2)} = \{\left(-,+,-,+\right)\}, \hspace{25pt} S_3^{(2)} = \{\left(-,+,+,-\right)\},\\
    S_1^{(1)} = \{\left(-,-,+,+\right), \left(-,+,-,+\right), \left(-,+,+,-\right)\}.
\end{gathered}
\end{equation}
The numbering scheme is arbitrary. Note that the subsets can overlap.

We further define $S^{(k)} \equiv \{S_j^{(k)}\}$ to be the set of all $S_j^{(k)}$ that
share the same $k$.
In the example of $n = 4$ there are two such sets of subsets 
\begin{equation}
\label{eq:neq4sets}
S^{(2)} = \{S_1^{(2)}, S_2^{(2)}, S_3^{(2)}\}, \qquad
S^{(1)} = \{S_1^{(1)}\},
\end{equation}
where we used the definitions of Eq.~\eqref{eq:S-defs-ex}.

We define $n_S^{(k)}$ to be the number of subsets $S_j^{(k)}$ in the set $S^{(k)}$. It is given by
\begin{equation}\label{eq:n_S_k}
    n_S^{(k)} = \binom{n-1}{k-1}.
\end{equation}
This result is a consequence of the first entry being fixed to be a minus sign.
Therefore only $k-1$ minus signs are left to be picked from the $n-1$ entries of the $n$-tuples. In the $n=4$ example we have 
\beq
n_S^{(2)}=\binom{3}{1}=3,
\qquad
n_S^{(1)}=\binom{3}{0}=1,
\eeq
in agreement with the explicit counting in Eq.~(\ref{eq:neq4sets}).

All subsets $S^{(k)}_j$ from a set $S^{(k)}$ consist of the same number of amplitude pairs. This number, $n_A^{(k)}$, is given by
\begin{equation}\label{eq:nA-k}
    n_A^{(k)} = \binom{n-k}{{n}/{2}-k}\,.
\end{equation}
In the $n=4$ example we have
\beq
n_A^{(2)}=\binom{3}{0}=1, \qquad
n_A^{(1)}=\binom{3}{1}=3, 
\eeq
in agreement with the explicit counting in Eq.~(\ref{eq:S-defs-ex}).

We are now ready to write all the sum rules. As we show in Appendix~\ref{app:ThII_doublets}, each $S^{(k)}_j$ corresponds to a different sum rule. The correspondence is as follows
\begin{itemize}
    \item 
For $S^{(k)}_j$ with
$b=n/2-k$ even, the sum rule is given by
\begin{equation}\label{eq:sym-form-of-SR-a}
    \sum_{a_i \in S^{(k)}_j} a_i = 0.
\end{equation}
\item
For $S^{(k)}_j$ with
$b=n/2-k$ odd, the sum rule is given by
\begin{equation}\label{eq:sym-form-of-SR-s}
    \sum_{s_i \in S^{(k)}_j} s_i = 0.
\end{equation}
\end{itemize}

Note the following regarding Eqs.~\eqref{eq:sym-form-of-SR-a} and \eqref{eq:sym-form-of-SR-s}:

\begin{enumerate}
\item
Each of the sum rules in Eqs.~(\ref{eq:sym-form-of-SR-a}) and ~(\ref{eq:sym-form-of-SR-s})
is valid up to order $b=n/2-k$.

\item
Each of the sum rules in Eqs.~(\ref{eq:sym-form-of-SR-a}) and ~(\ref{eq:sym-form-of-SR-s}) are broken at order $b+1$. 
\item
There could be, however, linear combinations of these sum rules that are not broken at order $b+1$. For example, the sum rules that  correspond to subsets $S_j^{(k - 2)}$ are the linear combinations of the sum rules that  correspond to subsets $S_j^{(k)}$that are valid up to order $b$. 
\item As we show at the end of Appendix~\ref{app:SR_counting_doublets}, the maximum order of breaking for which there is still a sum rule in the system is $b_\text{max} = n/2-1$. 
\item 
At  $b_\text{max}$ there is only one sum rule. The type of this sum rule alternates with the value of $b_\text{max}$. It is an $a$-type sum rule for odd
${n}/{2}$, that is even $b_\text{max}$. It is an $s$-type sum rule
for even ${n}/{2}$, that is odd $b_\text{max}$. This sum rule is given simply by a sum of all $a$-type or $s$-type amplitudes in the system.
\end{enumerate}

Given the correspondence between $S^{(k)}_j$ and the sum rules, we can find an algorithm to generate all the sum rules that are valid to order $b$ (with $b \le n/2 - 1$).
The algorithm is as follows. \begin{itemize}
    \item For $b$ even: use all $S^{(k)}_j$ with
$k=n/2-b$ to generate the relevant $a$-type sum rules. Then use $k=n/2-b-1$ to generate the relevant $s$-type sum rules. 
    \item For $b$ odd: use all $S^{(k)}_j$ with
$k=n/2-b$ to generate the relevant $s$-type sum rules. Then use $k=n/2-b-1$ to generate the relevant $a$-type sum rules. 
\end{itemize}

Table~\ref{tab:sum-rules-rule} in Appendix~\ref{app:ThII_doublets}
summarizes the counting and the form of sum
rules for $a$- and $s$-type amplitudes at even and odd
orders of breaking. One can readily use this table to write the sum
rules for a system of $n$ doublets in the final state at any order of
breaking.

In the $n=4$ example, the sum rules that are valid to $b=0$ come from $S_j^{(2)}$ and they are
\beq \label{eq:n-4-a}
a_3=0, \qquad a_5=0, \qquad a_6=0.
\eeq
The sum rule for 
$b=1$ corresponds to $S_1^{(1)}$ and it is given by 
\beq \label{eq:n-4-s}
s_3+s_5+s_6=0.
\eeq

\subsection{An example, $n=6$}

To demonstrate how the described algorithm works, we consider a $U$-spin set of processes such that the initial state and the Hamiltonian are $U$-spin singlets and the final state contains $n = 6$ doublets. Our goal is to use the algorithm to obtain all the sum rules for this system at any order of breaking.

The amplitudes for the system under consideration and the corresponding $n$-tuples are listed in Table~\ref{tab:n6-n-tuples}. The resulting sum rules at different orders of breaking and the counting of sum rules are summarized in Table~\ref{tab:sum-rules-n6-summary}. For the counting of sum rules we use the results derived in Appendix~\ref{app:SR_counting_doublets}, see Eq.~\eqref{eq:nSR_doublets}.
\begin{table}[t]
\centering
$\begin{array}{cc}
  & \\
 a_7, s_7 & (-,-,-,+,+,+)\\
 a_{11}, s_{11} & (-,-,+,-,+,+)\\
 a_{13}, s_{13} & (-,-,+,+,-,+)\\
 a_{14}, s_{14} & (-,-,+,+,+,-)\\
 a_{19}, s_{19} & (-,+,-,-,+,+)\\
\end{array}
\hspace{25pt}
\begin{array}{cc}
  & \\
 a_{21}, s_{21} & (-,+,-,+,-,+)\\
 a_{22}, s_{22} & (-,+,-,+,+,-)\\
 a_{25}, s_{25} & (-,+,+,-,-,+)\\
 a_{26}, s_{26} & (-,+,+,-,+,-)\\
 a_{28}, s_{28} & (-,+,+,+,-,-)\\
\end{array}$
\caption{$a$- and $s$-type amplitudes and their corresponding $n$-tuples for the case $n = 6$. The numbering scheme is described in Section~\ref{sec:An-tuples-doublets}.}\label{tab:n6-n-tuples}
\end{table}

\begin{table}[h!]
\centering
\begin{tabular}{|c|c|}
\hline
\multicolumn{2}{|c|}{6 doublets}\\
\hline
$a$-type & $s$-type \\
\hline
\multicolumn{2}{|c|}{$b = 0$, $n_{SR}^{(b = 0)} = 15$}\\
\hline
 $n_{SR-a}^{(b = 0)} = 10$ & $n_{SR-s}^{(b = 0)} = 5$\\
 \(\displaystyle ~~a_7 = a_{11} = a_{13} = a_{14}= a_{19} = a_{21} =\) ~~& \(\displaystyle s_7 + s_{11} + s_{13} + s_{14} = 0 \)\\
 \(\displaystyle = a_{22} = a_{25} = a_{26} = a_{28} = 0\) & \(\displaystyle s_7 + s_{19} + s_{21} + s_{22} = 0 \) \\
   & \(\displaystyle s_{11} + s_{19} + s_{25} + s_{26} = 0 \)\\
   & \(\displaystyle s_{13} + s_{21} + s_{25} + s_{28} = 0 \)\\
   & ~~\(\displaystyle s_{14} + s_{22} + s_{26} + s_{28} = 0 ~~\)\\
 \hline
\multicolumn{2}{|c|}{$b = 1$, $n_{SR}^{(b = 1)} = 6$}\\
\hline
$n_{SR-a}^{(b = 1)} = 1$ & $n_{SR-s}^{(b = 1)} = 5$\\
 \(\displaystyle a_7 + a_{11} + a_{13} + a_{14} +  a_{19} + a_{21} +\) & 
 same as for $b = 0$\\
 \(\displaystyle + a_{22} + a_{25} + a_{26} + a_{28} = 0\) & 
 \\
\hline
\multicolumn{2}{|c|}{$b = 2$, $n_{SR}^{(b = 2)} = 1$}\\
\hline
$n_{SR-a}^{(b = 2)} = 1$ & $n_{SR-s}^{(b = 2)} = 0$\\
 same as for $b = 1$ & no sum rules\\
 \hline
 \multicolumn{2}{|c|}{$b \ge 3$, $n_{SR}^{(b\ge3)} = 0$, no sum rules}\\
 \hline
\end{tabular}
\caption{Sum rules for a system of $n=6$ $U$-spin
  doublets in the final state. $n_{SR}^{(b)}$, $n_{SR-a}^{(b)}$ and $n_{SR-s}^{(b)}$ denote the total number of sum rules, the number of $a$-type sum rules and the number of $s$-type sum rules that are valid at the order of breaking $b$ respectively.\label{tab:sum-rules-n6-summary}}

\end{table}

In the symmetry limit, $b = 0$, there are $n_{SR}^{(b=0)} = 15$
sum rules between amplitudes of the system, see Eq.~\eqref{eq:nSR_doublets}.  Using $k=n/2-b$ we see that $k=3$ for $b=0$.
Out of the $15$ sum rules $n_S^{(3)} = 10$ are $a$-type sum rules, where each sum rule is a sum over $a$-type amplitudes for $U$-spin pairs in subsets $S^{(3)}_j$, with index $j = 1, 2, ..., 10$. The $S^{(3)}_j$ are defined via $n$-tuples that share at least $3$ minus signs at the same positions. According to Eq.~\eqref{eq:nA-k}, each such subset contains exactly one amplitude pair, that is $n^{(3)}_A = 1$. Thus the sum rules are trivial $a$-type sum rules that we already mentioned above, see Eq.~\eqref{eq:a-SR-LO}. The remaining 5 sum rules that hold at the leading order are $s$-type sum rules. They are given by the sums of $s$-type amplitudes over the subsets $S_j^{(2)}$, with $j = 1, 2, ..., 5$, which are defined via $n$-tuples that share at least $2$ minus signs at the same positions. There are $n_A^{(2)} = 4$ amplitude pairs in each such subset.

At the first order of breaking $b = 1$, there are $n_{SR}^{(b = 1)} = 6$ sum
rules. We see that 9 out of 15 leading order sum rules are broken by the order
$b=1$ corrections.
We know that moving from $b=0$ to $b=1$ can only break $a$-type sum rules.
Thus we learn that there is one $a$-type sum rule that is valid to
$b=1$, and thus must also be valid to $b=2$. This sum rule corresponds to $S_1^{(1)}$ and it is given by the
sum of all 10~$a$-type amplitudes of the system. 
Note, that this sum rule is also valid at the leading order, and can be expressed as a linear combination of the 10 sum rules that corresponds to $S^{(3)}$. 
The 5 $s$-type sum rules that were valid at $b=0$ are also valid at $b=1$, since corrections of odd orders $b$ do not contribute to $s$-type amplitudes.

Finally, we discuss the highest order of breaking at which there are
still sum rules in the system, $b = 2$. At this order there is only
one sum rule which is the $a$-type sum rule given by the sum of all $a$-type amplitudes of the $U$-spin set.
For $b=2$ all $s$-type sum rules are broken.

\subsection{Geometrical picture}\label{sec:halves-lattice}

In this section we provide the geometrical interpretation of Eqs.~\eqref{eq:sym-form-of-SR-a} and \eqref{eq:sym-form-of-SR-s}.

To proceed with the geometrical picture we first introduce a new 
notation for the $U$-spin pairs. We call this notation a {\it
 coordinate notation} (the reason for that name becomes clear below).
All the $U$-spin pairs can be one to one
mapped onto $n$-tuples starting with a minus sign. 
We then enumerate the elements of such $n$-tuples starting from $0$. That is, the first element of the $n$-tuple is assigned the index $0$, the second element is assigned the index $1$ and so on, up until the last element of an $n$-tuple is assigned the index $n-1$. In the coordinate notation  
any $n$-tuple is labeled by a string of ${n}/{2}-1$ numbers that
indicate the positions of the minuses in the $n$-tuple excluding the
first minus.
For example, in the coordinate notation
\begin{equation}
(\underset{0}{-},\underset{1}{-},\underset{2}{-},\underset{3}{+},\underset{4}{-},\underset{5}{+},\underset{6}{+},\underset{7}{+}) = \left(1,2,4\right).
\end{equation}
Note, that all permutations of the numerical labels denote the same $n$-tuple and thus the same $U$-spin pair, that is
\begin{equation}\label{eq:example-permutations-lattice-points}
    \left(1,2,4\right) = \left(1,4,2\right) = \left(2,1,4\right) = \left(2,4,1\right) = \left(4,1,2\right) = \left(4,2,1\right).
\end{equation}

With this notation in mind, the $U$-spin pairs can be represented by
nodes in a lattice with dimension
\begin{align}
d = \frac{n}{2}-1\,.
\end{align} 
The coordinates of the nodes are given by the numbers in the coordinate notation. Hence the name \lq\lq{}coordinate notation\rq\rq{}.
Note the following:
\begin{enumerate}
\item
Due to the permutation
symmetry, each $U$-spin pair appears in the lattice $d!$ times. 
\item
Not all the lattice points represent valid $U$-spin pairs. 
Nodes with any number of repeating
coordinates, for example, $\left(1,1,2\right)$ or $\left(1,1,1\right)$
do not represent any $U$-spin pair from the $U$-spin set under
consideration. Thus we exclude them from the lattice.
\item
The length of each dimension in the lattice is $n-1$.
\end{enumerate}

According to the algorithm described in the previous section, sum rules
are given by the sums of $a$-type or $s$-type amplitudes with $n$-tuples that
share a certain number of minuses at the same positions. Below we
consider a case when for a chosen order of breaking and a type of sum
rules the number of shared minuses is given by $k$.

Eqs.~\eqref{eq:sym-form-of-SR-a} and~\eqref{eq:sym-form-of-SR-s} can then be represented as follows. For a
given $k$, the sum rules correspond to the sums of all the nodes 
that share $k-1$ coordinates. That is, a $d - (k - 1) = b$ dimensional
subspace of the lattice with dimension $d$.  
For example, for $k=n/2$, $b = 0$ all the points give
the $a$-type sum-rules that are valid to $b=0$. For $k=n/2-1$, $b = 1$ all the lines
give us the $s$-type sum rules that are valid up to $b=1$.

Below we summarize the steps that are needed in order to write all the
sum rules for a $U$-spin system described by $n$ doublets in the geometrical
picture. We then work out explicitly the $n = 6$ case that is discussed
in the previous section.

For any $U$-spin system described by $n$ doublets we build a lattice according to the following rules:
\begin{enumerate}
\item 
The dimension of the lattice is $d = {n}/{2} - 1$.
\item 
Each node of the lattice is described by $d$ numbers (coordinates). If
all the numbers are different, that node represents a $U$-spin pair. Otherwise, it is removed from the lattice. In practice we just replace them by zeros for bookkeeping.
\item
Once the lattice is built we are ready to ``harvest'' the sum
rules. For this we consider all the $b$-dimensional subspaces of the lattice defined as follows. All the nodes that share $d-b = n/2-1-b$ coordinates form a $b$-dimensional subspace of interest. For $b=0,1,2$ the subspaces are given by nodes, lines, and planes, respectively. Sums of nodes of the lattice lying in such $b$ dimensional subspaces 
correspond to sum rules that are valid to order $b$ and are broken by corrections of order $b+1$. For even(odd) $b$, the $b$-dimensional subspaces of the lattice correspond to $a$($s$)-type sum rules.
\item
In order to get all the sum rules that are valid to order $b$, we need to combine those that come from the $b$ and $b+1$ dimensional subspaces.
\end{enumerate}

\begin{figure}[t]
\centering
\includegraphics[width=0.3\textwidth]{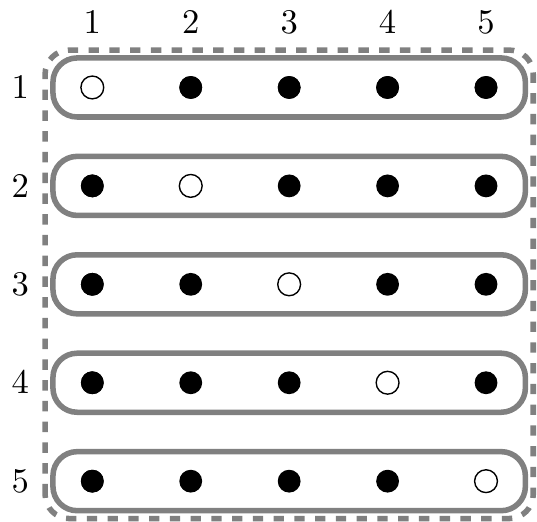}
\caption{The lattice for a system of $n = 6$ doublets. The black nodes
correspond to valid $U$-spin pairs, while the white nodes on the
diagonal are forbidden and do not describe any $U$-spin pair.
Harvesting of the sum rules is performed as follows. The black nodes
give the leading order, that is, $b = 0$, $a$-type sum rules. Sums of
nodes grouped by solid lines give the $b = 1$, $s$-type sum rules. The
sum of nodes within the dashed line gives the $a$-type sum rule that
is valid up to $b = 2$. Note that we do not explicitly show the
identical $b = 1$ sum rules that are obtained by vertical instead of
horizontal solid lines.}
\label{fig:n6-lattice}
\end{figure}

As an example consider the $n=6$ case. In that case the lattice is two-dimensional and thus it can be easily visualized, see Fig.~\ref{fig:n6-lattice}. 
The black nodes describe valid nodes of the lattice, while the white
nodes on the diagonal are forbidden nodes and do not correspond to any
$U$-spin pair.

The leading order $a$-type sum rules correspond to $b=0$ and thus are given by the nodes of the lattice (i.e.~the black nodes in the diagram in Fig.~\ref{fig:n6-lattice}). There are $10$ such sum rules:
\begin{equation}\label{eq:example-lattice-n6-LO-a-SR}
    a_{(1,2)} = a_{(1,3)} = a_{(1,4)} = a_{(1,5)} = a_{(2,3)} = a_{(2,4)} = a_{(2,5)} = a_{(3,4)} = a_{(3,5)} = a_{(4,5)} = 0,
\end{equation}
where the notation $a_{(x_1, x_2)}$ is used to represent an
  $a$-type combination of $U$-spin pair amplitudes with the coordinate notation
  $\left(x_1, x_2\right)$. 
In Eq.~\eqref{eq:example-lattice-n6-LO-a-SR} we also used the fact that nodes of the lattice related by permutations of coordinates represent the same $U$-spin pairs. We have chosen to use orderings of coordinates such that $x_1 < x_2$.

The $s$-type sum rules that are valid up to order $b = 1$ are given by lines. That is, they are given by the sums of lattice nodes in rows or columns of the lattice. These sum rules are represented by solid lines in Fig.~\ref{fig:n6-lattice}. We have
\begin{equation}\label{eq:example-lattice-SR}
\begin{gathered}
s_{\left(1,2\right)} + s_{\left(1,3\right)} + s_{\left(1,4\right)} + s_{\left(1,5\right)} = 0,\\
s_{\left(1,2\right)} + s_{\left(2,3\right)} + s_{\left(2,4\right)} + s_{\left(2,5\right)} = 0,\\
s_{\left(1,3\right)} + s_{\left(2,3\right)} + s_{\left(3,4\right)} + s_{\left(3,5\right)} = 0,\\
s_{\left(1,4\right)} + s_{\left(2,4\right)} + s_{\left(3,4\right)} + s_{\left(4,5\right)} = 0,\\
s_{\left(1,5\right)} + s_{\left(2,5\right)} + s_{\left(3,5\right)} + s_{\left(4,5\right)} = 0,
\end{gathered}
\end{equation}
where $s_{\left(x_1, x_2\right)}$ is used to represent an $s$-type combination of $U$-spin pair amplitudes with the coordinate notation $\left(x_1, x_2\right)$ and we use again the convention $x_1 < x_2$. 

Finally, the $a$-type sum rules that are valid up to order 
$b=2$ are represented by planes. In the case that we consider here there is just one
plane that corresponds to the sum of all the nodes of the lattice. This sum rule is represented by the dashed line in Fig.~\ref{fig:n6-lattice}:
\begin{equation}\label{eq:example-n6-NLO-a-SR}
     a_{(1,2)} + a_{(1,3)} + a_{(1,4)} + a_{(1,5)} + a_{(2,3)} + a_{(2,4)} + a_{(2,5)} + a_{(3,4)} + a_{(3,5)} + a_{(4,5)} = 0.
\end{equation}
To be explicit, because of $a_{(i,j)} = a_{(j,i)}$ we have also, completely equivalent to Eq.~(\ref{eq:example-n6-NLO-a-SR}),
\begin{align}
  a_{(2,1)} + a_{(3,1)} + a_{(4,1)} + a_{(5,1)} + a_{(3,2)} + a_{(4,2)} + a_{(5,2)} + a_{(4,3)} + a_{(5,3)} + a_{(5,4)} = 0\,.
\end{align}
Note, therefore, that Eq.~(\ref{eq:example-n6-NLO-a-SR}) really corresponds to a sum over the complete plane as shown in Fig.~\ref{fig:n6-lattice}, only that the corresponding factors of two resulting from $a_{(i,j)} = a_{(j,i)}$ have already been cancelled when writing Eq.~(\ref{eq:example-n6-NLO-a-SR}).

As we see the lattice representation reproduces the sum rules 
listed in Table~\ref{tab:sum-rules-n6-summary}.

We have determined the above sum rules also using the traditional method using Clebsch-Gordan coefficient tables and indeed, both methods give the same results, as they should.

\subsection{Generalization: doublets not only in the final state}

The proof of Eqs.~\eqref{eq:sym-form-of-SR-a} and~\eqref{eq:sym-form-of-SR-s} is given in
Appendix~\ref{app:ThII_doublets} for the case where all of the $n$
doublets are in the
final state. Here we generalize the result to the rest of the systems in the universality class, that is, to the cases where some of
the doublets are in the initial state and/or in the
Hamiltonian.

The generalization of the result can be done in two steps. First, we need to introduce a modified convention for constructing $n$-tuples. Second, we need to slightly modify the definitions of the $a$- and $s$-type amplitudes given in Eq.~(\ref{eq:as-comb-def}).

For $n$-tuples the convention becomes as follows. We order the doublets in an arbitrary but defined order. For the doublets that belong to the final state, we represent the upper component of a doublet as \lq\lq{}$+$\rq\rq{} and the lower component as \lq\lq{}$-$\rq\rq{}. The convention is, however, different for the doublets belonging to the initial state and the Hamiltonian. For them we represent the upper component of a doublet by \lq\lq{}$-$\rq\rq{} and the lower component by \lq\lq{}$+$\rq\rq{}. Note that with this convention all the $U$-spin sets in the same universality class, that is, with the same number of doublets, are described by the same sets of $n$-tuples.

To modify the definitions of the $a$- and $s$-type sum rules we use the results derived in Appendix~\ref{app:signs} where it is shown that the form of sum rules given in Eqs.~\eqref{eq:sym-form-of-SR-a} and \eqref{eq:sym-form-of-SR-s} is preserved if we define the $a$- and
$s$-type amplitudes with an overall factor of $(-1)^{q_i}$ as follows
\begin{equation}\label{eq:as-comb-def-app}
    a_i \equiv (-1)^{q_i} \left( A_i- (-1)^p A_\ell \right), \qquad
    s_i \equiv (-1)^{q_i} \left( A_i + (-1)^p A_\ell \right).
\end{equation}
The factor $(-1)^{q_i}$ is equal to $+1$ if the final state has an even number of minus signs and it is equal to $-1$ if the final state has an odd number of minus signs. More formally, the factor can be written as
\begin{equation}\label{eq:q-def}
(-1)^{q_i} = \prod_{j=1}^{n_f} \left(-1\right)^{{1/2} - m_j^{(i)}},
\end{equation}
where the $m_j^{(i)}$ are the $m$ QNs of the $n_f$ final
state doublets of the $A_i$ amplitude. 

The expression in Eq.~\eqref{eq:q-def} is identical to the product of the elements of the $n$-tuple that corresponds to the final state of the process with label $i$. Note that for the case where all the doublets are in the final state all the amplitudes gain the same factor $(-1)^{n/2}$ and thus it is cancelled in the sum rules.

We see that the factor $(-1)^{q_i}$ assigned to a specific $n$-tuple depends on the ordering chosen when defining the $n$-tuples. Particularly, it depends on the assignment of certain positions in the $n$-tuples to the initial state, final state, and the Hamiltonian. For example, in the $n=4$ system that has only one doublet in the final state, the $n$-tuple $(-,-,+,+)$ is assigned $(-1)^{q_i} = -1(+1)$ if the convention is such that the first (third) position of the $n$-tuple corresponds to the final state. This is a basis choice. 

The fact that the $(-1)^{q_i}$ factors depend on the basis might appear to be in contradiction with the general idea that sum rules are basis independent, but it is not. The solution to the seeming contradiction lies in the fact that our definitions of the $a$- and $s$-type amplitudes given in Eq.~\eqref{eq:as-comb-def-app} are in fact basis dependent. The two CKM-free amplitudes $A_i$ and $A_\ell$ form a $U$-spin pair no matter what is the basis. The assignment of the index $i$ to one of them and index $\ell$ to another however, is basis dependent. Thus, the sign with which the $a$- and $s$-type amplitudes enter the sum rules depends on the choice of the ordering of doublets that was used when defining the $n$-tuples. When we write the sum rules in terms of amplitudes, however, they do not depend on the ordering, as it should be.

\section{Systems with arbitrary representations}\label{sec:gen-arbitrary-irreps}

In this section we generalize the results obtained for processes described exclusively by $U$-spin doublets to the case when a system has arbitrary representations in its description. As in the previous section, we assume that all the irreps are distinguishable. Note that we use the terms \lq\lq{}representation\rq\rq{} and \lq\lq{}irrep\rq\rq{} interchangeably.

First we focus on the systems of arbitrary representations in the final state. We generalize the definition of the $n$-tuple, and then we show how the sum rules for an arbitrary $U$-spin system can be obtained from the sum rules for a system of doublets. 

Next, we generalize our geometric method in order to account for symmetrizations. This works in cases when there is at least one doublet in the system. When no doublets are present, we can use the lattice method for an auxiliary system with a doublet and then perform one additional step of symmetrization afterwards.

We close by generalizing the results of this section to the cases when $U$-spin representations also appear in the initial state and the Hamiltonian.

\subsection{Generalized $n$-tuples}\label{sec:n-tuples-generalized}

To generalize the definition of $n$-tuples introduced in
Section~\ref{sec:An-tuples-doublets} to the case of arbitrary irreps,
we consider a system with $r$ irreps $u_j$, where
$u_j$ is a positive integer or half-integer and $j = 0,1,...,r-1$. Note that only non-trivial irreps are included in this list, $U$-spin singlets are not relevant for sum rules. The addition of $U$-spin singlets to a system does not change the sum rules. As in Section~\ref{sec:n_doublet_system}, we first focus on systems that contain $U$-spin representations only in the final state.
Since each of the irreps $u_j$ can be built from $2 u_j$ doublets, we conclude that the complete system
can be constructed from combining $n$ doublets, where $n$ is given by
\begin{align}\label{eq:would-be-n-def}
 n = 2 \sum_{j=0}^{r-1} u_j\,.
\end{align}
Note that the number of would-be doublets does not change when an arbitrary number of $U$-spin singlets is added to the system. Furthermore, the order of the irreps is arbitrary, yet, in practice we usually sort them and assign 
$j=0$ to the lowest irrep. 

Consider an arbitrary multiplet $u_j$. The minimum number of doublets needed to construct the representation $u_j$ is given by $n_j = 2u_j$. A component of the multiplet $u_j$ with the $m$-type QN $m_j$ is denoted as $\ket{u_j; m_j}$ and can be represented as a string of $n_j$  pluses and minuses such that the corresponding total $m$-QN is equal to $m_j$. Thus for the component $\ket{u_j; m_j}$ we have $u_j -m_j$ minus signs and $u_j +m_j$ plus signs. The ordering of the signs is, in principle, arbitrary.
We adopt a convention in which we order the signs starting with all minuses. Thus for the component $\ket{u_j; m_j}$ of an arbitrary multiplet $u_j$ in the final state we write
\begin{equation}\label{eq:uj_m-comvention}
    \ket{u_j; m_j} \, :\,\, ( \underbrace{-\,...\,-}_{u_j -m_j}\,\underbrace{+\,...\,+}_{u_j + m_j})\,.
\end{equation}
For example, in the case of $u_j=3/2$ in the final state we have the following correspondence:
\begin{align}
\ket{\frac{3}{2},\frac{3}{2}}\hspace{9pt}:  &\hspace{10pt} (+++)\,, &
\ket{\frac{3}{2},\frac{1}{2}}\hspace{9pt}:  &\hspace{10pt} (-++)\,,\nonumber \\
\ket{\frac{3}{2},-\frac{1}{2}}: &\hspace{10pt} (--+)\,, &
\ket{\frac{3}{2},-\frac{3}{2}}: &\hspace{10pt} (---)\,.
\end{align}

We can represent any amplitude of an arbitrary $U$-spin system of $n$ would-be doublets via $n$ signs in the $n$-tuple. We do this by
letting the first $n_0$ signs of the $n$-tuple to represent a component of irrep $u_0$, the following $n_1$ signs to represent a component of irrep $u_1$ and so on. For the components of different irreps we use the convention in Eq.~\eqref{eq:uj_m-comvention}. We  separate the positions of the $n$-tuple describing different irreps by a comma, as we did in the case of only doublets in the system. We call such $n$-tuples \emph{generalized $n$-tuples} or, when there is no ambiguity, we refer to them simply as $n$-tuples.

For example, the following generalized $n$-tuple 
\begin{equation}\label{eq:gen-n-tuple-example}
 A_{11} =\left(-,-+,- ++\right),
\end{equation}
represents an amplitude from the $U$-spin system described by representations $u_0=1/2$, $u_1=1$ and $u_2=3/2$. If all three representations belong to the final state, this specific amplitude has $m_0 = -\frac{1}{2}$, $m_1 = 0$, and $m_2 = \frac{1}{2}$. Note that generalized $n$-tuples that correspond to valid amplitudes must have the same number of pluses and minuses, as it is in the case for doublets-only systems.

Similarly to the case of systems described exclusively by doublets, we define $U$-spin conjugation for generic systems described by arbitrary representations. The $U$-spin conjugation in the generic case is defined as flipping the signs of all the $m$-QNs of multiplet components describing an amplitude. In terms of generalized $n$-tuples the operation corresponds to
\begin{enumerate}
\item
Exchange of all plus and minus signs.
\item 
Reordering each set of signs corresponding to one representation such that it starts with minuses.
\end{enumerate}
For example, the $U$-spin conjugate of the amplitude $A_{11}$ given in Eq.~\eqref{eq:gen-n-tuple-example} is
\begin{equation}
 A_{41}=\left(+,-+,- -+\right).
\end{equation}
Note
that Eq.~\eqref{eq:cong-ind} still holds, that is, for $n=6$, $\ell = 2^6-i-1$, which for $i=11$ gives $\ell=52$ and implies that $A_{52}$ is the $U$-spin conjugate of $A_{11}$. This is not in contradiction to the above since $A_{41}\equiv A_{52}$.

As in the case of doublets-only systems, an amplitude and its $U$-spin conjugate amplitude form a $U$-spin pair. To represent the pair we use the $n$-tuple for which the first non-zero $m$-QN is negative. For the $U$-spin pairs we can also define the $a$- and $s$-types amplitudes the same way as in Eq.~\eqref{eq:as-comb-def}, with $p$ given in Eq.~\eqref{eq:p_def}.

There is, however, one subtlety that needs to be discussed when the system is described exclusively by integer representations. In this case there is one amplitude in each system which is $U$-spin self-conjugate, that is, it is its own $U$-spin conjugate. This amplitude is the one where the $m$-QNs of all the irreps are zero. 
Consider, for example, the system of two triplets. This system has $n=4$ would-be doublets. The amplitude with the following $n$-tuple is present in the $U$-spin set
\begin{equation}\label{eq:000}
    A_{5}=\left( -+, -+\right)\,,
\end{equation}
and it is its own $U$-spin conjugate since $m = 0$ for both multiplets. Another way to see it is to note that the would-be conjugate amplitude is $A_{10}$ and in this case $A_5 \equiv A_{10}$.

For amplitudes that are $U$-spin self-conjugates, which is possible only when all the irreps are integers, one out of the $a$-type or the $s$-type amplitudes identically vanishes.
Which of the two vanishes depends on the parity of $p$, where for the case of integer-only irreps $p=n/2$. 
For even $p$ we have  
\beq\label{eq:self-conj-n/2-even}
s_{j} = 2 A_{j}, \qquad a_{j}\equiv 0,
\eeq
while for odd $p$ 
we have
\beq\label{eq:self-conj-n/2-odd}
a_{j} = 2 A_{j}, \qquad s_{j}\equiv 0,
\eeq
where $j$ represents the index of the amplitude that is self-conjugate. We emphasize that in this case $a_{j} \equiv 0$ (for even $p$) and $s_j \equiv 0$ (for odd $p$) are identities and not sum rules.

\subsection{Symmetrization}
\label{sec:sym}
Given the fact that all higher representations can be constructed from doublets, we move to showing how to derive the sum rules for any generic system based on the underlying system of $n$ doublets.

The key idea for this is to perform a change of basis. This is similar to what we did when we talked about the rotation between the physical and the $U$-spin basis.

\subsubsection{An example}

To begin, let us consider an example of a system of $n$ would-be doublets in the final state. We denote the amplitudes of this system as $A^{(d)}(m_1, m_2, \dots, m_n)$, where the label $(d)$ indicates that the amplitude belongs to a system of doublets and $m_j$, $j=1,\dots,n$ are the specific $m$-type QNs of these doublets that describe the amplitude. We assume we have used the algorithm described in Section~\ref{sec:n_doublet_system} and found all the sum rules for this system. Then we can use this result to write the sum rules for a system of $n-2$ doublets and a triplet. We recall that
\begin{equation}\label{eq:1/2times1/2}
    \frac{1}{2}\otimes \frac{1}{2} = 0 \oplus 1
\end{equation}
and perform the basis rotation for the last two doublets of the system of doublets according to Eq.~\eqref{eq:1/2times1/2}. The result is as follows
\begin{align}\label{eq:doubleds_and_1_decomp}
    A^{(d)}(m_1, m_2, \dots, m_n) = \mathop{C_{1/2, m_{n-1}}}_{\hspace{0pt} 1/2, m_n}^{\hspace{-14pt}1, M} A^{(1)}(m_1, m_2, \dots,m_{n-2},M) \nonumber \\
    + \mathop{C_{1/2, m_{n-1}}}_{\hspace{0pt} 1/2, m_n}^{\hspace{-14pt}0, M} A^{(0)}(m_1, m_2, \dots,m_{n-2},M).
\end{align}
In Eq.~\eqref{eq:doubleds_and_1_decomp} we introduced the notation $A^{(1)}(m_1, m_2, \dots,m_{n-2},M)$ for amplitudes that pick up the triplet component from the tensor product of two doublets. 

Our notation is such that the label $(1)$ stands for triplet, $m_j$, $j = 1, \dots, n-2$ are the $m$-QNs of the $n-2$ doublets and $M = m_{n-1}+m_n$ is the $m$-QN of the triplet. Similarly, the amplitudes $A^{(0)}(m_1,m_2,\dots,m_{n-2},M)$ denote the amplitudes that pick up the singlet component from the tensor product, hence the label $(0)$. Note, that the second term in Eq.~\eqref{eq:doubleds_and_1_decomp} is present only if $M=0$.

Once we performed the basis rotation, the next step in writing the sum rules for the $U$-spin set of interest is to take the sum rules for the system of doublets and plug in the expressions in Eq.~\eqref{eq:doubleds_and_1_decomp}. This allows us to rewrite the sum rules of the system of doublets in terms of the $A^{(1)}$ and $A^{(0)}$ amplitudes. After we do this, we need to rearrange the amplitudes such that we obtain the sum rules that only involve the amplitudes $A^{(1)}$. 
As we show in Appendix~\ref{app:mu-factor}, it is guaranteed that the sum rules for $A^{(1)}$ and $A^{(0)}$ decouple. The decoupling also means that instead of using the full expression in the RHS of Eq.~\ref{eq:doubleds_and_1_decomp} we can simply do the substitution
\begin{equation}\label{eq:substitution-example}
    A^{(d)}(m_1, m_2, \dots, m_n) \, \rightarrow \, \mathop{C_{1/2, m_{n-1}}}_{\hspace{0pt} 1/2, m_n}^{\hspace{-14pt}1, M} A^{(1)}(m_1, m_2, \dots,m_{n-2},M).
\end{equation}
The sum rules that we obtain after this substitution give the full set of sum rules for the system of $n-2$ doublets and a triplet.

\subsubsection{Generalization}

Above we have considered a simple example of a system of many doublets and a triplet. The result in Eq.~\eqref{eq:substitution-example} can be generalized to the case of a system of arbitrary representations. 

Consider a system of $r$ irreps $u_0, u_1, \dots, u_{r-1}$. 
Each of the representations that are obtained from a symmetrization is the highest possible representation in the tensor product of $2 u_j$ would-be doublets. In this case, as shown in Appendix~\ref{app:mu-factor}, the substitution analogous to the one in Eq.~\eqref{eq:substitution-example} takes the following form
\begin{equation}\label{eq:symmetrization_gen}
    A^{(d)}(m_1, m_2, \dots, m_n) \, \rightarrow \, \left(\prod_{j=0}^{r-1}  \frac{1}{\sqrt{C_\text{sym}(u_j,M_j)}} \right) A(M_0, M_1, \dots, M_{r-1}),
\end{equation}
where we use $A(M_0, M_1, \dots, M_{r-1})$ to represent the amplitudes of the system described by representations $u_0, u_1, \dots, u_{r-1}$, and $M_j$, $j=0,\dots,r-1$ are the $m$-type QNs of the representations describing the amplitude. 
Note that the symmetry factors $C_\text{sym}(u_j, M_j)$ do not depend on $m_j$ but only on $u_j$ and $M_j$.
Furthermore, $C_\text{sym}(u_j, M_j)$ can be written in terms of binomial coefficients as follows
\begin{equation}\label{eq:C_sym}
    C_\text{sym}(u_j,M_j) = C(2u_j, u_j-M_j) \equiv \binom{2u_j}{u_j-M_j}\,,
\end{equation}
see Appendix~\ref{app:mu-factor} for details.

\subsubsection{Iterative approach}

On the fundamental level the symmetry factors in Eq.~\eqref{eq:C_sym} come from the products of the relevant Clebsch-Gordan coefficients. In some cases it becomes important that we understand how the symmetry factors are build iteratively.

Assume we know the sum rules for a system of representations $u_0, u_1, \dots, u_{r-1}$, where $u_0 = 1/2$, and the rest of the irreps are arbitrary. We denote the amplitudes of this system as $A^{(1/2, u_1)}(m_0,m_1,\dots,m_{r-1})$. We want, however, to obtain the sum rules for a system of representations $u_+, u_2, \dots, u_{r-1}$, where $u_+ = u_1 + 1/2$. We denote the amplitudes of the later system as $A^{(+)}(m, m_2, \dots, m_{r-1})$. 

As above, we are building the higher representation as a component of the tensor product of doublets. When the construction is done iteratively we can focus on taking a tensor product of the arbitrary representation $u_1$ with the representation $u_0=1/2$. Using 
\beq
u_1 \otimes 1/2 = u_+ \oplus u_-, \qquad u_\pm = u_1 \pm 1/2,
\eeq
we write for the amplitudes
\begin{align}\label{eq:iterative_decompose}
    A^{(1/2, u_1)}(m_0,m_1,\dots,m_{r-1}) = C_+ A^{(+)}(m, m_2, \dots, m_{r-1})\nonumber \\ + C_- A^{(-)}(m, m_2, \dots, m_{r-1})\,,
\end{align}
where $m = m_1 + m_2$ and $C_+$ and $C_-$ are the appropriate CG coefficients 
\beq
C_+ = \mathop{C_{u_1, m_1}}_{\hspace{10pt} 1/2, m_0}^{\hspace{6pt}u_+, m}\, ,
\qquad
C_- = \mathop{C_{u_1, m_1}}_{\hspace{10pt} 1/2, m_0}^{\hspace{6pt}u_-, m}\,.
\eeq
Note that the coefficients $C_+$ and $C_-$ depend on the $m$-QNs and thus are different for different amplitudes.

Once we have the decomposition in Eq.~\eqref{eq:iterative_decompose}, we can plug it into the sum rules for the system with representations $u_0 = 1/2$ and $u_1$ and obtain the sum rules for the system of interest that is described by $u_+$. 
Using the fact that the sum rules for the two types of amplitudes $A^{(+)}$ and $A^{(-)}$ decouple, see Appendix~\ref{app:mu-factor}, we can just perform the substitution
\begin{equation}\label{eq:symmetrization-iterative}
    A^{(1/2, u_1)}(m_0,m_1,\dots,m_{r-1}) \, \rightarrow \, C_+ A^{(+)}(m, m_2, \dots, m_{r-1}).
\end{equation}

If instead of combing the representations $1/2$ and $u_1$, we consider two arbitrary representations $u_0$ and $u_1$ from which we want to obtain the representation $u_+=u_0 + u_1$, we need to perform the following substitution
\begin{equation}\label{eq:substitution-iterative-gen}
    A^{(u_0, u_1)}(m_0,m_1,\dots,m_{r-1}) \, \rightarrow \, C_+ A^{(+)}(m, m_2, \dots, m_{r-1}),
\end{equation}
where the coefficient $C_+$ needs to be modified as follows 
\begin{equation}
    C_+ = \mathop{C_{u_1, m_1}}_{\hspace{10pt} u_0, m_0}^{\hspace{6pt}u_+, m}.
\end{equation}

\subsubsection{Summary of the symmetrization process}

We refer to the substitutions described in Eqs.~\eqref{eq:substitution-example},~\eqref{eq:symmetrization_gen},~\eqref{eq:symmetrization-iterative}, and \eqref{eq:substitution-iterative-gen} as \emph{symmetrization}. When building higher representations we pick the highest components of the tensor products, which are totally symmetric, hence the name of the procedure. Note that when the symmetrization is performed the number $n$ of would-be doublets of the system stays the same.

In practice the task that we encounter is as follows. We are given the sum rules for a system of representations $u_0, u_1,\dots, u_{r-1}$. We refer to this system as \lq\lq{}original system\rq\rq{}. We want to obtain the sum rules for a system where some of the representations are symmetrized. We refer to the latter system as \lq\lq{}new system\rq\rq{}.

Assuming that $u_0, u_1,\dots, u_{r-1}$ are ordered such that the representations that we symmetrize are grouped together. In order to solve the problem at hand we proceed with the following two-step algorithm:
\begin{enumerate}
\item 
Replace the $n$-tuples of the original system with $n$-tuples for the new system. For this take the $n$-tuples of the original system and remove all the commas between the components that correspond to the irreps that we symmetrize. Then rearrange the signs such that in entries corresponding to individual irreps minuses precede the pluses.
\item Take the sum rules written in terms of the new $n$-tuples and multiply each of the $n$-tuples with an appropriate symmetry factor.
\end{enumerate}
Note that we can perform the above for any $n$-tuple, that is, individual amplitudes $A_j$ as well as the $a$- and $s$- type amplitudes. For the case of the $a$- and $s$-type amplitude we can do it as long as $u_0$ is a doublet and it is not a part of the symmetrization process.

\subsubsection{Example: symmetrization of systems with $n=4$ would-be doublets} \label{sec:exam-n-4}

To demonstrate the idea of symmetrization we consider three different systems with $n=4$ would-be doublets. For simplicity we consider the case when all representations are in the final state.
\begin{itemize}
\item System I: 4 doublets, that is $u_0=u_1=u_2=u_3=1/2$. The six amplitudes of this system
are listed in Eq.~\eqref{eq:n4_example_notation}, and we rewrite them below with a slightly different notation where we add a superindex to indicate that an amplitude belongs to system~I
\begin{equation}\label{eq:n4_example_notation-again}
\begin{gathered}
A_3^{(\text{I})} = (-, -, +, +)\,, \qquad A_{12}^{(\text{I})} = (+, +, -, -)\,,\\
A_{5}^{(\text{I})} = (-, +, -, +)\,, \qquad A_{10}^{(\text{I})} = (+, -, +, -)\,, \\
    A_{6}^{(\text{I})}=(-, +, +, -)\,, \qquad A_{9}^{(\text{I})} = (+, -, -, +)\,.
\end{gathered}
\end{equation}
The sum rules for this system in terms of $a$- and $s$-type amplitudes are given in Eqs.~\eqref{eq:n-4-a} and \eqref{eq:n-4-s}.
\item System II: 2 doublets and a triplet, which we order such that $u_0=u_1=1/2$, $u_2 = 1$. The amplitudes are given by
\begin{equation}\label{eq:n4_example_notation-again-2}
\begin{gathered}
    A_3^{(\text{II})} = (-, -, + +)\,, \qquad A_{12}^{(\text{II})} = (+, +, - -)\,,\\
    A_{5}^{(\text{II})} = (-, +, - +)\,, \qquad A_{10}^{(\text{II})} = (+, -, -+)\,.
\end{gathered}
\end{equation}
\item 
System III: 2 triplets, that is $u_0=u_1=1$. In this case the amplitudes are
\beq
\begin{gathered}
A_3^{(\text{III})} = (--,++)\,, \qquad  A_{12}^{(\text{III})}= (++,--)\,, \\
A_5^{(\text{III})}= (-+,-+)\,. \qquad
\phantom{A_9^{(\text{III})}= (-+,-+)} \end{gathered}
\eeq
Note that since $A_5^{(\text{III})}$ is a $U$-spin self conjugate amplitude, we have $s_5^{(\text{III})} \equiv 2 A_5^{(\text{III})}$ and $a_5^{(\text{III})}\equiv 0$.
\end{itemize}

We start by obtaining the sum rules of system II given the sum rules for system I. We do this by performing the steps outlined in the previous subsection. We only show it for one amplitude out of each $U$-spin pair. We have the following replacements
\begin{align}
A_3^{(\text{I})} &= (-, -, +, +) \to (-,-,++) = A_3^{(\text{II})}\,, \nn\\
A_{5}^{(\text{I})} &= (-, +, -, +) \to {1 \over \sqrt{2}}(-, +, -+) = {1 \over \sqrt{2}} A_{5}^{(\text{II})}\,, \label{eq:sub-rule} \\
A_{6}^{(\text{I})} &= (-, +, +,-) \to {1 \over \sqrt{2}}(-, +, -+) = {1 \over \sqrt{2}} A_{5}^{(\text{II})}\,. \nn
\end{align}
Note the following regarding Eq.~\eqref{eq:sub-rule}
\begin{enumerate}
\item The symmetry factor for amplitude $A_3^{(\text{II})}$ is 1 and we do not write it explicitly.
\item 
The symmetry factor for amplitudes $A_5^{(\text{II})}$ and $A_6^{(\text{II})}$ is ${1/ \sqrt{2}}$.
\item
In the case of the $n$-tuple for amplitude $A_6^{(\text{I})}$ we had to rearrange the signs after dropping the comma, and thus it corresponds to $A_5^{(\text{II})}$.
\end{enumerate}

Similar substitutions work for the $a$ and
$s$-type amplitudes, that is
\beq  \label{eq:sub-rule-a-s}
\begin{gathered}
a_3^{(\text{I})} \to a_3^{(\text{II})}, \qquad
a_5^{(\text{I})} \to \frac{a_5^{(\text{II})}}{\sqrt{2}}, \qquad 
a_{6}^{(\text{I})} \to \frac{a_{5}^{(\text{II})}}{\sqrt{2}}, \\
s_3^{(\text{I})} \to s_3^{(\text{II})}, \qquad
s_5^{(\text{I})} \to \frac{s_5^{(\text{II})}}{\sqrt{2}}, \qquad 
s_{6}^{(\text{I})} \to \frac{s_{5}^{(\text{II})}}{\sqrt{2}}.
\end{gathered}
\eeq
Substituting Eq.~\eqref{eq:sub-rule} into the $a$-type sum rules of Eq.~\eqref{eq:n-4-a},
we obtain
\beq  \label{eq:n-4-t1-a}
a_3^{(\text{II})}= 0 ,\qquad
a_5^{(\text{II})} = 0, 
\eeq
where the last equation appears twice.
Substituting Eq.~\eqref{eq:sub-rule} into the $s$-type sum rules of Eq.~\eqref{eq:n-4-s},
we obtain
\beq \label{eq:n-4-t1-s}
s_3^{(\text{II})} + {1 \over \sqrt{2}} s_5^{(\text{II})} + {1 \over \sqrt{2}} s_5^{(\text{II})} 
=  s_3^{(\text{II})} + \sqrt{2} s_5^{(\text{II})} = 0.
\eeq

Next, we derive the sum rules for system III from the sum rules for system II. For that we symmetrize the first two doublets and perform the following substitutions for the amplitudes:
\begin{equation}\label{eq:sub-rule-2}
\begin{aligned}
&A_3^{(\text{II})} = (-, -, ++) \to (--,++) = A_3^{(\text{III})}\,, \\
&A_{5}^{(\text{II})} = (-, +, -+) \to {1 \over \sqrt{2}}(-+, -+) = {1 \over \sqrt{2}} A_{5}^{(\text{III})} .
\end{aligned}
\end{equation}
In terms of the $a$- and $s$- type amplitude we have
\beq  
\begin{aligned}
a_3^{(\text{II})} &\to a_3^{(\text{III})}\,, &\qquad s_{3}^{(\text{II})} &\to s_{3}^{(\text{III})}\,, \\
a_5^{(\text{II})} &\to a_5^{(\text{III})} \equiv 0\,, &\qquad 
s_{5}^{(\text{II})} &\to \frac{s_{5}^{(\text{III})}}{\sqrt{2}}\,.
\end{aligned}
\eeq
Recall that $a_5^{(\text{III})}\equiv 0$ identically.

From the sum rules in Eqs.~\eqref{eq:n-4-t1-a} and 
\eqref{eq:n-4-t1-s} we obtain
\beq
a_3^{(\text{III})} = 0, \qquad
s_3^{(\text{III})} + s_5^{(\text{III})} = 0,
\eeq
where the first one is valid to zeroth order in the breaking and the second one to first order.
Writing the above in terms of amplitudes we have
\beq
A_3^{(\text{III})}= 
A_{12}^{(\text{III})}, \qquad
A_3^{(\text{III})}+
A_{12}^{(\text{III})}+ 2 A_5^{(\text{III})}=0.
\eeq

We have demonstrated how to obtain the sum rules for all the different $n=4$ systems from the system of 4 doublets.

\subsection{Generalization of the geometrical picture}
\label{sec:gen_1d}

We move to the discussion of the case of systems with at least one doublet. In this case, we can define a lattice in a way similar to the doublets-only case. Then, we can harvest the sum rules directly from the lattice without the need to perform the symmetrization explicitly. 

\subsubsection{Generalized coordinate notation}

We start by generalizing the coordinate notation introduced in
Section~\ref{sec:halves-lattice} to the case that we discuss here. We order the multiplets such that the first one is a doublet, that is $u_0 = {1}/{2}$.
We then label every $n$-tuple by a string of ${n}/{2} - 1$
numbers as follows. 
Out of the $n/2$ minus signs in the $n$-tuple we ignore the first one and for each of the rest we write the index $j$ of the irrep $u_j$ that it belongs to. For example,
\begin{equation}\label{eq:arbitrary-rep-notation}
\begin{gathered}
    (\underset{0}{-},\underset{1}{--},\underset{2}{-+},\underset{3}{-++++}) = \left(1,1,2,3\right),\\
    (\underset{0}{-},\underset{1}{-+},\underset{2}{-+},\underset{3}{--+++}) = \left(1,2,3,3\right),\\
    (\underset{0}{-},\underset{1}{++},\underset{2}{-+},\underset{3}{---++} ) = \left(2,3,3,3\right).
 \end{gathered}
\end{equation}
As in Section~\ref{sec:halves-lattice} the order of indices is
unimportant, that is,
all permutations describe the same amplitude. 
With this generalized notation we see that similarly to the case of
doublets only, also for the case of arbitrary irreps, $U$-spin pairs can
be represented as nodes of a $d = n/2 - 1$ dimensional lattice.
The length of each dimension of the lattice is $r-1$. 
Recall that $r$ denotes the number of irreps in the system. (Note that in the case of all doublets we have $r = n$ and the length of each dimension is $r-1 = n-1$.) The lattice is built by assigning each $U$-spin pair to a node of the lattice based on its coordinate notation.

We finish the construction of the lattice by assigning a multiplication factor to each node of the lattice. These factors account for the symmetrization process, which we then do not need to perform explicitly, and are denoted as $\mu$-factors. We show in the next subsection how the $\mu$-factors are calculated.

Once the lattice is built and the proper $\mu$-factors are assigned, the sum rules can be harvested from the lattice in a similar  way to the way they are harvested in the case of doublets-only systems.

\subsubsection{The $\mu$-factor}

To write the explicit expressions for the $\mu$-factors we introduce yet another auxiliary notation for $U$-spin pairs. We call this notation the $y_j$ \emph{notation}. In this notation each amplitude is described by $r-1$ numbers $\left[y_1, y_2, ..., y_{r-1}\right]$, where $y_j$ is the number of times that representation $u_i$ enters the coordinate description of the amplitude pair. Square brackets are used to distinguish the $y_j$ notation from the coordinate notation. Equivalently, $y_j$ is the number of minuses in the $n$-tuple at positions that correspond to the representation $u_j$. For example, the amplitudes in Eq.~\eqref{eq:arbitrary-rep-notation} are denoted by
\begin{equation}
\begin{gathered}
    (\underset{0}{-},\underset{1}{--},\underset{2}{-+},\underset{3}{-++++}) = (1,1,2,3) = [2,1,1] ,\\
    (\underset{0}{-},\underset{1}{-+},\underset{2}{-+},\underset{3}{--+++}) = (1,2,3,3) = [1,1,2] ,\\
    (\underset{0}{-},\underset{1}{++},\underset{2}{-+},\underset{3}{---++} ) = (2,3,3,3) = [0,1,3] .
 \end{gathered}
\end{equation}
Note that also here we have omitted the very first minus sign.

We denote the $\mu$-factor that corresponds to a certain node as $\mu[y_1,y_2,\dots,y_r]$. There are two sources that contribute to the $\mu$-factors. The first  is the symmetry factors introduced in Section~\ref{sec:sym}. The second comes from the fact that several amplitude pairs of the underlying doublets-only system may correspond to only one pair of the system under consideration. 

We write the $\mu$-factor for a node  as a product of factors corresponding to each of the $r$ representations
\beq\label{eq:mu_product}
\mu[y_1,y_2,...,y_{r-1}]=\prod_{j=0}^{r-1} \mu_j,
\eeq
where $\mu_j$ depends only on $u_j$ and $y_j$. In Appendix~\ref{app:mu-factor} we show that
\begin{equation}\label{eq:mu_j}
    \mu_j = \sqrt{C(2u_j,y_j)}\times y_j!\,, \qquad C(2u_j,y_j) = \binom{2u_j}{y_j},
\end{equation}
where $C(2u_j, y_j)$ is a binomial coefficient. 

Note that $C(2u_k,2u_k) = 1$ and for $y_k > 2u_k$, $C(2u_k,y_k)=0$. This means that for the doublets-only systems the $\mu$-factors are equal to $1$ for the allowed nodes and to $0$ for the nodes that do not correspond to a valid amplitude. This is a generalization of the empty and filled nodes in the lattices of Section~\ref{sec:n_doublet_system}.

\subsubsection{Harvesting the sum rules and the multiplication factor}

Once we constructed the lattice with the associated $\mu$-factors we are ready to harvest the sum rules. For order $b$, the sum rules correspond to the different sums over all the $b$-dimensional lattice subspaces. For even(odd) $b$ these sums correspond to $a$-type($s$-type) sum rules.

There is one subtle point that arises for $b \ge 2$. In that case some of the off-diagonal nodes are redundant. For example, for the two-dimensional case in coordinate notation we have $(x_{1}, x_{2}) \equiv (x_2, x_1)$. Thus, when we sum all the nodes in the lattice, the amplitudes that correspond to the off-diagonal nodes enter the sum rules more then once.

While one can manually collect all the identical nodes, we can also do it in the following, more straightforward way:
\begin{enumerate}
\item
Sum over subspaces without duplicating nodes. That is consider only nodes with $x_{1} \le x_{2} \le... \le x_{d}$.
\item
When harvesting a sum rule each node is multiplied by a corresponding $\mu$-factor and a multiplication factor $M_b$ that accounts for the redundancy of the lattice. Note that unlike the $\mu$ factor, $M_b$ depends on $b$.
\end{enumerate}

To calculate $M_b$, we account for the symmetry of the lattice by counting the number of nodes that correspond to the given amplitude in the $b$-dimensional subspace. To write an expression for this number, recall that the $b$-dimensional subspaces over which we sum the amplitudes, are defined via fixing certain coordinates in the coordinate notation. We define 
\begin{equation}\label{eq:yjprime-def}
    y_j = y_j^{\text{fix}} + y_j^{\prime},
\end{equation}
where $y_j^{\text{fix}}$ is the number of fixed coordinates that are equal to $j$ and thus correspond to representation $u_j$, and $y_j^\prime$ is the number of coordinates that can take value $j$ among the $b$ ``free'' coordinates that define the subspace. Recall that $y_j$ is the number of coordinates of the node that are equal to $j$. Using this notation, the number of the lattice nodes in the $b$-dimensional subspace that describe the same amplitude is given by the following multinomial coefficient
\begin{equation}
M_b[y_1^\prime, y_2^\prime, \dots, y_{r-1}^\prime] = \binom{b}{y_{1}^\prime, y_{2}^\prime, \dots, y_{r-1}^\prime} = \frac{b!}{y_{1}^\prime! y_{2}^\prime! \dots y_{r-1}^\prime!}.
\end{equation}
Note that the sum of all $y_j^\prime$ is equal to $b$
\begin{equation}
    \sum_{j=1}^{r-1} {y_j^\prime}=b.
\end{equation}

We see that any $a$($s$)-type sum rule that corresponds to a $b$-dimensional subspace of the lattice can be found as a weighted sum over all the amplitudes that correspond to the nodes in the subspace (each amplitude accounted for only once). The weights are denoted by $W_b$ and they are the product of the $\mu$ and $M_b$ factors. They are given by
\begin{equation}\label{eq:Wb-def}
W_b[y_1, y_2, \dots, y_{r-1}] = \mu[y_1, y_2, \dots, y_{r-1}] \times M_b[y_1^\prime, y_2^\prime, \dots, y_{r-1}^\prime] =  b! \prod_{j=1}^{r-1} \frac{\mu_j}{y_j^\prime!}.
\end{equation}
Note that the $M_b[y_1^\prime, y_2^\prime, \dots, y_{r-1}^\prime]$ depend on the amplitude and the subspace,~i.e.,~for the same $b$ they may be different for different sum rules.

\subsubsection{An example}\label{sec:gen-examples}

\begin{figure}[t]
\centering
\includegraphics[width=0.2\textwidth]{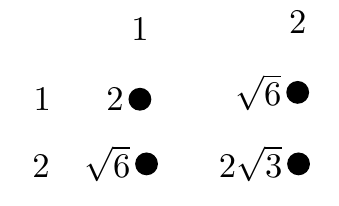}
\caption{Lattice for a system with $n=6$, $u_0=1/2$, $u_1 = 1$, $u_2 = 3/2$.}
\label{fig:n6-1d-1t-32}
\end{figure}

As an example we consider the case of $n=6$, $r=3$ with $u_0=1/2$, $u_1=1$, and $u_2=3/2$. For this system $d = n/2-1= 2$ and thus the lattice is two-dimensional. Each of the dimensions has a length of $r-1=2$. There are three $U$-spin pairs that we write below in the $n$-tuple, index, coordinate, and  $y_j$ notations
\begin{align}
&(-,--,+++) = A_7\,\, =  (1,1) = [2,0], \\
&(-,-+,-++) = A_{11} = (1,2) = [1,1], \\
&(-,++,--+) = A_{25} = (2,2) = [0,2].
\end{align}
Using Eqs.~\eqref{eq:mu_product} and~\eqref{eq:mu_j} we calculate the $\mu$-factors for all the amplitudes in the $U$-spin system
\begin{align}
&\mu[2,0]= \left(\sqrt{C(2,2)} \times 2!\right) \times 
\left(\sqrt{C(3,0)} \times 0!\right) = 2, \\
&\mu[1,1]= \left(\sqrt{C(2,1)} \times 1!\right) \times 
\left(\sqrt{C(3,1)} \times 1!\right) = \sqrt{6} ,\\
&\mu[0,2]= \left(\sqrt{C(2,0)} \times 0!\right) \times 
\left(\sqrt{C(3,2)} \times 2!\right) = 2\sqrt{3}.
\end{align}
The resulting lattice is shown in Fig.~\ref{fig:n6-1d-1t-32}.

We are now ready to harvest the sum rules. The $a$-type sum rules that are valid to zeroth order are trivial
\beq
a_{(1,1)}= a_{(1,2)}= a_{(2,2)}=0.
\eeq
The $s$-type sum rules that correspond to the lines of the lattice are
\beq
2\, s_{(1,1)}+ \sqrt{6}\, s_{(1,2)} = 0, \qquad
2 \sqrt{3}\, s_{(2,2)}+ \sqrt{6} \,s_{(1,2)} = 0. \eeq

For the $a$-type sum rule that is obtained from the plane we need to calculate the $M_b$ factors
\begin{align}
(1,1): \quad M_b[2,0] &= \frac{2}{2!\times 0!} = 1, \\
(2,2): \quad M_b[0,2] &= \frac{2}{0!\times 2!} = 1, \\
(1,2): \quad M_b[1,1] &= \frac{2}{1!\times 1!} = 2.
\end{align}
We then find the sum rule that is valid up to $b=2$ to be
\beq
2\,a_{(1,1)}+ 2\sqrt{6}\,a_{(1,2)} +
2\sqrt{3} \,a_{(2,2)} = 0. \label{eq:a-type-sum-rule-plane-Mb}
\eeq

\begin{figure}[t]
\centering
\includegraphics[width=0.4\textwidth]{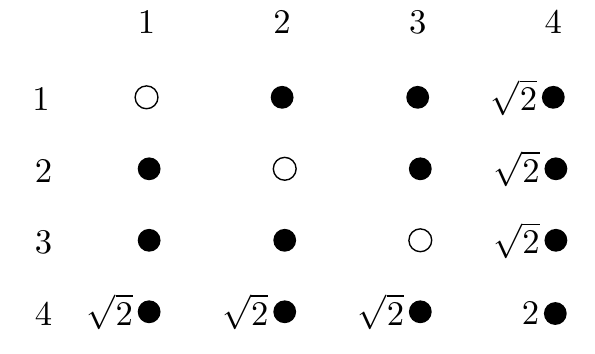}
\caption{Lattice for a system with $n=6$, $u_0=u_1=u_2=u_3=1/2$ and $u_4 = 1$.}
\label{fig:n6-4d-1t}
\end{figure}

\begin{figure}[t]
\centering
\includegraphics[width=0.3\textwidth]{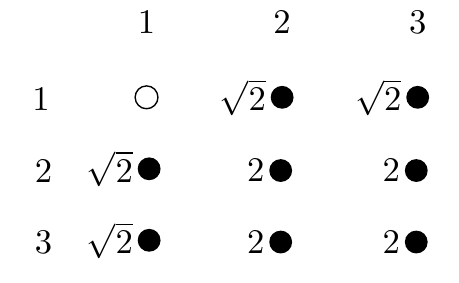}
\caption{Lattice for a system with $n=6$, $u_0=u_1=1/2$ and $u_2=u_3 = 1$.}
\label{fig:n6-2d-2t}
\end{figure}

\begin{figure}[t]
\centering
\includegraphics[width=0.3\textwidth]{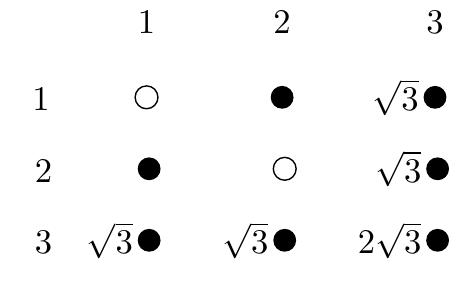}
\caption{Lattice for a system with $n=6$, $u_0=u_1=u_2=1/2$ and $u_3=3/2$.}
\label{fig:n6-3d-32}
\end{figure}

For completeness, in Figs.~\ref{fig:n6-4d-1t},~\ref{fig:n6-2d-2t}, and~\ref{fig:n6-3d-32} we show the lattices for other non-trivial $n=6$ systems with at least one doublet.

\subsection{Generalization for irreps also in the initial state and the Hamiltonian} 

\label{sec:gen-gen}
Finally, we summarize how the results of this section are generalized to the case when $U$-spin representations are present not only in the final state, but also in the initial state and the Hamiltonian.

First, we discuss the convention for building $n$-tuples. As in the case of doublets, the $m$-QNs of the components of the multiplets belonging to the initial state and the Hamiltonian are inverted in the $n$-tuple. For example, the amplitude in Eq.~\eqref{eq:gen-n-tuple-example}, which is described by representations $u_0 = 1/2$, $u_1 = 1$ and $u_2=3/2$, would have $m_0 = 1/2$, $m_1 = 0$ and $m_2=-1/2$, if all the representations belong to the initial state or the Hamiltonian.

Second, we need to use the modified definitions of $a$- and $s$-type amplitudes given in Eq.~\eqref{eq:as-comb-def-app}. Note that both factors $(-1)^p$ and $(-1)^{q_i}$ are the same for a system and its underlying system of doublets. The general expression for $p$ is given in Eq.~\eqref{eq:p_def}. The $(-1)^{q_i}$ factor for each amplitude can be read-off the corresponding $n$-tuple as a product of all the signs in the final state. The generalized definition for $(-1)^{q_i}$ can be written as follows
\begin{equation}\label{eq:qi-gen}
    (-1)^{q_i} = \prod_{j=1}^{r_f} (-1)^{u_j - m_j^{(i)}},
\end{equation}
where the product goes over the $r_f$ representations in the final state, and $m_j^{(i)}$ is the $m$-QN of the representation $u_j$ in $A_i$.

\section{Physical systems}\label{sec:gen_algo}

We are now ready to show how the mathematical results that we discuss above can be applied to physical systems. For that we first discuss how to map the physical systems into the mathematical ones, and then summarize the algorithm for writing the sum rules for physical systems. We then provide several examples of sum rules for physical systems. For some of the examples we also compare our results to the results obtained in the standard approach using Clebsch-Gordan coefficient tables. While the results are, of course, the same, these examples demonstrate that our novel approach provides a significant reduction of complexity of the calculation.

\subsection{Understanding the group-theoretical structure of physical $U$-spin sets}\label{sec:physical-to-Uspin-mapping}

The first step in obtaining amplitude sum rules for physical systems is to understand what is the group-theoretical structure of the system of interest. For that, one needs to list the $U$-spin representations that appear in the initial and final states as well as the  group-theoretical properties of the operators in the Hamiltonian. Once the $U$-spin structure of the system is understood and the amplitudes of the physical system are mapped into the abstract mathematical amplitudes, the sum rules are derived by following the procedures discussed in Sections~\ref{sec:n_doublet_system} and~\ref{sec:gen-arbitrary-irreps}.

To understand the group-theoretical structure of the $U$-spin set we assign the physical particles in the initial and final state of the system, as well as the Hamiltonian that generates the processes, to $U$-spin multiplets. This is done based on the fundamental $U$-spin doublets defined in Eq.~\eqref{eq:sd-uspin}. We rewrite these definitions here for convenience
\begin{equation} \label{eq:sd-uspin-again}
    \begin{bmatrix}
    \,d\,\, \\
    s
    \end{bmatrix} =
    \begin{bmatrix}
    \ket{\frac{1}{2}, +\frac{1}{2}}\\
    \ket{\frac{1}{2}, -\frac{1}{2}}
    \end{bmatrix}, \hspace{25pt}
    \begin{bmatrix}
    \,\bar{s}\,\, \\
    -\bar{d}
    \end{bmatrix} =
    \begin{bmatrix}
    \ket{\frac{1}{2}, +\frac{1}{2}}\\
    \ket{\frac{1}{2}, -\frac{1}{2}}
    \end{bmatrix}\,.
\end{equation}
Note, that the lower component of the anti-doublet is defined as $-\bar d$.

The particles in the initial and final state are assigned to $U$-spin multiplets based on their quark content. This is a straightforward procedure, except for one subtle point. When arranging hadrons into $U$-spin multiplets, there is a freedom in the overall phase of the hadron. We adopt the convention such that all hadrons enter the multiplets with a plus sign. For example, we define $\pi^+ =-\ket{u \bar d}$ 
and then we write $P^+$ and $P^-$, which are pseudoscalar doublets, as
\begin{equation} \label{eq:Pp-Pm-def}
P^+ = \begin{bmatrix}
K^+\\
\pi^+
\end{bmatrix}=
\begin{bmatrix}
\ket{u\bar s}\\
-\ket{u \bar d} \end{bmatrix}=
\begin{bmatrix}
\ket{\frac{1}{2}, +\frac{1}{2}}\\
\ket{\frac{1}{2}, -\frac{1}{2}}
\end{bmatrix}, \hspace{25pt}
P^- = \begin{bmatrix}
\pi^-\\
K^-
\end{bmatrix} = \begin{bmatrix}
\ket{d \bar u}\\
\ket{s \bar u} \end{bmatrix}=
\begin{bmatrix}
\ket{\frac{1}{2}, +\frac{1}{2}}\\
\ket{\frac{1}{2}, -\frac{1}{2}}
\end{bmatrix}\,.
\end{equation}
Note that this convention is different from some other sign conventions present in the literature. For example comparing to   Ref.~\cite{Soni:2006vi}, we have 
\begin{align}\label{eq:sign-convention-relation}
\ket{\pi^+}_{\text{this work}}  = -\ket{\pi^+}_{\text{Ref.~\cite{Soni:2006vi} }}.
\end{align}
Since amplitudes of physical processes are defined by the particles in the initial and final state, with this convention, the mapping of the amplitudes of the physical processes into abstract amplitudes does not introduce any relative phases. Thus, when writing the sum rules for the physical system, it is enough to derive the sum rules for the corresponding abstract system of $U$-spin representations and then just replace all the amplitudes with the corresponding CKM-free amplitudes of the physical system.

\subsection{The algorithm}\label{sec:algorithm}

In this subsection we summarize the step-by-step algorithm for writing all the amplitude sum rules for any $U$-spin set that is described by $r$ non-singlet representations and satisfies the assumptions introduced in Sections~\ref{sec:Uspin},~\ref{sec:comments}. As we discuss in Section~\ref{sec:n-tuples-generalized} the addition of $U$-spin singlets to the system does not affect the sum rules and thus all the singlet states and operators can be ignored when we describe the group-theoretical structure of the system.

The algorithm is organized into three main steps. First, we describe the group-theoretical structure of the physical system of interest and thus set up the mathematical problem. Then we find the sum rules for the abstract system of $U$-spin representations. Finally, we map the abstract amplitudes into the amplitudes of the physical system of interest.

The algorithm goes as follows (we repeat key formulas from the text for convenience):
\renewcommand{\labelenumii}{\arabic{enumi}.\arabic{enumii}}
\renewcommand{\labelenumiii}{\arabic{enumi}.\arabic{enumii}.\arabic{enumiii}}
\renewcommand{\labelenumiv}{\arabic{enumi}.\arabic{enumii}.\arabic{enumiii}.\arabic{enumiv}}

\begin{enumerate}
\item {\bf Set up the mathematical problem. }
\begin{enumerate}
\item \textit{Describe the group-theoretical structure of the system.} \\
Arrange the physical states and the operators in the Hamiltonian into $U$-spin multiplets according to the conventions in Section~\ref{sec:physical-to-Uspin-mapping}. List all the non-singlet $U$-spin multiplets $u_0, \dots, u_{r-1}$ that describe the system of interest. Order the representations such that $u_0$ is the lowest (or one of the lowest) representations.
\item \textit{List all the $n$-tuples and calculate the $(-1)^{q_i}$ factors for each $n$-tuple.} \\
The procedure of writing the generalized $n$-tuples is described in detail in Sections~\ref{sec:n-tuples-generalized} and~\ref{sec:gen-gen}. The length of the $n$-tuple is given by the number of would-be doublets in the system
\begin{equation}
    n = \sum_{j=0}^{r-1} 2u_j.
\end{equation}
The $m$-QNs of the representations in the initial state and the Hamiltonian are inverted when generating the $n$-tuples. The calculation of the $(-1)^{q_i}$ factors for the generic systems is described in Section~\ref{sec:gen-gen}. They are given as the products over the representations in the final state
\begin{align}
(-1)^{q_i} = \prod_{j=1}^{r_f} (-1)^{u_j - m_j^{(i)}}\,.
\end{align}
Note that since for sum rules only relative minus signs are important, one could equivalently define the factor $(-1)^{q_i}$ as a product over the initial state and the Hamiltonian. This corresponds to multiplying all sum rules of a system by a factor $(-1)^{n/2}$.
\item \textit{Find $p$ and list all the $a$- and $s$-type amplitudes.}\\
The $p$ factor for the system can be found using Eq.~\eqref{eq:p_def},
\begin{align}
p = 2\sum_{j = 1}^{g_F} u^{F}_j - {n \over 2}\,,
\end{align}
where the sum goes over the representations in the final state. $p$ does not depend on the specific $n$-tuple,~\emph{i.e.}~it is system-universal, and only its parity is relevant. Furthermore, for integer-only systems and for doublet-only systems with all the doublets in the final state, one can choose $p=n/2$.

The $a$- and $s$-type amplitudes are defined in Eq.~\eqref{eq:as-comb-def-app},
\begin{align}
a_i \equiv (-1)^{q_i} \left( A_i- (-1)^p A_\ell \right), \qquad
    s_i \equiv (-1)^{q_i} \left( A_i + (-1)^p A_\ell \right)\,.
\end{align}
Note that if the system contains only integer irreps, for the self-conjugated amplitude these definitions take the form given in Eqs.~\eqref{eq:self-conj-n/2-even} and \eqref{eq:self-conj-n/2-odd}.  
\end{enumerate}
\item 
{\bf Harvest the sum rules.}
\begin{enumerate}
\item \textit{System with at least one doublet.}\\
Build the lattice according to Section~\ref{sec:gen_1d} and calculate the relevant $\mu$-factors using Eqs.~\eqref{eq:mu_product} and~\eqref{eq:mu_j},
\begin{align}
 \mu[y_1,y_2,...,y_r]=\prod_{j=0}^{r-1} \mu_j, \qquad
 \mu_j = \sqrt{C(2u_j,y_j)}\times y_j!\,, \qquad C(2u_j,y_j) = \binom{2u_j}{y_j}\,.
\end{align}
\item
{\it Harvest the $a$- and $s$-type sum rules.} \\
The way to do this is described in
Section~\ref{sec:halves-lattice}. 
For all even dimensional subspaces the sum rules are
\begin{equation}
\sum_{\text{$b$-dim subspace}} W_{b}[y_1, \dots, y_{r-1}] a_i = 0.
\end{equation}
For all odd dimensional subspaces the sum rules are
\begin{equation}
\sum_{\text{$b$-dim subspace}} W_{b}[y_1, \dots, y_{r-1}] s_i = 0.
\end{equation}
The weight factors are given in Eq.~\eqref{eq:Wb-def},
\begin{equation}
W_b[y_1, y_2, \dots, y_{r-1}] = \mu[y_1, y_2, \dots, y_{r-1}] \times M_b[y_1^\prime, y_2^\prime, \dots, y_{r-1}^\prime] =  b! \prod_{j=1}^{r-1} \frac{\mu_j}{y_j^\prime!}.
\end{equation}
Note that the weights $W_b$ for $b=0$ and $b=1$ are simply given by the corresponding $\mu$-factors. 

\item \textit{System without doublets.}\\
First, construct an auxiliary $U$-spin system such that
\begin{equation}
    u_0^{\text{aux}} = \frac{1}{2}, \qquad u_1^{\text{aux}} = u_0 - \frac{1}{2}, \qquad u_j^{\text{aux}} = u_{j-1},\, \qquad \text{for } j = 2,\,\dots,\,r,
\end{equation}
and then write the sum rules for this system following the steps 1.2 to 2.2 of this algorithm. Then perform the last symmetrization
between $u_0^{\text{aux}}$ and   $u_1^{\text{aux}}$ to get the sum rules for the system of interest as explained in Section~\ref{sec:sym}.
\end{enumerate}
\item {\bf Write the sum rules for the physical system.}
\begin{enumerate}
\item \textit{Obtain the sum rules for the CKM-free amplitudes of the physical system.}\\
Replace all the amplitudes of the abstract system of $U$-spin representations with the corresponding CKM-free amplitudes of the physical system. The mapping is performed according to Section~\ref{sec:physical-to-Uspin-mapping}. In our sign convention the CKM-free amplitudes and the amplitudes of the abstract system map identically.
\item \textit{Restore the CKM dependence.} \\
Write the sum rules in terms of the amplitudes with the CKM factors included.
\end{enumerate}
\end{enumerate}

Once all the sum rules are obtained, note that due to the alternating nature of the sum rules, in order to get the complete set of linear independent sum rules to a given order $b$, one has to harvest all the sum rules that are valid up to order $b$ and $b+1$.

\subsection{Examples}
\label{sec:examples}

\subsubsection{$D^0\to P^+ P^-$ decays} \label{sec:D-to-PP}

As our first example we consider the $U$-spin set of the $D^0\to P^+ P^-$ decay processes, where $D^0$ denotes the neutral $D$-meson, which is a $U$-spin singlet, and $P^\pm$ are the $U$-spin doublets of pseudoscalar mesons, which are defined in Eq.~\eqref{eq:Pp-Pm-def}. This system had been studied before, see, for example,
Refs.~\cite{Brod:2012ud, Grossman:2019xcj, Muller:2015lua, Grossman:2006jg, Hiller:2012xm, Grossman:2013lya, Grossman:2012ry}.

We already discuss this system in Section~\ref{sec:Uspin} using the traditional method. 
Here we repeat the analysis using our novel approach.

The Hamiltonian that realizes the process in this $U$-spin set is a sum of a $U$-spin singlet and a triplet. 
\begin{equation} \label{eq:D-Ham}
    \mathcal{H}_\text{eff}^{(0)} = f_{0,0} H^0_0 + \sum_{m = -1}^{1} f_{1,m} H^1_m,
\end{equation}
where $H_0^0$ and $H^1_m$ are given in Eqs.~\eqref{H0-charm}
and \eqref{H1-charm} that we rewrite below 
\begin{align} 
	H^0_0 &=  {(\bar{u} s) (\bar{s} c)+(\bar{u} d) (\bar d c)\over
  \sqrt{2}}, \label{H0-charm-copy} \\
H^1_{1} =  (\bar{u} s) (\bar d c),\qquad
H^1_{-1} &= -(\bar{u} d) (\bar s c), \qquad
H^1_0 =  {(\bar{u} s) (\bar s c)-(\bar{u} d) ( \bar d c)\over \sqrt{2}}, \label{H1-charm-copy}
\end{align}
The corresponding CKM-factors are given in Eqs.~\eqref{eq:charmCKM-f00} and \eqref{eq:charmCKM-f1m} and we repeat them here
\begin{align}
f_{0,0} &= \frac{V_{cs}^* V_{us} + V_{cd}^* V_{ud}}{2}\approx 0, \label{eq:charmCKM-f00-again} \\
f_{1,1} = V_{cd}^* V_{us}, \qquad 
f_{1,-1} &= -V_{cs}^* V_{ud}, \qquad 
f_{1,0} = \frac{V_{cs}^* V_{us} - V_{cd}^* V_{ud}}{\sqrt{2}}\approx \sqrt{2}\,\left(V_{cs}^* V_{us}\right)\,,\label{eq:charmCKM-f1m-again}
\end{align}
where the approximation used for $f_{0,0}$ and $f_{1,0}$ holds up to $O(\lambda^4)$.
In the following we use this approximation and, as a result, we only keep the triplet part of the Hamiltonian, $H^1$, and not the singlet part, $H^0$. Adopting this approximation, we can define the CKM-free amplitudes using Eq.~\eqref{eq:def-ckm-free-amp}.

Note that the standard convention in the literature is
\beq
H^1_0 = (\bar{u} s) (\bar s c)-(\bar{u} d) ( \bar d c), \qquad
f_{1,0} = \frac{V_{cs}^* V_{us} - V_{cd}^* V_{ud}}{2}\approx \left(V_{cs}^* V_{us}\right)
\,,
\eeq
and similarly for $H^0_0$ and $f_{0,0}$. The reason why we prefer to use the definitions in Eqs.~\eqref{H0-charm-copy}-\eqref{eq:charmCKM-f1m-again} is to keep the $H_0^1$ to be a part of a triplet, where the normalization is $\sqrt{2}$. The final result, of course, does not depend on this normalization choice.

Now we are ready to describe the $U$-spin structure of the system of interest. The number of non-trivial representations in the system is $r=3$. We order the representations as follows:
\begin{equation}
 u_0 = u_1 = \frac{1}{2},\qquad u_2 = 1,
\end{equation}
where $u_0$ corresponds to $P^+$, $u_1$ to $P^-$ and $u_2$ to the Hamiltonian. The mapping of the four CKM-free amplitudes of the physical system into generalized $n$-tuples, as well as the $U$-spin paring is summarized in 
Table~\ref{tab:map-c-mesons}. Note that we choose to show the results for $D^0$ decays, which contain a $c$ quark. Additionally, Table~\ref{tab:map-c-mesons} lists the coordinate notation for the amplitude pairs (``Nodes''), the values of the $\mu$-factors, and $(-1)^{q_i}$. The last column ``Indices'' lists the indices of the amplitudes from the $U$-spin pair using the notation introduced in Section~\ref{sec:An-tuples-doublets}. 
Recall that the smaller index corresponds to the amplitude of the pair where the corresponding $n$-tuple starts with a minus sign.

\begin{table}[t]
\centering
\begin{tabular}{|c|c|c|c|c|c|c|}
\hline
Decay &  $U$-spin conjugate & $n$-tuple & Node & ~~$\mu$-factor~~ &~~$(-1)^{q_i}$~~ & ~~Indices~~  \\
\hline
~~~$D^0 \rightarrow \pi^+ K^-$~~~ & 
~~~$D^0 \rightarrow K^+\pi^-$~~~& ~~$(-,-,++)$~~ &~~$(1)$~~ & 1 & $+1$ & 3, 12 \\
~~~$D^0 \rightarrow \pi^+ \pi^- $~~~ & 
~~~$D^0 \rightarrow K^+ K^-$~~~& ~~$(-,+,-+)$~~ &~~$(2)$~~ & $\sqrt{2}$ & $-1$ & 5, 10\\
\hline
\end{tabular}
\caption{The mapping of the $U$-spin amplitude pairs of $D^0\to P^+ P^-$ decays into generalized $n$-tuples.\label{tab:map-c-mesons}}
\end{table}

Next, we calculate $p$ and find that $p = 0$. Using this and the $(-1)^{q_i}$ factors listed in Table~\ref{tab:map-c-mesons} we define the $a$- and $s$-type amplitudes using Eq.~\eqref{eq:as-comb-def-app}. We have in the index notation
\begin{align}
    a_3 &= A_3 - A_{12},& s_3 &= A_3 + A_{12}\nonumber\\
    a_5 &= -(A_5 - A_{10}),& s_5 &= -(A_5 + A_{10}).
\end{align}

The resulting sum rules in terms of $a$- and $s$-type amplitudes take the following form
\beq
a_3=0,\qquad
a_5=0,\qquad
s_3+\sqrt{2} s_5=0,
\eeq
where the first two $a$-type sum rules hold only at zeroth order, while the last $s$-type sum rule holds up to order $b=1$ and is broken at order $b=2$.

In terms of the CKM-free amplitudes of the physical system, we have the following  sum rules. The $a$-type sum rules that hold up to $b=0$ are
\beq
A(D^0 \rightarrow \pi^+ K^-) = 
A(D^0 \rightarrow K^+\pi^-), \qquad
A(D^0 \rightarrow \pi^+\pi^-) = 
A(D^0 \rightarrow K^+ K^-).
\eeq
The $s$-type sum rule that holds up to $b=1$ is given by 
\beq \label{eq:DPP-s}
A(D^0 \rightarrow \pi^+ K^-) + 
A(D^0 \rightarrow K^+\pi^-)-
\sqrt{2}A(D^0 \rightarrow \pi^+\pi^-) - 
\sqrt{2}A(D^0 \rightarrow K^+ K^-)= 0.
\eeq

Our last step is to restore the CKM dependence. We find
\begin{align}
\frac{\mathcal{A}(D^0 \rightarrow \pi^+ K^-)}{ -V_{cs}^* V_{ud} } = 
\frac{\mathcal{A}(D^0 \rightarrow K^+\pi^-)}{ V_{cd}^* V_{us} }, \qquad
\frac{\mathcal{A}(D^0 \rightarrow \pi^+\pi^-)}{\sqrt{2} V_{cs}^* V_{us}} = 
\frac{\mathcal{A}(D^0 \rightarrow K^+ K^-)}{\sqrt{2} V_{cs}^* V_{us} }.
\end{align}
and
\begin{align}
\frac{\mathcal{A}(D^0 \rightarrow \pi^+ K^-)}{ -V_{cs}^* V_{ud} } + 
\frac{\mathcal{A}(D^0 \rightarrow K^+\pi^-)}{ V_{cd}^* V_{us} }-
\sqrt{2}\frac{\mathcal{A}(D^0 \rightarrow \pi^+\pi^-)}{\sqrt{2} V_{cs}^* V_{us} } - 
\sqrt{2}\frac{\mathcal{A}(D^0 \rightarrow K^+ K^-)}{\sqrt{2} V_{cs}^* V_{us} }= 0.
\end{align}
A few remarks are in order:
\begin{enumerate}
\item 
The results agree with the known results.
\item
The conventional factors of $\sqrt{2}$ consistently cancel out in the final result. 
\item 
Recall that we work in the approximation $V_{cd}^*V_{ud}\approx - V_{cs}^* V_{us}$. 
\item
Any sign difference between the literature and our result is due to the sign convention for the operator $H^1_{-1}$ and the corresponding coefficient $f_{1,-1}$.
\item
The difference between the literature and our result regarding the minus sign in $-V_{cs}^* V_{ud}$ is due to the conventional minus sign in the Hamiltonian Eq.~(\ref{H1-charm-copy}), and that the above CKM matrix elements are the ones for $D^0$ decays.
\end{enumerate}

\subsubsection{Semileptonic $K \to \pi$ decays}

All of the results presented in this work are also valid for $SU(2)$ isospin. The fundamental doublets of isospin are
\begin{equation} \label{eq:ud-isospin}
\begin{bmatrix}
\,u\,\, \\
d
\end{bmatrix} =
\begin{bmatrix}
\ket{\frac{1}{2}, +\frac{1}{2}}\\
\ket{\frac{1}{2}, -\frac{1}{2}}
\end{bmatrix}, \hspace{25pt}
\begin{bmatrix}
\,\bar{d}\,\, \\
-\bar{u}
\end{bmatrix} =
\begin{bmatrix}
\ket{\frac{1}{2}, +\frac{1}{2}}\\
\ket{\frac{1}{2}, -\frac{1}{2}}
\end{bmatrix}\,.
\end{equation}
In this example we consider the isospin system of $K \to \pi$ decays, where $K$ and $\pi$ stand for the following isospin doublet and triplet, respectively
\begin{equation}
    K = \begin{bmatrix}
    K^+\\
    K^0
    \end{bmatrix} =
    \begin{bmatrix}
    \ket{\frac{1}{2}, +\frac{1}{2}}\\
    \ket{\frac{1}{2}, -\frac{1}{2}}
    \end{bmatrix}, \hspace{25pt}
    \pi = \begin{bmatrix}
    \pi^+\\
    \pi^0\\
    \pi^-
    \end{bmatrix} =
    \begin{bmatrix}
    \ket{1, +1}\\
    \ket{1, 0}\\
    \ket{1, -1}
    \end{bmatrix}.
\end{equation}
The processes in the $K \to \pi$ isospin-set are realized via an isospin doublet Hamiltonian. We write the Hamiltonian as 
\begin{equation}
    \mathcal{H}_\text{eff}^{(0)} = \sum_{m=-1/2}^{1/2} f^u_m H_m^{1/2},
\end{equation}
where 
\beq \label{eq:H-k-sl}
H_{1/2}^{1/2} = 
(\bar{u} s) (\bar e \nu),\qquad
H_{-1/2}^{1/2} = 
(\bar{d} s) ( \bar \nu\nu),
\eeq
where by $\nu$ we refer to $\nu_e$.
The CKM factors are given by
\begin{align}
f_{1/2}^{1/2} = V_{us}, \qquad
f_{-1/2}^{1/2} = \sum_{i=u,c,t} V_{id}^* V_{is} f(m_i^2),
\end{align}
where $f(m^2)$ is a known loop factor that can be found, for example, in Ref.~\cite{Buchalla:1998ba} and we do not write it explicitly here.

From the group theoretical point of view this system is described by $r=3$ representations: one doublet in the initial state, one triplet in the final state and a Hamiltonian that transforms as a doublet. We order the representations as follows:
\begin{equation}
    u_0 = u_1 = \frac{1}{2},\qquad u_2 = 1,
\end{equation}
where $u_0$ corresponds to $K$, $u_1$ to $H$, and $u_2$ to $\pi$. We see that this system is in the same universality class as the $D^0\to P^+P^-$ system we consider in section~\ref{sec:D-to-PP}.

The mapping of the four CKM-free amplitudes of the set into the generalized $n$-tuples is presented in Table~\ref{tab:map-K-sl}. The value of $p$ for this system is the same as for $D^0\to P^+P^-$, that is, $p = 0$.

\begin{table}[t]
\centering
\begin{tabular}{|c|c|c|c|c|c|c|}
\hline
Decay &  $U$-spin conjugate & $n$-tuple & Node & ~~$\mu$-factor~~ &~~$(-1)^{q_i}$~~ & ~~Indices~~  \\
\hline
~~~$K^+ \to \pi^+ \nu \bar \nu  $~~~ & 
~~~$K^0 \to \pi^- \nu e^+$~~~& ~~$(-,-,++)$~~ &~~$(1)$~~ & 1 & $+1$ & 3, 12 \\
~~~$K^+ \to \pi^0  \nu e^+$~~~ & 
~~~$K^0 \to \pi^0  \nu \bar \nu$~~~& ~~$(-,+,-+)$~~ &~~$(2)$~~ & $\sqrt{2}$ & $-1$ & 5, 10\\
\hline
\end{tabular}
\caption{The mapping of the $U$-spin pairs of semileptonic $K\to \pi$ decays into generalized $n$-tuples.\label{tab:map-K-sl}}
\end{table}

Comparing Tables~\ref{tab:map-c-mesons} and~\ref{tab:map-K-sl} we write the sum rules for the $K \to \pi$ system. The $a$-type sum rules that hold in the isospin limit are
\beq \label{eq:Kpi-sum-rules-1}
A(K^+ \rightarrow \pi^+ \nu \bar \nu) = 
A(K^0 \rightarrow \pi^- \nu e^+), \qquad
A(K^+ \rightarrow \pi^0 \nu e^+) = 
A(K^0 \rightarrow \pi^0 \nu \bar \nu).
\eeq
The $s$-type sum rule that holds up to $b=1$ order of breaking is given by
\beq \label{eq:Kpi-sum-rules-2}
A(K^+ \to \pi^+ \nu \bar \nu)  + 
A(K^0 \to \pi^- \nu e^+)-
\sqrt{2}A(K^+ \to \pi^0  \nu e^+) - 
\sqrt{2}A(K^0 \to \pi^0  \nu \bar \nu) = 0.
\eeq

Note the following:
\begin{enumerate}
\item 
From the group theoretical point of view, the $D^0 \to P^+P^-$ and semileptonic $K \to \pi$ decays are identical. 
\item
Often, the isospin relations above are written in terms of the hadronic part of the decay, which is parametrized by a form factor function of $q^2$. For example, a common notation is $f^{K^i\pi^j}$.
\item
The relation in Eq.~\eqref{eq:Kpi-sum-rules-2} may be of limited use, as there are also electromagnetic corrections that are not captured by the group theoretical treatment. These corrections are expected to be of the order of the first order isospin breaking.
\end{enumerate}
Our result is in agreement with the explicit calculation given in Eq.~(17) of Ref.~\cite{Mescia:2007kn}. 
In order to see it we have to neglect electromagnetic corrections, account for the conventional factors of $\sqrt{2}$ and take the isospin-breaking meson mass splitting into account for the power counting. The latter is needed in order to see that the loop functions consistently cancel each other in the combinations appearing in the sum rules.

\subsubsection{Baryonic charm decays}\label{sec:CbtoLbPP}

Consider the following system
\beq
{C_b} \to {L_b} P^- P^+
\eeq
where $P^+$ and $P^-$ are defined in Eq.~\eqref{eq:Pp-Pm-def}, $C_b$ stands for a doublet of charmed baryons and $L_b$ for a doublet of light baryons defined as follows:
\begin{equation}
C_b = \begin{bmatrix}
\Lambda_c^+ \\ \Xi_c^+
    \end{bmatrix} =
    \begin{bmatrix}
    \ket{cud}\\
    \ket{cus}
    \end{bmatrix} =
    \begin{bmatrix}
    \ket{\frac{1}{2}, +\frac{1}{2}}\\
    \ket{\frac{1}{2}, -\frac{1}{2}}
    \end{bmatrix}\,, 
    \qquad
L_b = \begin{bmatrix}
    p \\
    \Sigma^+
    \end{bmatrix} =
    \begin{bmatrix}
    \ket{uud}\\
    \ket{uus}
    \end{bmatrix} =
    \begin{bmatrix}
    \ket{\frac{1}{2}, +\frac{1}{2}}\\
    \ket{\frac{1}{2}, -\frac{1}{2}}
    \end{bmatrix}\,.
\end{equation}
The Hamiltonian and CKM factors for this system are the same as for $D \to P^+ P^-$ and are given in Eqs.~\eqref{eq:D-Ham}--\eqref{eq:charmCKM-f1m-again}. As in Section~\ref{sec:D-to-PP} we only consider the $H^1$ operators in the Hamiltonian.

The $C_b \to L_b P^- P^+$ system is described by $r = 5$ irreps, among which four are doublets and one is a triplet. We order the representations as follows
\begin{equation}
    u_0 = u_1 = u_2 = u_3 = \frac{1}{2},\qquad u_4 = 1,
\end{equation}
where we choose the order to be $(C_b,L_b,P^-,P^+,H^1)$.
The mapping of the decay processes to $n$-tuples is given in Table~\ref{tab:map-c-baryons}. For this system $p=0$.

\begin{table}[t]
\centering
\begin{tabular}{|c|c|c|c|c|}
\hline
Decay &  $U$-spin conjugate & $n$-tuple & Node & ~~$(-1)^{q_i}$~~ \\
\hline
~~~$\Lambda^+_c \to \Sigma^+ K^- K^+$~~~ & ~~~$\Xi_c^+ \to p\pi^- \pi^+$~~~ & ~~$(-, -, -, +, + +)$~~ & ~~$(1,2)$~~ & $+1$ \\
$\Lambda_c^+ \to \Sigma^+ \pi^- \pi^+$ & $\Xi_c^+ \to p K^- K^+$ & $(-,-,+,-,++)$ & $(1,3)$ & $+1$ \\
$\Lambda^+_c \to \Sigma^+ \pi^- K^+$ & $\Xi_c^+ \to p K^- \pi^+$ & $(-,-,+,+,-+)$ & $(1,4)$ & $-1$ \\
$\Lambda_c^+ \to p K^- \pi^+$ & $\Xi_c^+ \to \Sigma^+ \pi^- K^+$ & $(-,+,-,-,++)$ & $(2,3)$ & $+1$ \\
$\Lambda_c^+ \to p K^- K^+$ & $\Xi_c^+ \to \Sigma^+ \pi^- \pi^+$ & $(-,+,-,+,-+)$ & $(2,4)$ & $-1$ \\
$\Lambda_c^+ \to p \pi^- \pi^+$ & $\Xi_c^+ \to \Sigma^+ K^- K^+$ & $(-,+,+,-,-+)$ & $(3,4)$ & $-1$ \\
$\Lambda_c^+ \to p \pi^- K^+$ & $\Xi_c^+ \to \Sigma^+ K^- \pi^+$ & $(-,+,+,+,--)$ & $(4,4)$ & $+1$ \\
\hline
\end{tabular}
\caption{The mapping of the $U$-spin pairs of baryonic charm decays $C_b \to L_b P^- P^+$ decays into generalized $n$-tuples. \label{tab:map-c-baryons} }
\end{table}

We consider the system of four doublets and one triplet in Section~\ref{sec:gen-examples}. The  lattice is shown in Fig.~\ref{fig:n6-4d-1t}.  Harvesting the sum rules from the lattice we obtain seven trivial $a$-type sum rules that are valid to order $b = 0$. They are
\begin{equation}
    a_{(1,2)} = a_{(1,3)} = a_{(1,4)} = a_{(2,3)} = a_{(2,4)} = a_{(3,4)} = a_{(4,4)} = 0.
\end{equation}
In terms of CKM-free amplitudes these $b=0$ sum rules take the following form
\begin{align}
A\left(\Lambda^+_c \to \Sigma^+ K^- K^+\right) &= A\left(\Xi_c^+ \to p\pi^- \pi^+\right),\\
A\left(\Lambda_c^+ \to \Sigma^+ \pi^- \pi^+\right) &= A\left(\Xi_c^+ \to p K^- K^+\right),\\
A\left(\Lambda^+_c \to \Sigma^+ \pi^- K^+\right) &= A\left(\Xi_c^+ \to p K^- \pi^+\right),\\
A\left(\Lambda_c^+ \to p K^- \pi^+\right) &= A\left(\Xi_c^+ \to \Sigma^+ \pi^- K^+\right),\\
A\left(\Lambda_c^+ \to p K^- K^+\right) &= A\left(\Xi_c^+ \to \Sigma^+ \pi^- \pi^+\right),\\
A\left(\Lambda_c^+ \to p \pi^- \pi^+\right)& = A\left(\Xi_c^+ \to \Sigma^+ K^- K^+\right),\\
A\left(\Lambda_c^+ \to p \pi^- K^+\right)& = A\left(\Xi_c^+ \to \Sigma^+ K^- \pi^+\right).
\end{align}
The $s$-type sum rules that are valid up to order $b = 1$ of $U$-spin breaking are read off the lines of the lattice in Fig.~\ref{fig:n6-4d-1t} and are given by
\begin{align}
s_{(1,2)} + s_{(1,3)} + \sqrt{2} s_{(1,4)} &= 0, \\
s_{(1,2)} + s_{(2,3)} + \sqrt{2} s_{(2,4)} &= 0, \\
s_{(1,3)} + s_{(2,3)} + \sqrt{2} s_{(3,4)} &= 0, \\
s_{(1,4)} + s_{(2,4)} + s_{(3,4)} + \sqrt{2} s_{(4,4)} &= 0.
\end{align}
The above four $s$-type sum rules take the following form in terms of CKM-free amplitudes of the physical system
\begin{align}
+A\left(\Lambda^+_c \to \Sigma^+ K^- K^+\right)+A\left(\Xi_c^+ \to p\pi^- \pi^+\right) + A\left(\Lambda_c^+ \to \Sigma^+ \pi^- \pi^+\right) & \nonumber \\
+A\left(\Xi_c^+ \to p K^- K^+\right) - \sqrt{2}A\left(\Lambda^+_c \to \Sigma^+ \pi^- K^+\right) - \sqrt{2} A\left(\Xi_c^+ \to p K^- \pi^+\right) & = 0,\\
A\left(\Lambda^+_c \to \Sigma^+ K^- K^+\right)+A\left(\Xi_c^+ \to p\pi^- \pi^+\right) + A\left(\Lambda_c^+ \to p K^- \pi^+\right) & \nonumber \\ 
+ A\left(\Xi_c^+ \to \Sigma^+ \pi^- K^+\right) - \sqrt{2} A\left(\Lambda_c^+ \to p K^- K^+\right) - \sqrt{2} A\left(\Xi_c^+ \to \Sigma^+ \pi^- \pi^+\right) & = 0,
\\
+ A\left(\Lambda_c^+ \to \Sigma^+ \pi^- \pi^+\right) + A\left(\Xi_c^+ \to p K^- K^+\right) + A\left(\Lambda_c^+ \to p K^- \pi^+\right) & \nonumber \\ 
+ A\left(\Xi_c^+ \to \Sigma^+ \pi^- K^+\right) - \sqrt{2}A\left(\Lambda_c^+ \to p \pi^- \pi^+\right) - \sqrt{2} A\left(\Xi_c^+ \to \Sigma^+ K^- K^+\right) & = 0,
\\
- A\left(\Lambda^+_c \to \Sigma^+ \pi^- K^+\right) - A\left(\Xi_c^+ \to p K^- \pi^+\right) - A\left(\Lambda_c^+ \to p K^- K^+\right) \nonumber \\ 
- A\left(\Xi_c^+ \to \Sigma^+ \pi^- \pi^+\right) - A\left(\Lambda_c^+ \to p \pi^- \pi^+\right) - A\left(\Xi_c^+ \to \Sigma^+ K^- K^+\right) \nonumber \\
+\sqrt{2} A\left(\Lambda_c^+ \to p \pi^- K^+\right) + \sqrt{2} A\left(\Xi_c^+ \to \Sigma^+ K^- \pi^+\right) & = 0.
\end{align}
The $a$-type sum rule that holds at $b=2$ is
\beq
a_{(1,2)} + a_{(1,3)} + a_{(2,3)} + a_{(4,4)} + \sqrt{2} a_{(1,4)} +
\sqrt{2} a_{(2,4)} + \sqrt{2} a_{(3,4)} =0.
\eeq
In terms of CKM-free amplitudes of the physical system this sum rule becomes 
\begin{align}
&+\left[A\left(\Lambda^+_c \to \Sigma^+ K^- K^+\right) - A\left(\Xi_c^+ \to p\pi^- \pi^+\right)\right] \nonumber\\&
+ \left[A\left(\Lambda_c^+ \to \Sigma^+ \pi^- \pi^+\right) - A\left(\Xi_c^+ \to p K^- K^+\right)\right]\nonumber\\&
+\left[A\left(\Lambda_c^+ \to p K^- \pi^+\right) - A\left(\Xi_c^+ \to \Sigma^+ \pi^- K^+\right)\right]
\nonumber\\&
+\left[A\left(\Lambda_c^+ \to p \pi^- K^+\right) - A\left(\Xi_c^+ \to \Sigma^+ K^- \pi^+\right)\right]
\nonumber\\&
-\sqrt{2}\left[A\left(\Lambda^+_c \to \Sigma^+ \pi^- K^+\right)  -A\left(\Xi_c^+ \to p K^- \pi^+\right)\right]
\nonumber\\ &-
\sqrt{2}\left[A\left(\Lambda_c^+ \to p K^- K^+\right) - A\left(\Xi_c^+ \to \Sigma^+ \pi^- \pi^+\right)\right]
\nonumber\\&
-\sqrt{2}\left[A\left(\Lambda_c^+ \to p \pi^- \pi^+\right)- A\left(\Xi_c^+ \to \Sigma^+ K^- K^+\right)\right] =0.
\end{align}

For comparison, in Appendix~\ref{app:CbtoLbPP} we show the decompositions of the CKM-free amplitudes of the $C_b \to L_b P^+ P^-$ system in terms of RME. One can check explicitly that the sum rules we obtain here using our novel method are indeed satisfied.

\subsubsection{$D^0 \to P^0 P^{*0}$ decays}

In this section we obtain the sum rules for a system that is not easy to probe as the final states contain
particles which are not mass eigenstates. Still it is an instructive example in order to illustrate our formalism because it contains only triplet representations and no doublets. Specifically, we obtain the sum rules for the $U$-spin set $D^0\to P^0  P^{*0}$, where $D^0$ is a $U$-spin singlet, $P^0$ is the neutral pseudoscalar triplet, and $P^{*0}$ is the neutral vector triplet. Explicitly the $P^0$ and $P^{*0}$ triplets are
\begin{equation} 
P^0 = \begin{bmatrix}
K^0\\
\eta_u \\ \bar K^0
\end{bmatrix}=
\begin{bmatrix}
\ket{1, +1}\\
\ket{1,0} \\ \ket{1,-1}
\end{bmatrix},
\qquad
P^{*0} = \begin{bmatrix}
K^{*0}\\
\phi_u \\ \bar K^{*0}
\end{bmatrix}=
\begin{bmatrix}
\ket{1, +1}\\
\ket{1,0} \\ \ket{1,-1}
\end{bmatrix},
\end{equation}
where  we use the notation
\beq
\eta_u, \phi_u \equiv \frac{s \bar s - d \bar d}{\sqrt{2}}.
\eeq
While the $\eta_u$ and $\phi_u$ states are not mass eigenstates, they do have definite transformation behavior under $U$-spin and are therefore well-suited for an illustration of our methodology. 
In terms of physical states $\eta_u$ is a mixture of $\pi^0$, $\eta$, and $\eta'$ while $\phi_u$ is a mixture of $\rho$, $\omega$ and $\phi$. The Hamiltonian and CKM factors for $D\to P^0  P^{*0}$ are the same as for $D^0 \to P^+ P^-$ and $C_b \to L_b P^- P^+$ decays and are summarized in Eqs.~\eqref{eq:D-Ham}--\eqref{eq:charmCKM-f1m-again}.

The $U$-spin system under consideration is thus described by $r = 3$ representations all of which are triplets
\begin{equation}
    u_0 = u_1 =u_2 = 1.
\end{equation}
We choose to order the irreps as $(H,P^0,P^{*0})$. The mapping of the CKM-free amplitudes of the set into generalized $n$-tuples is presented in Table~\ref{tab:map-3-triplets}. Note that the last amplitude $A\left(D^0 \to \eta_u \phi_u\right)$ in the table is $U$-spin self conjugate. For this system $p=1$.

\begin{table}[t]
\centering
\begin{tabular}{|c|c|c|c|c|}
\hline
Decay & $U$-spin conjugate & $n$-tuple & $(-1)^{q_i}$ & Indices\\
\hline
$D^0 \to \eta_u K^{*0}  $ & $D^0 \to \eta_u \bar K^{*0} $ & $(--,-+,++)$ & $-1$ &7, 52  \\
$D^0 \to K^0 \phi_u  $& $D^0 \to \bar K^0 \phi_u $&  $(--,++,-+)$ & $-1$ & 13, 49 \\
~~~$D^0 \to \bar K^0  K^{*0}  $~~~ & $D^0 \to K^0 \bar K^{*0} $ & ~~$(-+,--,++)$~~ & $+1$ & 19, 28  \\
~~~$D^0 \to \eta_u \phi_u  $~~~ & $D^0 \to \eta_u \phi_u  $ & ~~$(-+,-+,-+)$~~ & $+1$ & $21$ \\
\hline
\end{tabular}
\caption{The mapping of the $U$-spin pairs of $D^0\to P P^*$ decays into generalized $n$-tuples. \label{tab:map-3-triplets}}
\end{table}

The sum rules for a system of three triplets are derived in Appendix~\ref{app:3t}. Using Eqs.~\eqref{eq:3t-b0}--\eqref{eq:3t-b2} and taking into account the values of $(-1)^{q_i}$ listed in Table~\ref{tab:map-3-triplets}, we have the following $b=0$ sum rules
\begin{align} 
A(D^0 \to \eta_u K^{*0}) &= -
A(D^0 \to \eta_u \bar K^{*0}),
\\
A(D^0 \to K^0 \phi_u) &= -
A(D^0 \to \bar K^0 \phi_u),
\\
A(D^0 \to \bar K^0 K^{*0}) &= -
A(D^0 \to K^0 \bar K^{*0}),
\\
A(D^0 \to \eta_u \phi_u) &=0.
\end{align}
For $b=1$ the sum rules are
\begin{align} \label{eq:3-tri-1}
& -A(D^0 \to \eta_u K^{*0}) +
A(D^0 \to \eta_u \bar K^{*0}) - A(D^0 \to K^0 \phi_u) +
A(D^0 \to \bar K^0 \phi_u) = 0, \\
& -A(D^0 \to \eta_u K^{*0}) +
A(D^0 \to \eta_u \bar K^{*0}) + 
A(D^0 \to \bar K^0 K^{*0}) -
A(D^0 \to K^0 \bar K^{*0}) = 0,
\end{align}
and, finally, for $b=2$ we have
\begin{align}
-A(D^0 \to \eta_u K^{*0}) -
A(D^0 \to \eta_u \bar K^{*0}) - A(D^0 \to K^0 \phi_u) -
A(D^0 \to \bar K^0 \phi_u) &+ \nn \\
A(D^0 \to \bar K^0 K^{*0}) +
A(D^0 \to K^0 \bar K^{*0}) 
+2 A(D^0 \to \eta_u \phi_u) &=0.
\end{align}

\subsubsection{$\bar B^0 \to D^0 \Omega_B^- \bar\Omega_B^+$ decays}

We conclude this section with an example of a system with $n=8$ and $r = 4$. We consider $B$ decays into a $D^0$ meson and two baryons that belong to the multiplets $\Omega_B^-$ and $\Omega_B^+$ each with $U$-spin $3/2$. This is an example of a system that was not studied theoretically and is yet to be measured experimentally. The baryon multiplets $\Omega_B^-$ and $\Omega_B^+$ are defined as follows:
\begin{equation}
\Omega_B^- = \begin{bmatrix}
\Delta^- \\ \Sigma^{*-} \\\Xi^{*-} \\ \Omega^-
\end{bmatrix} =
\begin{bmatrix}
\ket{ddd}\\
\ket{dds} \\ \ket{dss} \\ \ket{sss}
\end{bmatrix} =
\begin{bmatrix}
\ket{\frac{3}{2}, +\frac{3}{2}}\\
\ket{\frac{3}{2}, +\frac{1}{2}} \\\ket{\frac{3}{2}, -\frac{1}{2}}\\
\ket{\frac{3}{2}, -\frac{3}{2}}
\end{bmatrix}\,, 
\qquad
\Omega_B^+ = \begin{bmatrix}
\bar\Omega^+ \\\bar\Xi^{*+} \\ \bar \Sigma^{*+} \\
\bar \Delta^+ 
\end{bmatrix} =
\begin{bmatrix}
\ket{\bar{s} \bar{s} \bar{s}}\\
-\ket{\bar d \bar s \bar s}\\
\ket{\bar d \bar d \bar s}\\
-\ket{\bar d \bar d \bar d}
\end{bmatrix} =
\begin{bmatrix}
\ket{\frac{3}{2}, +\frac{3}{2}}\\
\ket{\frac{3}{2}, +\frac{1}{2}} \\\ket{\frac{3}{2}, -\frac{1}{2}}\\
\ket{\frac{3}{2}, -\frac{3}{2}}
\end{bmatrix}\,.
\end{equation}
The $\bar B^0$ doublet is given by
\beq
\bar B^0 = \begin{bmatrix}
\bar B_s^0 \\
\bar B_d^0
\end{bmatrix} =
\begin{bmatrix}
\ket{b \bar s}\\ -\ket{b \bar d}
\end{bmatrix} =
\begin{bmatrix}
 \ket{\frac{1}{2}, +\frac{1}{2}}  \\ \ket{\frac{1}{2}, -\frac{1}{2}}
\end{bmatrix}\,.
\eeq
The $D^0$ meson is a $U$-spin singlet.

The leading order effective Hamiltonian for the $\bar{B}^0 \to D^0 \Omega_B^- \bar{\Omega}^+$ system is a doublet
\begin{equation}
    \mathcal{H}^{(0)}_\text{eff} = \sum_{m=-1/2}^{1/2} f_{1/2,m} H_m^{1/2}.
\end{equation}
with
\beq \label{eq:H-b-to-c}
H_{1/2}^{1/2} = 
(\bar c b) (\bar s u),\qquad
H_{-1/2}^{1/2} = 
(\bar c b) (\bar d u),
\eeq
and the CKM factors
\beq
f_{1/2,1/2} = V_{cb}V_{us}^*,  \qquad
f_{1/2, -1/2} = 
V_{cb}V_{ud}^*.
\eeq

We order the $U$-spin representations that describe the system as follows
\beq
u_0=u_1={1 \over 2}, \qquad u_2=u_3={3 \over 2},
\eeq
where we choose the order to be $(\bar B^0,H,\Omega_B^-,\bar\Omega_B^+)$. The mapping of the amplitudes of the physical system into generalized $n$-tuples is then given in \cref{tab:map-b-baryons}.

\begin{table}[t]
\centering
\begin{tabular}{|c|c|c|c|c|c|}
\hline
Decay & $U$-spin conjugate & $n$-tuple & Node  &~$\mu$-factor~& $(-1)^{q_i}$\\
\hline
~$\bar B^0_s\rightarrow D^0\, \Xi^{*-} \bar \Omega^+ $ ~&
~$\bar B^0_d\rightarrow D^0 \,\Sigma^{*-} \bar\Delta^+ $~& 
~~$(-,-,- -+,+++)$~~ & 
~~$(1,2,2)$~~ & $2\sqrt{3}$ & $+1$\\
~$\bar B^0_s\rightarrow D^0\, \Sigma^{*-} \bar\Xi^{*+} $ ~&~$\bar B^0_d\rightarrow D^0 \,\Xi^{*-} \bar\Sigma^{*+} $~&
~~$(-,-,- ++,-++)$~~ &
~~$(1,2,3)$~~ & 3 & $+1$\\
~$\bar B^0_s\rightarrow D^0\, \Delta^{-} \bar\Sigma^{*+} $~&~$\bar B^0_d\rightarrow D^0 \,\Omega^{-} \bar\Xi^{*+} $~&
~~$(-,-,+++,--+)$~~ &
~~$(1,3,3)$~~ & $2 \sqrt{3}$ & $+1$\\
~$\bar B^0_s\rightarrow D^0 \,\Delta^{-} \bar\Delta^+ $ ~&~$\bar B^0_d\rightarrow D^0\, \Omega^{-} \bar\Omega^+ $  ~&
~~$(-,+,+++,---)$~~ &
~~$(3,3,3)$~~ & 6 & $-1$\\
~$\bar B^0_s\rightarrow D^0\, \Omega^{-} \bar\Omega^+ $~&~$\bar B^0_d\rightarrow D^0\, \,\Delta^{-} \bar\Delta^+$ ~&
~~$(-,+,---,+++)$~~ &
~~$(2,2,2)$~~ & 6 & $-1$\\
~$\bar B^0_s\rightarrow D^0\, \Sigma^{*-} \bar\Sigma^{*+} $~&~$\bar B^0_d\rightarrow D^0\, \,\Xi^{*-} \bar\Xi^{*+} $~&
~~$(-,+,-++,--+)$~~ &
~~$(2,3,3)$~~ & 6 & $-1$\\
~$\bar B^0_s\rightarrow D^0\, \Xi^{*-} \bar\Xi^{*+} $~&~$\bar B^0_d\rightarrow D^0\, \,\Sigma^{*-} \bar\Sigma^{*+} $~& 
~~$(-,+,--+,-++)$~&
~~$(2,2,3)$~~ & 6 & $-1$\\
\hline
\end{tabular}
\caption{The mapping of the $U$-spin pairs of $\bar B^0 \to D^0 \Omega_B^- \bar\Omega_B^+$ decays into generalized $n$-tuples. \label{tab:map-b-baryons}}
\end{table}

The $a$-type sum rules that are valid up to $b=0$ are the trivial ones, and we do not write them explicitly. The $s$-type sum rules that are valid up to $b=1$ are given by 
\begin{align}
2 s_{(1,2,2)}+ \sqrt{3} s_{(1,2,3)} &=0\,, \\
2 s_{(1,3,3)}+ \sqrt{3} s_{(1,2,3)} &=0\,, \\
s_{(1,2,2)}+\sqrt{3}s_{(2,2,2)}+ \sqrt{3}s_{(2,2,3)} &=0\,,\\
s_{(1,2,3)}+ 2s_{(2,2,3)}+ 2s_{(2,3,3)}&=0\,,\\
s_{(1,3,3)}+\sqrt{3}s_{(3,3,3)}+ \sqrt{3}s_{(2,3,3)} &=0\,.
\end{align}
For the $a$-type sum rules that are valid up to $b=2$, we need to calculate the $W_b=M_b\times \mu$ factors. We write the results with explicit product of the $M_b$ and $\mu$-factors:
\begin{align}
1 \times 2 \sqrt{3} a_{(1,2,2)}+ 2 \times 3a_{(1,2,3)}+ 1 \times 2 \sqrt{3} a_{(1,3,3)}&=0, \\
2 \times 2\sqrt{3} a_{(1,2,2)}+ 2 \times 3 a_{(1,2,3)}+ 1\times 6 a_{(2,2,2)}+ 2 \times 6 a_{(2,2,3)} + 1 \times 6 a_{(2,3,3)} &=0, \\
2 \times 2\sqrt{3} a_{(1,3,3)}+ 2 \times 3 a_{(1,2,3)}+ 1\times 6 a_{(3,3,3)}+ 1 \times 6 a_{(2,2,3)} + 2 \times 6 a_{(2,3,3)} &=0.
\end{align}
simplifying we can write the sum rules as
\begin{align}
a_{(1,2,2)}+ 
\sqrt{3} a_{(1,2,3)}+  a_{(1,3,3)}&=0, \\
2\sqrt{3} a_{(1,2,2)}+  3 a_{(1,2,3)}+ 3 a_{(2,2,2)}+ 6 a_{(2,2,3)} + 3 a_{(2,3,3)} &=0, \\
2\sqrt{3} a_{(1,3,3)}+ 3 a_{(1,2,3)}+ 3 a_{(3,3,3)}+ 3 a_{(2,2,3)} + 6 a_{(2,3,3)} &=0.
\end{align}

The $s$-type sum rule for $b=3$ is given by
\begin{align}
&
3 \times 2 \sqrt{3} s_{(1,2,2)}+ 3 \times 2 \sqrt{3} s_{(1,3,3)}+
6 \times 3 s_{(1,2,3)}+ \nonumber
\\ & ~~~~~~ 1 \times 6 s_{(2,2,2)}+ 1\times 6 s_{(3,3,3)}+ 3 \times 6 s_{(2,2,3)} + 3 \times 6 s_{(2,3,3)} =0.
\end{align}
Simplifying we get
\begin{align}
\sqrt{3} s_{(1,2,2)}+ \sqrt{3} s_{(1,3,3)}+
3 s_{(1,2,3)}+ s_{(2,2,2)}+ s_{(3,3,3)}+ 3 s_{(2,2,3)} + 3 s_{(2,3,3)} &=0.
\end{align}

\section{Conclusion and discussion \label{sec:conclusions}}

We have studied the general group-theoretical properties of $SU(2)$ amplitude sum rules, with particular emphasis on $U$-spin amplitude sum rules, and have found that there is a rich mathematical structure underlying them.

We have found that the basis of $a$- and $s$-type amplitudes is particularly useful for writing $SU(2)$ flavour sum rules. All the sum rules at any order of breaking can be written in terms of $a$- and $s$-type amplitudes. The mathematical structure that we found allowed us to formulate a straightforward algorithm to derive all the sum rules to all orders in symmetry breaking without the need to explicitly express the amplitudes in terms of reduced matrix elements (see Section~\ref{sec:algorithm}). The formulated method is computationally much simpler than the standard method of deriving the sum rules.

The sum rules for most of the systems that have been studied experimentally have been already derived using the standard approach. Yet, for some of them the higher order corrections were not studied before. We derived the sum rules, including  all higher order corrections for them, see Section~\ref{sec:examples} for the examples. While the experimental precision for these systems does not require it yet, our hope is that the higher order corrections will eventually become relevant as the experimental precision is getting better.

There are several future directions we plan to take. Below we list them.

\myben
\myitt
In this work we only discuss the amplitude sum rules. It is not trivial how to relate them to physical observables such as decay rates and CP asymmetries. This is in particular problematic when we go beyond the leading order. The reason is that the amplitudes are functions of kinematic variables, while the measurements are done at some specific kinematic points. This issue is most problematic in the presence of narrow resonances and on the boundary of phase space.

\myitt
One of the main simplifying assumptions in this work is that we work with processes where the Hamiltonian is given by only one $U$-spin representation. It is interesting, both theoretically and practically, to generalize our results to the case where the Hamiltonian is given by a sum of several different representations.

\myitt 
One other assumption that we have made, and would like to relax in the future, is that the initial and final state particles have well defined properties under $U$-spin. Physical particles, however, at times are given by mixtures of several representations. 

\myitt 
One more point to study is the case of
processes with identical particles. In practice, this situation emerges only in two-body decays. For more particles in the final state the momentum is in general different and thus the particles are not truly identical. Yet, it is interesting to discuss this also from the theoretical point of view.

\myitt
In this paper we only consider the $SU(2)$ flavor group. It seems plausible that similar results can be obtained for $SU(3)$.

\myitt
One other point that we did not discuss in this work is the freedom to redefine amplitudes, in particular by a phase. In this work, we have chosen one convention and follow it everywhere. In future work we would like to understand however which phases are physical and which are not.

To conclude, our hope is that a deeper understanding of flavor symmetries and higher order amplitude sum rules would eventually lead to performing precise measurements of fundamental parameters that are related to flavor physics.

\begin{acknowledgements}
We thank Saquib Hassan, Zoltan Ligeti, Wee Hao Ng,
Dean Robinson, and Yotam Soreq for helpful discussions. 
The work of YG is supported in part by the NSF grant PHY1316222.
S.S. is supported by a Stephen Hawking Fellowship from UKRI under reference EP/T01623X/1 and the Lancaster-Manchester-Sheffield Consortium for Fundamental Physics, under STFC research grant ST/T001038/1.
For the purpose of open access, the authors have applied a Creative Commons Attribution (CC BY) licence to any Authors Accepted Manuscript version arising.
This work uses existing data which is available at locations cited in the bibliography.

\end{acknowledgements}

\begin{appendix}

\section{Rotation between Physical and $U$-spin bases}\label{app:physical_vs_Uspin_basis}

We start our analysis with \emph{a set of states} related by $U$-spin (not to be confused with a $U$-spin set of amplitudes). Such a $U$-spin set of states is fully defined by listing the representations forming the states. If we talk specifically about the initial or final state this would be a list of $U$-spin multiplets of particles in the initial/final state.

Consider a state that is described by $g\ge 2$ QNs. We distinguish between two types of QNs: $U$-type and $m$-type. The $U$-type QNs are the ones that describe the total $U$-spin of states and $m$-type QNs are the ones that describe the $m$-type QNs. Specific states can be expressed in different bases. We are in particular using the following two:
\begin{enumerate}
\item 
{\it Physical basis:} This basis is what is commonly referred to as a ``product state basis.'' In this basis all the $g$ QNs are $m$-type.
\item
{\it $U$-spin basis:}
This basis is commonly referred to as a ``basis of definite value of total $U$-spin.'' In this basis the states are defined by one $m$-type QN and $g-1$ $U$-type QNs. 
\end{enumerate}

The physical basis is unique up to reordering of multiplets and their corresponding $m$-QNs, which does not change the states, while, as we explain later, there might exist many different $U$-spin bases.

In what follows we elaborate on these definitions and formulate how to perform the basis rotation between the two bases. Let us consider a system of states described by $g$ irreps $u_1, u_2, \dots, u_g$. For shortness we denote the entire set of representations forming the system as $\bar{u}$. That is we define
\begin{equation}
    \bar{u} = \{u_1, u_2, \dots, u_g\},
\end{equation}
where we use the bar to represent a set of similar objects. (While this is the same notation as for anti-up quark the context makes it clear what is meant.)

Now, once the system of states is defined via listing the representations $\bar{u}$ we move to describing the states themselves. 

In the physical basis each state in the $U$-spin set of states that we consider can be described by $g$ $m$-QNs: $m_1, m_2, \dots, m_g$, which are the third-component projections of each of the representations in the set $\bar{u}$. We denote such set as $\bar{m}$ and we write
\begin{equation}
    \bar{m} = \{m_1, m_2, \dots, m_g\}.
\end{equation}
We emphasize that the set $\bar u$ describes the system and thus is the same for all the states of the set of states, while different sets $\bar m$ describe different states within the set. 

In the following we adopt the notation $\ket{*;*}$ for states where the semicolon divides the QNs that describe the system and the QNs that describe a specific state of the system. At times, however, for shortness, we omit the part that describes the system and simply use $\ket{*}$ to describe the states.

To represent a state that belongs to the system of states $\bar{u}$ and is described by the set $\bar m$ in the physical basis we use $\ket{\bar u;\bar m}$, which is given explicitly by the following tensor product
\begin{equation}\label{eq:um_state_in_physical_basis}
    \ket{\bar{u};\bar{m}} \equiv \ket{u_1;m_1}\otimes\ket{u_2; m_2}\otimes \dots \otimes \ket{u_g;m_g},
\end{equation}
thus the alternative name ``product state basis.'' At times the $\bar u$ in the above is implicit and we write the state as $\ket{\bar m}$.

In the $U$-spin basis each state of the system under consideration is described by one $m$-type QN which we denote as $M$ and by a set of $g-1$ $U$-type QNs that we denote as 
\beq
\bar U = \{U_1, U_2, \dots,U_{g-1}\}.
\eeq
Note that we use capital $U$ here to distinguish from the $u$ that describes the system. The $m$-type QN $M$ is given by
\begin{equation}
    M = \sum_{j=1}^g m_j,
\end{equation}
and it is the total $m$-QN of the state. With this, each state in the $U$-spin basis can be denoted as $\ket{\bar{u};\bar U, M} \equiv \ket{\bar U, M}$, where we omit the $\bar u$ for brevity of notation.

The difference between the two bases $\ket{\bar{m}}$ and $\ket{\bar{U}, M}$ is that the former one is written as a product of states, while the latter one arises from explicitly taking the tensor products in Eq.~\eqref{eq:um_state_in_physical_basis}. There are $g-1$ tensor products in Eq.~\eqref{eq:um_state_in_physical_basis} and the $g-1$ elements of the set $\bar{U}$ are the values of the total $U$-spin taken one from each tensor product. The last element of the set $\bar{U}$, that is $U_{g-1}$, gives the total $U$-spin of the state in the $U$-spin basis.

In general, for $g\ge3$, there are many basis choices that result in different $U$-spin bases. In these different bases each state has the same $M$ and total $U$-spin, but the other $U$-spin QNs could be different depending on the order of the tensor product.

To better understand the definitions of the physical and $U$-spin bases, consider as an example the tensor product of three doublets:
\begin{equation}
    \left(\frac{1}{2} \otimes \frac{1}{2}\right)\otimes \frac{1}{2} = \left(0 \oplus 1\right)\otimes \frac{1}{2} = \left(\frac{1}{2}\right)_0\oplus \left(\frac{1}{2}\right)_1\oplus\left(\frac{3}{2}\right)_1.
\end{equation}
The subscripts $0$ and $1$ on the RHS indicate from which intermediate representations the representations $(1/2)$, $(1/2)$ and $(3/2)$ are combined. Note that we distinguish between representations coming from different intermediate terms even if their total $U$-spin is the same. In this example of three doublets there are three possible sets of $\bar U= \{U_1,U_2\}$:
\begin{equation}\label{eq:eq:ubar-options}
    \left\{0, \frac{1}{2}\right\}, \qquad \left\{1, \frac{1}{2}\right\}, \qquad \left\{1,\frac{3}{2}\right\}.
\end{equation}
There are 8 states in this system. In the physical basis they are  
\beq
\ket{m_1,m_2,m_3},
\eeq
where each $m_j$ can assume a value of $\pm 1/2$. In the $U$-spin basis the states are
\beq
\ket{0,1/2,M}, \qquad
\ket{1,1/2,M}, \qquad
\ket{1,3/2,M}.
\eeq
with $-U_2 \le M \le U_2$ and $U_2$ is the total $U$-spin of the state. As we see the number of states in the $U$-spin basis is also equal to 8 as it should be.

The rotation between the physical basis and the $U$-spin basis can be written as follows
\begin{equation}\label{eq:basis_rot_def}
    \ket{\bar u; \bar m} = \sum_{\bar U} C^*(\bar u; \bar{m}, \bar U) \ket{\bar U, M},
\end{equation}
where the coefficients $C^*(\bar u; \bar{m}, \bar U)$ are given by products of Clebsch-Gordan coefficients.
Note that the sum in Eq.~(\ref{eq:basis_rot_def}) goes over the different $\bar{U}$ sets, not the elements of one particular $\bar{U}$. In the example above that would, for instance, be a sum over three sets listed in Eq.~\eqref{eq:eq:ubar-options}.

To write the coefficients $C^*(\bar u; \bar{m}, \bar U)$ explicitly one needs to specify a concrete $U$-spin basis, that is, to specify the order of the tensor product. When the tensor product is taken iteratively in the order in which the representations are listed in the set $\bar u$, then the coefficients $C^*(\bar u; \bar{m}, \bar U)$ are given by
\begin{align}\label{eq:C*_def}
C^*(\bar u; \bar m, \bar U) = \mathop{C_{u_1, m_1}}_{\hspace{8pt} u_2, m_2}^{\hspace{8pt} U_1, M_1} \times \mathop{C_{U_1, M_1}}_{\hspace{8pt} u_3, m_3}^{\hspace{8pt} U_2, M_2} \times
...\times \mathop{C_{U_{r-2}, M_{r-2}}}_{\hspace{-10pt} u_r, m_r}^{\hspace{-4pt} U_{r-1}, M},
\end{align}
where 
\begin{align}
	\mathop{C_{u_j, m_j}}_{\hspace{8pt} u_k, m_k}^{\hspace{8pt} U, M} = \braket{u_j\, m_j\, u_k\, m_k}{U\, M} 
\end{align}
are the Clebsch-Gordan coefficients and
\beq
M_j = \sum_{k=1}^{j+1} m_k, \qquad M=M_{g-1}=\sum_{k=1}^{g} m_k.
\eeq
Recall that the parameters before the semicolon describe the system and can be omitted. Thus, sometimes we simply use $C^*(\bar u; \bar m, \bar U)\equiv C^*(\bar{m}, \bar{U})$.\\

Several remarks are in order. First, above we only discuss the states described by $g \ge 2$ QNs. If $g < 2$ there are two possibilities:
\begin{enumerate}
    \item[(i)] $g = 1$: the state is described by one non-trivial irrep and thus a single $m$-type QN,
    \item[(ii)] $g = 0$: the state is a $U$-spin singlet.
\end{enumerate}
For both of these cases, the physical basis and the $U$-spin basis are the same. Thus for both $g=1$ and $g=0$ we say that the set of $U$-type QNs for the state is an empty set $\bar U = \emptyset$ (as in this case no product can be formed). For the case of $g=1$ we have for the $C^*$ coefficients
\begin{equation}
    C^*(u_1; m_1, \emptyset) \equiv 1, \qquad \forall u_1, -u_1 \le m_1 \le u_1.
\end{equation}
Note $u_1$ and $m_1$ are single elements of $\bar u$ and $\bar m$, and for simplicity, we write in these cases directly the single elements instead of the corresponding sets of one element.

In the case of a singlet state $g=0$ we have $\bar u = \bar m = \emptyset$ and thus we define
\begin{equation}
    C^*(\emptyset;\emptyset,\emptyset) \equiv 1.
\end{equation}

Second, some care has to be taken when there are identical irreps in the system. Consider, for example, two identical doublets. In the physical basis the fact that they are identical can be written as 
\beq
\ket{+1/2,-1/2} = \ket {-1/2,+1/2}\,.
\eeq
In the $U$-spin basis the fact that they are identical implies only the triplet combination is possible and the singlet combination is identically zero.

More generally, the presence of identical irreps in the system leads to some states being identical in the physical basis.
In the $U$-spin basis this implies that some sets $\bar U$ that occur for distinguishable irreps are identically zero.

We conclude this section with a remark about physical and $U$-spin bases for amplitudes. We say that an amplitude is written in the physical ($U$-spin) basis when the initial state, final state and the Hamiltonian are in the physical ($U$-spin) basis. 

\section{Decomposition of amplitudes in terms of RMEs \label{app:RMEdecomposition}}

In this appendix we show how basis rotations together with the Wigner-Eckart theorem allow to express amplitudes of physical processes in terms of RMEs, that is, we derive  Eq.~\eqref{eq:factorization}, which we rewrite below for convenience:
\begin{equation}\label{eq:decomposAi}
    \mathcal{A}_j = f_{u,m} \sum_\alpha C_{j\alpha} X_\alpha\,.
\end{equation}
In what follows we give explicit definitions for the coefficients $C_{j\alpha}$ and the RMEs $X_\alpha$.

\subsection{Defining the $U$-spin set}\label{app:def-u-set}

To define a generic $U$-spin set it is necessary to describe the $U$-spin structure of the initial state, the final state, and the Hamiltonian. In this appendix we consider a $U$-spin set which is described by $g_I$ $U$-spin irreps in the initial state and $g_F$ irreps in the final state, where the irreps can be arbitrary. (Note that in practice we are usually interested in decays and thus $g_I=1$. Yet here we consider the general case.) For the Hamiltonian, the number of irreps in the $U$-spin limit is one, see Sec.~\ref{sec:definitions}. This irrep can also be the singlet operator.

To describe the $U$-spin structure of the initial and final states of the system we use
\beq \label{eq:irreps_in_out}
\bar u^I  = \{u_1^I, u_2^I, ... u^I_{g_I}\}, \qquad
\bar u^F  = \{u_1^F, u_2^F, ... u^F_{g_F}\},
\eeq
where $u^I_j$($u^F_j$) denote the sets of $g_I$($g_F$) representations in the initial(final) state. The $U$-spin of the Hamiltonian is denoted as $u$. The $U$-spin system is fully described by $\bar u^I$, $\bar u^F$, and $u$.

Each amplitude from the $U$-spin set under consideration can be described by a set of $(g_I+g_F)$ QNs. In the physical basis they are the $m$-type QN of the specific components of the multiplets in Eq.~\eqref{eq:irreps_in_out}. We denote the corresponding sets of $m$-QNs as $\bar{m}^I$ and $\bar{m}^F$, where
\begin{equation}
    \bar{m}^I = \{m_1, m_2, \dots, m_{g_I}\}, \qquad \bar{m}^F = \{m_1, m_2, \dots, m_{g_F}\}.
\end{equation}
With this, according to the notation we introduced in Appendix~\ref{app:physical_vs_Uspin_basis} the initial and final states of the $U$-spin system in the physical basis are denoted as
\begin{equation}\label{eq:m_of_irreps_in_out}
\ket{\bar m^I} =\ket{ m^I_1, m^I_2,..., m^I_{g_I}}, \qquad   
\ket{\bar m^F} = \ket{m^F_1, m^F_2,..., m^F_{g_F}}.
\end{equation}
Each amplitude in the physical basis is then described by $\bar m^I$ and $\bar m^F$. We use $m$ to denote the $m$ of the Hamiltonian that contributes to the amplitude. Note that for non-zero amplitudes $m$ is not an independent QN and is given by
\beq\label{eq:mH-def}
m = \sum_{j=1}^{g_I} m^I_j - \sum_{j=1}^{g_F} m^F_j.
\eeq
Thus, for a given $U$-spin set, each amplitude can be indexed by a pair $(\bar m^I, \bar m^F)$. 
This allows us to denote the amplitudes from any $U$-spin set as $\mathcal{A}_{(\bar m^I, \bar m^F)}$.

\subsection{The Wigner-Eckart theorem }
The Wigner-Eckart theorem states that for any spherical operator $O(u,m)$ and two states of angular momentum $\ket{u_1;m_1}$ and $\ket{u_2;m_2}$, the matrix element of the operator between the two states can be written as a product of a factor that depends on $m$-QNs and a factor that only depends on the values of the total $U$-spin. Formally the statement of the theorem is given by the following relation
\begin{equation}\label{eq:WE}
    \mel{u_2; m_2}{O(u,m)}{u_1; m_1} = \mathop{C_{u_1, m_1}}_{\hspace{2pt} u, m}^{\hspace{8pt} u_2, m_2} \mel{u_2}{O(u)}{u_1},
\end{equation}
where the $m$-dependent factor is given by the Clebsch-Gordan coefficient. The $m$-independent factor, $\mel{u_2}{O(u)}{u_1}$, is called Reduced Matrix Element (RME).

The situation becomes more complicated when the states that we work with are given by products of several different states. In terms of the amplitudes in the physical basis this would correspond to $g_I >1$ and/or $g_F>1$. To apply the Wigner-Eckart theorem to such states we need to work in the $U$-spin basis where each state has definite value of total $U$-spin.

Consider two states in the $U$-spin basis
\beq\label{eq:U_basis_states_WE}
\ket{\bar U^I,M^I}, \qquad
\ket{\bar U^F,M^F}.
\eeq
The last elements in the sets $\bar U^{I}$ ($\bar U^F$) is $U_{g_I-1}^{I}$ ($U^F_{g_F-1}$) and it gives the total $U$-spin of the states. Introducing the notations
\begin{equation}\label{eq:UT_notation}
    U^I_T = U^I_{g_I-1}, \qquad U^F_T = U^F_{g_F-1},
\end{equation}
where the label \lq\lq{}$T$\rq\rq{} stands for the ``total'' value of $U$-spin, the Wigner-Eckart theorem can be rewritten for the case of the states in Eq.~\eqref{eq:U_basis_states_WE} as follows 
\begin{equation}\label{eq:WE-gen}
\mel{\bar U^F, M^F}{O(u,m)}{\bar U^I, M^I } = \mathop{C_{U^I_T, M^I}}_{\hspace{-4pt} u, m}^{\hspace{12pt} U_T^F, M^F} \times \mel{\bar U^F}{O(u)}{\bar U^I}.
\end{equation}
That is, the matrix element in the $U$-spin basis is proportional to a RME that does not depend on the $m$-type QNs, but depends on all the $U$-type QNs. The CG coefficient depends on the $m$-QNs as well as the total $U$-spin of the initial state, final state, and the operator. Note that there is only one CG coefficient and not a product of them.

We conclude that in order to write the decomposition of the amplitudes
$\mathcal{A}_{(\bar m^I, \bar m^F)}$ in terms of RMEs,
we need to, first, perform the basis rotation from the physical basis to the $U$-spin basis and then apply the Wigner-Eckart theorem according to Eq.~\eqref{eq:WE-gen}. In the following subsections we elaborate on  these two steps. 

\subsection{Basis rotation}

First, we perform the basis rotation for the states from the physical basis to the $U$-spin basis using Eq.~\eqref{eq:basis_rot_def}. For the initial state we have
\begin{equation}\label{eq:in_decomp}
\ket{\text{in}} = \ket{\bar m^I} = \sum_{\bar{U}^{I}} C^*(\bar{u}^{I}; \bar{m}^I, \bar U^I) \ket{\bar U^I, M^I}, \qquad M^I=\sum_{j=1}^{g_I} m^I_j,
\end{equation}
and similarly for the final state
\begin{equation}\label{eq:out_decomp}
\ket{\text{out}} = \ket{\bar m^F} = \sum_{\bar{U}^{F}} C^*(\bar{u}^{F}; \bar{m}^F, \bar U^F) \ket{\bar U^F, M^F}, \qquad M^F=\sum_{j=1}^{g_F} m^F_j.
\end{equation}
In the equations above, $\bar U^I$ and $\bar U^F$ are the sets of $U$-type QNs for the initial and final states, respectively.

Next, we consider the Hamiltonian. The Hamiltonian in the $U$-spin limit takes the following form
\begin{equation}
\mathcal{H}_{\text{eff}}^{(0)} \, = \, \sum_{m} f_{u, m} \, H(u,m)\,.
\end{equation}
where for clarity we use $H^u_m \equiv H(u,m)$ and as everywhere in this paper we focus on the Hamiltonians with only one fixed value of $U$-spin, in this case it is denoted as $u$.

Taking into account $U$-spin breaking results in an effective Hamiltonian as given in Eq.~\eqref{eq:Heff}, which we rewrite here in a slightly modified notation as 
\begin{equation}
    \mathcal{H}_\text{eff} = \sum_{m,b} f_{u,m} \, H(u,m)\otimes H_\varepsilon(1,0)^{\otimes b}.
\end{equation}
We use $H(u,m,b)$ to denote an operator of order $b$ in the Hamiltonian above, that is we define
\begin{equation}
    H(u,m,b) \equiv H(u,m)\otimes H_\varepsilon(1,0)^{\otimes b}.
\end{equation}
This term in the Hamiltonian is written in the physical basis. Now, we would like to perform a basis rotation for this operator. Using Eq.~\eqref{eq:basis_rot_def} we obtain
\beq
H(u,m,b) = \sum_{\bar U} C^*(\bar u^H;\bar m^H,\bar U) H(\bar U,m,b),
\eeq
where the sets $\bar u^H$ and $\bar m^H$ are both of length $b+1$ and their elements are given by
\begin{align}\label{eq:uH-mH-de}
    u^H_1 = u, \qquad u^H_{j+1} = 1, \qquad 1 \le j \le b, \nonumber\\
    m^H_1 = m, \qquad m^H_{j+1} = 0, \qquad 1 \le j \le b.
\end{align}
$\bar U$ is a set of $U$-type QNs for the Hamiltonian to order $b$.
Note that the $b$ in $H(\bar U,m,b)$ is redundant as it is included in $\bar U$ that has a length of $b$. Yet, due to its importance we keep it explicitly. Finally we write the effective Hamiltonian in the $U$-spin basis as
\begin{equation} \label{eq:H_decomp}
\mathcal{H}_{\text{eff}}\, = \, \sum_{m,b} f_{u,m} \, 
\left(\sum_{\bar U} C^*(\bar u^H;\bar m^H,\bar U) H(\bar U,m,b)\right)\,.
\end{equation}

\subsection{Applying the Wigner-Eckart theorem}

Equipped with the decompositions in Eqs.~\eqref{eq:in_decomp}, \eqref{eq:out_decomp}, and \eqref{eq:H_decomp} and the Wigner-Eckart theorem, Eq.~\eqref{eq:WE-gen}, we finally can write the amplitudes in terms of RMEs and recover the expression in Eq.~\eqref{eq:decomposAi} with amplitudes
\beq\label{eq:Aj=AmImF}
\A_j \equiv
\A_{(\bar m^{I},\bar m^{F})},
\eeq
the RMEs
\begin{equation}\label{eq:X-def}
    X_\alpha \equiv \mel{\bar U^F}{H(\bar U,b)}{\bar U^I},
\end{equation}
and the multi-index $\alpha$ given by
\beq \label{eq:def-alpha}
\alpha \equiv \{\bar U^I,\bar U^F, \bar U, b\}.
\eeq
The coefficients $C_{j\alpha}$ are given by
\beq\label{eq:Cjalpha}
C_{j\alpha} \equiv C^*(\bar{u}^{I}; \bar{m}^I, \bar U^I) \times
C^*(\bar{u}^{F}; \bar{m}^F, \bar U^F) \times C^*(\bar u^H;\bar m^H,\bar U) \times
\mathop{C_{U_T^I, M^I}}_{\hspace{2pt} U_T, m}^{\hspace{8pt} U_T^F, M^F}\, ,
\eeq
where we used the notation for the total $U$-spin of states in the $U$-spin basis from Eq.~\eqref{eq:UT_notation} and introduced $U_T$ for the total $U$-spin of operators in the Hamiltonian, which is equal to the last element of $\bar{U}$, that is $U_T \equiv U_b$.

Using $\alpha$ from Eq.~\eqref{eq:def-alpha} and 
$j= \{\bar m^I,\bar m^F\}$ at times we also write the $m$-QN dependence explicitly
\beq\label{eq:Cjalpha-notation}
C_{j\alpha} =  
C(\bar m^I,\bar m^F, \alpha). 
\eeq

\section{Relation between decomposition of amplitudes forming a $U$-spin pair}\label{app:Upair_relation}

In this appendix we prove the relation between the RME decompositions of the amplitudes in Eqs.~\eqref{eq:factorization-again} and \eqref{eq:u-pair}. To establish the relation between the two amplitudes that form a $U$-spin pair, we use the following symmetry property of the CG coefficients
\begin{equation}\label{eq:CG_sym}
\mathop{C_{u_1, m_1}}_{\hspace{8pt} u_2, m_2}^{\hspace{25pt} u_3, m_1 + m_2} = (-1)^{u_1 + u_2 - u_3} \times \!\!\!\!\!\!\!\mathop{C_{u_1, -m_1}}_{\hspace{8pt} u_2, -m_2}^{\hspace{25pt} u_3, -m_1 - m_2}.
\end{equation}
For a given $U$-spin set we consider the CKM-free amplitude 
\begin{equation}\label{eq:Aif}
A_i\equiv A_{(\bar m^{I}, \bar m^{F})} = \sum_\alpha C(\bar m^I,\bar m^F,  \alpha) X_\alpha\,,
\end{equation}
where we use the notation introduced in Appendix~\ref{app:RMEdecomposition}. The $U$-spin pair amplitude is expressed through the same set of RMEs, since they do not depend on the $m$-QNs, but with different coefficients:
\begin{equation}\label{eq:Aifpair}
A_\ell\equiv A_{(-\bar m^{I}, -\bar m^{F})} = \sum_\alpha C(-\bar m^I,-\bar m^F,  \alpha) X_\alpha\,,
\end{equation}
Using Eqs.~\eqref{eq:Cjalpha},~\eqref{eq:C*_def}, and the symmetry property in Eq.~\eqref{eq:CG_sym} we find that
\begin{equation}\label{eq:C_i_l_relation}
C(\bar m^{I},\bar m^{F},\alpha) = (-1)^{p + b} C(-\bar m^{I},-\bar m^{F},\alpha),
\end{equation}
where the parity $(-1)^p$ can be defined as follows
\begin{equation}\label{eq:p_1st_version}
(-1)^p = (-1)^{-\sum_{j = 1}^{g_I} u^{I}_j - \sum_{j = 1}^{g_F} u^{F}_j + 2U^{F}_{g_F-1} - u}.
\end{equation}
As we see the factor $(-1)^p$ is the same for all the amplitudes of a system and thus we find the following expression for the $U$-spin pair amplitude $A_\ell$ of the amplitude $A_i$:
\begin{align}
\label{eq:Upair_decomp_theorem}
A_\ell &= 
(-1)^p \sum_\alpha (-1)^b  C(\bar m^I,\bar m^F,  \alpha) X_\alpha\,.
\end{align}

Next, several comments about Eq.~\eqref{eq:p_1st_version} are in order. First, even though when deriving the results in Eqs.~\eqref{eq:C_i_l_relation} and~\eqref{eq:p_1st_version} we referred to Eq.~\eqref{eq:C*_def}, which gives the expression of $C^*$ coefficients for a specific basis choice, the results are in fact basis independent. The basis independence, i.e. the independence on the specific order of the tensor products is achieved since all the intermediate representations are always bound to cancel due to the minus sign in front of $u_3$ in Eq.~\eqref{eq:CG_sym}.

Second, note the following:
\begin{enumerate}
\item 
The power in Eq.~\eqref{eq:p_1st_version} is always integer and thus $(-1)^p = \pm 1$.
\item
The expressions $\sum_{j = 1}^{g_F} u^{F}_j$ and $U^{F}_{g_F-1}$ are both either half-integer or integer and therefore $2\sum_{j = 1}^{g_F} u^{F}_j$ and $2 U^{F}_{g_F-1}$ have the same parity. Thus we can write
\begin{align}
(-1)^p = (-1)^{-\sum_{j = 1}^{g_I} u^{I}_j  + \sum_{j = 1}^{g_F} u^{F}_j-u} \,.
\end{align}
\item
Recalling that $n$ is the number of would-be doublets for the system we have
\beq
\sum_{j = 1}^{g_I} u^{I}_j + \sum_{j = 1}^{g_F} u^{F}_j  + u = {n \over 2}.
\eeq
\end{enumerate}
These properties allow us to introduce the following definition for $p$
\begin{equation}\label{eq:p_def}
p = \sum_{j = 1}^{g_F} u^{F}_j-u -\sum_{j = 1}^{g_I} u^{I}_j   = 
 2\sum_{j = 1}^{g_F} u^{F}_j - {n \over 2}.
\end{equation}
This, of course, is only one of many possible definitions. For consistency, everywhere in this work we use Eq.~\eqref{eq:p_def} as the definition for $p$.

Next, consider two special cases:
\begin{enumerate}
\item [$(i)$]
The process is entirely described by $n$ $U$-spin doublets in the final state. In this case Eq.~\eqref{eq:p_def} results in $p=n/2$.
\item [$(ii)$] 
All the irreps of the system are integers. In this case the sum $2\sum_{j = 1}^{g_F} u^{F}_j$ is an even number and thus the parity of $p$ is determined by $n/2$, so we can set $p = n/2$.
\end{enumerate}
We thus conclude that in both of the special cases above the $p$-factor can be chosen as
\begin{equation}\label{eq:p_n_doublets}
p = \frac{n}{2}.
\end{equation}

\section{Universality of sum rules}\label{app:signs}

In this appendix we compare two $U$-spin sets that belong to the same universality class, see Section~\ref{sec:universality}. We first show that one can establish a one-to-one correspondence between the amplitudes for any two $U$-spin sets from the same universality class. We then show that there is also a one-to-one correspondence between the RMEs of the two systems. With these correspondences in mind, we derive the relation between the coefficients that enter the group theoretical decompositions of the amplitudes for the two systems. This explicit result shows that the sum rules for any two $U$-spin sets from the same universality class are the same up to relative signs between the amplitudes.

We consider the following two systems:
\begin{enumerate}
\item \textbf{System I} is very general and is described by a set of $k$ representations $\bar{u}^{(1)}$ in the initial state, a set of $l$ representations $\bar u^{(2)}$ in the final state and a Hamiltonian with $U$-spin $u^{(3)}$. Using the notation of Appendix~\ref{app:def-u-set}, we have for System~I
\begin{align}
g_I = k, \qquad u^I_j &= u^{(1)}_j, \qquad 1 \le j \le k, \nonumber \\
g_F = l, \qquad u^F_j &= u^{(2)}_j, \qquad 1 \le j \le l, \nonumber \\
u &= u^{(3)}.
\end{align}
For the sets of irreps in the initial and final states we write
\begin{equation}
    \bar{u}^I = \bar u^{(1)}, \qquad \bar{u}^F = \bar u^{(2)}.
\end{equation}
The CKM-free amplitudes of the System I are denoted as $A^{(\text{I})}_j$.

\item \textbf{System II} is described by the same $k+l+1$ representations as System I, but all of them belong to the final state (the Hamiltonian and the initial state are given by singlet states). Using the notation of Appendix~\ref{app:def-u-set} we write for the initial state and the Hamiltonian of System II
\begin{equation}
    g_I = 0, \qquad \bar u^I = \emptyset, \qquad u = 0,
\end{equation}
and for the final state $g_F = k+l+1$ and the elements of the set $\bar u^F$ are given by
\begin{align}
u^F_j = u^{(1)}_j, & \qquad 1 \le j \le k,\nonumber \\
u^F_{j+k} = u^{(2)}_{j}, & \qquad 1 \le j \le l, \nonumber\\
u^{F}_{k+l+1} = u^{(3)}.
\end{align}
The set of the final state irreps can be written as a union of sets as follows
\begin{equation}
    \bar{u}^F = \bar{u}^{(1)}\cup\, \bar{u}^{(2)} \cup\, u^{(3)}.
\end{equation}
The CKM-free amplitudes of System~II are denoted as $A^{(\text{II})}_j$.
\end{enumerate}

Systems I and II are described by the same sets of irreps and the only difference is the assignment of the irreps to the initial/final state and the Hamiltonian. Establishing the one-to-one correspondence between the amplitudes of the two systems is trivial. This can be done using $n$-tuples. If we construct the $n$-tuples for the two systems using the same assignment of $n$-tuple positions to representations, the two systems are described by two sets of $n$-tuples that are exactly the same. Thus we can say that an amplitude $A_{j_1}^{(\text{I})}$ is mapped into an amplitude $A_{j_2}^{(\text{II})}$ if and only if the two are described by the same $n$-tuple. In the index notation that we introduce in Section~\ref{sec:An-tuples-doublets} and generalize in Section~\ref{sec:n-tuples-generalized} this also implies that $j_1 = j_2$.

Next we move to establishing the one-to-one correspondence between the RME of System~I and System II.

\subsection{System I}
First, we consider System I. To describe an amplitude from this system it is enough to list the $m$-QNs of the representations in the initial and final states. We consider an amplitude that is described by a set of $m$-QNs $\bar m^{(1)}$ in the initial state and a set $\bar m^{(2)}$ in the final state:
\begin{align}
    m^I_{j} &= m^{(1)}_j, \qquad 1 \le j\le k, \nonumber \\
    m^{F}_{j} &= m^{(2)}_j, \qquad 1 \le j \le l,
\end{align}
that is for the amplitude under consideration we have
\begin{equation}\label{eq:SystemI-QNs}
    \bar m^I = \bar m^{(1)}, \qquad \bar m^{F} = \bar m^{(2)}.
\end{equation}

Introducing
\begin{equation}
    M^{(1)} = \sum_{j = 1}^{k} m^{(1)}_j, \qquad  M^{(2)} = \sum_{j = 1}^{l} m^{(2)}_j,
\end{equation}
we can write for the amplitude
\begin{equation}\label{eq:Mrelations}
    M^I = M^{(1)}, \qquad M^F = M^{(2)}, \qquad m = m^{(3)} = M^{(2)} - M^{(1)},
\end{equation}
where $m$, $M^I$, and $M^F$ are defined in Eqs.~\eqref{eq:mH-def}, \eqref{eq:in_decomp}, and \eqref{eq:out_decomp} respectively.

The sets $\bar u^H$ and $\bar m^H$ for a given order of breaking are defined in Eq.~\eqref{eq:uH-mH-def}. For the example we consider, the elements of these sets take the following values
\begin{align}\label{eq:uH-mH-def}
    u^H_1 = u^{(3)}, \qquad u^H_{j+1} = 1, \qquad 1 \le j \le b, \nonumber\\
    m^H_1 = m^{(3)}, \qquad m^H_{j+1} = 0, \qquad 1 \le j \le b.
\end{align}
Introducing two sets of $b$ elements $\bar{u}^{(\text{br})}$ and $\bar{m}^{(\text{br})}$ (where the label (br) stands for breaking) such that
\begin{align}
    \bar{u}^{(\text{br})}: \qquad u^{(\text{br})}_{j} = \bar{u}^H_{j+1}, \qquad 1 \le j \le b, \nonumber \\
    \bar{m}^{(\text{br})}: \qquad m^{(\text{br})}_j = m^H_{j+1}, \qquad 1 \le j \le b
\end{align}
we rewrite the sets $\bar u^H$ and $\bar m^H$ as the following unions
\begin{equation}\label{eq:uH-mH-def-v2}
    \bar{u}^H = u^{(3)} \cup \bar u^{(\text{br})}, \qquad \bar{m}^H = u^{(3)} \cup \bar m^{(\text{br})}.
\end{equation}

For the $U$-type QNs $\bar U^I$ and $\bar U^F$ of the initial and final state, respectively, we use the following notation
\begin{align}
    U^I_{j} = U^{(1)}_j, \qquad 1 \le j \le k-1, \nonumber \\
    U^F_j = U^{(2)}_j, \qquad 1 \le j \le l-1,
\end{align}
where the sets $\bar U^{(1)}$ and $\bar U^{(2)}$ are the sets of $U$-type QNs for the tensor products of representations in sets $\bar u^{(1)}$ and $\bar u^{(2)}$.

For the $U$-type QNs $\bar U$ of the operators in the $b$th order Hamiltonian we write
\begin{align}
    U_j &= U^{(\text{br})}_j, \qquad 1 \le j \le b-1, \nonumber \\
    U_{b} &= U^\prime,
\end{align}
where $U'$ is the result of the tensor product of the breaking operators with the $U$-spin limit Hamiltonian. 
We define $\bar U^{(\text{br})}$ to be a set of $b-1$ elements
\begin{align}
\bar{U}^{\text{(br)}} = \{ U^{(\text{br})}_1, \dots, U^{(\text{br})}_{b-1} \}.
\end{align}
Thus we can rewrite $\bar U$ as a union
\begin{equation}
    \bar U = \bar U^{(\text{br})} \cup U^\prime. \label{eq:basis-choice-hamiltonian}
\end{equation}
Note that Eq.~(\ref{eq:basis-choice-hamiltonian}) corresponds to a specific basis choice. The basis choice consists in choosing in which order the tensor products of the $b+1$ representations ($b$ insertions of the $U$-spin breaking spurion and a $U$-spin limit Hamiltonian) are performed. In Eq.~(\ref{eq:basis-choice-hamiltonian}) the set $\bar U^{(\text{br})}$ is a set of $U$-type QNs for the tensor product of $b$ spurions and $U^\prime$ is the total $U$-spin of a term in the $b$th order Hamiltonian.

Using Eqs.~\eqref{eq:Aj=AmImF}-\eqref{eq:def-alpha}, we write the amplitudes of System I as
\begin{equation}
    A_j^{(\text{I})} \equiv A_{(\bar{m}^{(1)}, \bar{m}^{(2)})},
\end{equation}
the RMEs as
\begin{equation}
    X_\alpha^{(\text{I})} \equiv \mel{\bar{U}^{(2)}}{H(\bar{U}^{(3)},b)}{\bar{U}^{(1)}},
\end{equation}
and the multi-index $\alpha$ as
\begin{equation}
    \alpha \equiv \{\bar{U}^{(1)}, \bar{U}^{(2)}, \bar{U}^{(3)}, b\}.
\end{equation}
The group-theoretical decomposition of an amplitude described by the $m$-QNs in Eq.~\eqref{eq:SystemI-QNs} is given by
\begin{equation}
    A_j^{\text{(I)}} = \sum_\alpha C^{(\text{I})}_{j \alpha} X_\alpha^{\text{(I)}},
\end{equation}
where we have introduced
\begin{equation}
    C^{(\text{I})}_{j \alpha} \equiv C(\bar{m}^{(1)}, \bar{m}^{(2)}, \alpha).
\end{equation}
Using the definition for the coefficients $C^*$ given in Eq.~\eqref{eq:basis_rot_def}, we can write the following expression for the coefficients $C^{(\text{I})}_{j \alpha}$:
\begin{align}\label{eq:SystemI_C}
C^{(\text{I})}_{j \alpha}  = C^*(\bar{u}^{(1)};\bar{m}^{(1)}, \bar{U}^{(1)})\times C^*(\bar{u}^{(2)};\bar{m}^{(2)}, \bar{U}^{(2)}) \times C^*(\bar{u}^{(\text{br})}; \bar m^{(\text{br})}, \bar{U}^{(\text{br})}) \nonumber \\
\times \hspace{0pt}\mathop{C_{u^{(3)}, m^{(3)}}}_{\hspace{-4pt} U^{(\text{br})}_{b-1}, 0}^{\hspace{2pt} U^\prime, m^{(3)}} \times \hspace{-2pt}\mathop{C_{U^{(1)}_{k-1}, M^{(1)}}}_{\hspace{-4pt} U^\prime, m^{(3)}}^{\hspace{4pt} U^{(2)}_{l-1}, M^{(2)}}.
\end{align}
Recall that the coefficients $C^{(\text{I})}_{j \alpha}$ and the RMEs $X_\alpha^{(\text{I})}$ are basis dependent, that is, they depend on the specific choice of the order of the tensor product. In Eq.~\eqref{eq:SystemI_C}, for the tensor product of the operators in the Hamiltonian we made a choice to first take a tensor product of all the $U$-spin breaking spurions and only then to multiply the result with the representation $u^{(3)}$. Nevertheless, the sum rules are the same for any choice of basis. 

\subsection{System II}
Next, we focus on System~II. We consider an amplitude $A^{(\text{II})}_j$ that is described by the same $n$-tuple as the amplitude $A^{(\text{I})}_{j}$ that we consider above. The amplitude of interest is then described by the following sets of $m$-QNs:
\begin{equation}\label{eq:SystemII-QNs}
    \bar{m}^I_1 = \emptyset, \qquad \bar{m}^F = \left(-\bar{m}^{(1)}\right)\cup \, \bar{m}^{(2)}\cup\, \left(-m^{(3)}\right),
\end{equation}
where we use $-\bar m^{(1)}$ to denote a set made of all the elements of $\bar m^{(1)}$ multiplied by $(-1)$. The change of sign for the set $\bar m^{(1)}$ and $m^{(3)}$ is due to our convention for constructing generalized $n$-tuples as described in Section~\ref{sec:gen-gen}. In the convention that we use, the $m$-QNs of the representations in the initial state and the Hamiltonian are inverted in the $n$-tuple.

Note that the amplitudes $A_j^{(\text{II})}$ of System~II are described by the same number of independent $m$-QNs as the amplitudes $A_j^{(\text{I})}$ of System~I. This is the case since the total $m$-QN of the final state must be zero. That is we have
\begin{equation}
    -M^{(1)} + M^{(2)} - m^{(3)} = 0,
\end{equation}
which is the same as the last relation in Eq.~\eqref{eq:Mrelations}.
For the total $m$-QNs of the initial state, final state and the Hamiltonian of System~II we have
\begin{equation}
    M^I = M^F = m = 0.
\end{equation}
The sets $\bar u^H$ and $\bar m^H$ are defined as in Eqs.\eqref{eq:uH-mH-def}-\eqref{eq:uH-mH-def-v2}. The only difference are the new values of $u$ and $m$ for System~II, $u = m =0$.

In order to define the $U$-type QNs for System~II we choose to perform the tensor product in the final state of the system in the following order:
\begin{equation}
    \left(\left(u_{1}^{(1)} \otimes ... \otimes u_{k}^{(1)}\right) \otimes \left(u_{1}^{(2)} \otimes ... \otimes u_{l}^{(2)}\right)\right)\otimes u^{(3)}.
\end{equation}
As a result we can introduce the following notation for the $U$-type QNs
\begin{align}
    \bar{U}^{I} = \emptyset, \qquad \bar{U}^{F} = \bar{U}^{(1)}\cup\bar{U}^{(2)}\cup \{U^{\prime \prime}, U^F_{k+l}\}, \qquad
    \bar U = \bar{U}^{(\text{br})}\cup U_{b},
\end{align}
where $U^F_{k+l}$ and $U_b$ are the last elements of the sets $\bar U^F$ and $\bar U$ respectively, and $U^{\prime \prime}$ is used to represent a tensor product of the final irreps in the sets $\bar U^{(1)}$ and $\bar U^{(2)}$. $U_b$ results from the multiplication of the last element of $\bar{U}^{(\text{br})}$, \emph{i.e.} $U^{(\text{br})}_{b-1}$, with the $U$-spin limit Hamilton operator. Since the $U$-spin limit Hamiltonian for System~II is a singlet,
\begin{equation}
    U^{F}_{k+l} = U_b = U^{(\text{br})}_{b-1}.
\end{equation}

Using Eqs.~\eqref{eq:Aj=AmImF}-\eqref{eq:def-alpha}, we write for the amplitudes of System II
\begin{equation}
    A_j^{(\text{II})} \equiv A_{(0, (-\bar{m}^{(1)})\cup\bar m^{(2)}\cup(-m^{(3)}))},
\end{equation}
for the RMEs we write
\begin{equation}
    X_\beta^{(\text{II})} \equiv \mel{\bar{U}^{(1)}\cup\bar{U}^{(2)}\cup U^{\prime\prime}\cup U_{k+l}^{F}}{H(\bar{U}^{(\text{br})}\cup U_b,0,b)}{0},
\end{equation}
where the multi-index $\beta$ is given as
\begin{equation}
    \beta \equiv \{\bar{U}^{(1)}, \bar{U}^{(2)}, \bar{U}^{(\text{br})}, U^{\prime\prime}, b\}.
\end{equation}
The group-theoretical decomposition of an amplitude described by the $m$-QNs in Eq.~\eqref{eq:SystemII-QNs} is given by
\begin{equation}\label{eq:SystemII-decomp}
    A_j^{\text{(II)}} = \sum_{\beta} C^{(\text{II})}_{j\beta} X_\beta^{(\text{II})},
\end{equation}
where we have introduced
\begin{equation}
    C^{(\text{II})}_{j\beta} \equiv C(0,(-\bar{m}^{(1)})\cup \bar{m}^{(2)} \cup (-m^{(3)}), \beta).
\end{equation}
Using the definition for the coefficients $C^*$ in Eq.~\eqref{eq:basis_rot_def} we can write the following expression for the coefficients in Eq.~\eqref{eq:SystemII-decomp} 
\begin{align}\label{eq:SystemII-C}
      C^{(\text{II})}_{j\beta} = C^*(\bar{u}^{(1)};-\bar{m}^{(1)}, \bar{U}^{(1)})\times C^*(\bar{u}^{(2)};\bar{m}^{(2)}, \bar{U}^{(2)}) \times C^*(\bar{u}^{(\text{br})}; \bar{m}^{(\text{br})}, \bar{U}^{(\text{br})}) \nonumber \\
    \times \hspace{0pt}
    \mathop{C_{U^{(2)}_{l-1}, M^{(2)}}}_{\hspace{16pt} U^{(1)}_{k-1}, -M^{(1)}}^{\hspace{0pt} U^{\prime \prime}, m^{(3)}}
    \times \hspace{-10pt} \mathop{C_{U^{\prime\prime}, m^{(3)}}}_{\hspace{20pt} u^{(3)}, -m^{(3)}}^{\hspace{2pt} U^{(\text{br})}_{b-1}, 0} \times\hspace{-14pt} \mathop{C_{0, 0}}_{\hspace{22pt} U^{(\text{br})}_{b-1}, 0}^{\hspace{20pt} U^{(\text{br})}_{b-1}, 0},
\end{align}
where the last CG coefficient is coming from the Wigner-Eckart theorem and is equal to one.

Comparing Eqs.~\eqref{eq:SystemI_C} and~\eqref{eq:SystemII-C} we see that for non-zero RMEs $U^\prime$ takes the same values as $U^{\prime \prime}$ and thus the two can be identified $U^\prime \equiv U^{\prime \prime}$. Most importantly, this implies that $\alpha \equiv \beta$ and thus we have shown that one can establish a one-to-one correspondence between the RMEs of the two systems $X^{(\text{I})}_\alpha$ and $X^{(\text{II})}_\alpha$. 

Finally, Eq.~\eqref{eq:SystemII-C} can be rewritten as follows
\begin{align}\label{eq:SystemII-C-mod}
C^{(\text{II})}_{j\alpha} = C^*(\bar{u}^{(1)};-\bar{m}^{(1)}, \bar{U}^{(1)})\times C^*(\bar{u}^{(2)};\bar{m}^{(2)}, \bar{U}^{(2)}) \times C^*(\bar{u}^{(\text{br})}; \bar{m}^{(\text{br})}, \bar{U}^{(\text{br})}) \nonumber \\
\times \hspace{0pt}
\mathop{C_{U^{(2)}_{l-1}, M^{(2)}}}_{\hspace{16pt} U^{(1)}_{k-1}, -M^{(1)}}^{\hspace{0pt} U^{\prime}, m^{(3)}}
\times \hspace{-10pt} \mathop{C_{U^{\prime}, m^{(3)}}}_{\hspace{20pt} u^{(3)}, -m^{(3)}}^{\hspace{2pt} U^{(\text{br})}_{b-1}, 0}.
\end{align}

\subsection{Universality of sum rules}
As we show above, even though the two systems are indeed described by different RMEs, the matrix elements still carry the same indices and thus there is a one-to-one correspondence between the RMEs of the two systems.
Next, using the following symmetry properties of the CG coefficients
\begin{equation}
    \mathop{C_{j_1, m_1}}_{\hspace{8pt} j_2, m_2}^{\hspace{8pt} j_3, m_3} = (-1)^{j_1 + j_2 - j_3} \mathop{C_{j_1, -m_1}}_{\hspace{8pt} j_2, -m_2}^{\hspace{8pt} j_3, -m_3},
\end{equation}
\begin{equation}
    \mathop{C_{j_1, m_1}}_{\hspace{8pt} j_2, m_2}^{\hspace{8pt} j_3, m_3} = (-1)^{j_1-m_1} \sqrt{\frac{2j_3 + 1}{2j_2 + 1}} \mathop{C_{j_3, m_3}}_{\hspace{16pt} j_1, -m_1}^{\hspace{8pt} j_2, m_2},
\end{equation}
we find
\begin{align}
    C^*(\bar{u}^{(1)}; \bar m^{(1)}, \bar U^{(1)}) = (-1)^{\left(\sum_{i=1}^k u^{(1)}_i\right) - U^{(1)}_{k-1}}  C^*(\bar{u}^{(1)}; -\bar m^{(1)}, \bar U^{(1)} ), \label{eq:C*-CG-relations-1} \\
    \mathop{C_{u^{(3)}, m^{(3)}}}_{\hspace{-4pt} U^{(\text{br})}_{b-1}, 0}^{\hspace{2pt} U^\prime, m^{(3)}} = (-1)^{u^{(3)} - m^{(3)}} \sqrt{\frac{2U^\prime+1}{2U^{(\text{br})}_{b-1} +1}} \mathop{C_{U^{\prime}, m^{(3)}}}_{\hspace{20pt} u^{(3)}, -m^{(3)}}^{\hspace{2pt} U^{(\text{br})}_{b-1}, 0}, \label{eq:C*-CG-relations-2}\\
    \mathop{C_{U^{(1)}_{k-1}, M^{(1)}}}_{\hspace{-4pt} U^\prime, m^{(3)}}^{\hspace{4pt} U^{(2)}_{l-1}, M^{(2)}} = (-1)^{U^{(1)}_{k-1}-M^{(1)}} \sqrt{\frac{2U^{(2)}_{l-1}+1}{2U^\prime + 1}} \mathop{C_{U^{(2)}_{l-1}, M^{(2)}}}_{\hspace{16pt} U^{(1)}_{k-1}, -M^{(1)}}^{\hspace{0pt} U^{\prime}, m^{(3)}}.\label{eq:C*-CG-relations-3}
\end{align}
Combining Eqs.~\eqref{eq:C*-CG-relations-1}-\eqref{eq:C*-CG-relations-3} we obtain the following relation between the coefficients in the group-theoretical decompositions for the two systems:
\begin{equation}\label{eq:CI-CII-relation}
    C^{(\text{I})}_{j \alpha} = (-1)^{\tilde{q}_j} Q\times C^{(\text{II})}_{j \alpha},
\end{equation}
with
\begin{equation}
    (-1)^{\tilde{q}_j} = (-1)^{\left(\sum_{i=1}^k u^{(1)}_i\right) - M^{(1)} + u^{(3)} - m^{(3)}}, \qquad Q = \sqrt{\frac{2U^{(2)}_{l-1}+1}{2U^{(\text{br})}_{b-1} +1}}. 
\end{equation}
For a fixed $n$-tuple the factor $(-1)^{\tilde{q}_j}$ is the same for all RMEs. 
The factor $Q$ does not depend on the $m$-QNs and thus for a fixed RME it is the same for all amplitudes. Note that $Q > 0$.

Let us assume that we have found a sum rule for System~II that holds up to a certain order of breaking. This means that there exists a linear combination of amplitudes such that the coefficients $C^{(\text{II})}_{j\alpha}$ in front of all RMEs cancel. The cancellation still holds for the coefficients $Q\times C^{(\text{II})}_{j\alpha}$ because this is just a multiplication with a global factor. 
Thus knowing a sum rule for System~II one can  write the corresponding sum rule for System~I by multiplying the amplitudes in the sum rule by factors $(-1)^{\tilde{q}_j}$. Equivalently, one can redefine the amplitudes of System~I by multiplying them by these factors. We choose the latter convention. In this case the amplitudes of System~I are redefined as follows
\begin{equation}\label{eq:AIj-redef-v1}
    A^{(\text{I})}_j \rightarrow (-1)^{\tilde{q}_j} A^{(\text{I})}_j
\end{equation} and the sum rules for the two systems take exactly the same form. The factor $(-1)^{\tilde{q}_j}$ can be rewritten as follows
\begin{equation}\label{eq:q-factor-intro-1}
(-1)^{\tilde{q}_j}=(-1)^{\sum_{i=1}^k u^{(1)}_i - M^{(1)} + u^{(3)} - m^{(3)}} = (-1)^{u^{(3)}-m^{(3)}} \left(\prod_{i=1}^k (-1)^{u_i^{(1)}-m^{(1)}_i}\right),
\end{equation}
which in terms of $u$, $m$, $u^I_i$ and $m^I_i$ of System~I becomes
\begin{equation}\label{eq:q-factor-intro}
(-1)^{\tilde{q}_j}= (-1)^{u-m} \left(\prod_{i=1}^k (-1)^{u_i^I-m^{I}_i}\right).
\end{equation}
Note that even though $u$ and $\sum_i u^I_i$ are fixed for all the amplitudes of the system it is important to keep them to ensure that the powers in Eq.~\eqref{eq:q-factor-intro} are integer numbers.

The factors $u-m$ and $u^I_i - m^I_i$ in Eq.~\eqref{eq:q-factor-intro} give the numbers of plus signs in the $n$-tuple that correspond to the representation of the Hamiltonian and the representation $u^I_i$ from the initial state respectively, see the discussion at the end of Appendix~\ref{eq:deriving-the-symmetry-factor}. They give the numbers of plus signs and no minus signs, since in our convention the $m$-QNs are inverted when we build $n$-tuples. Thus, according to Eq.~\eqref{eq:q-factor-intro}, the factor $(-1)^{\tilde{q}_j}$ can be found as the parity of the total number of the plus signs that correspond to the initial state and the Hamiltonian in the $n$-tuple.

The parity of the number of the plus signs for the initial state and the Hamiltonian can also be found as the product of the parity of the number of all the pluses in the $n$-tuple and the parity of the number of pluses corresponding to the final state. That is
\begin{equation}\label{eq:q-factor-2}
    (-1)^{\tilde{q}_j} = (-1)^{n/2} \prod_{i=1}^l (-1)^{u_i^F+m^{F}_i},
\end{equation}
where $n$ is the total number of elements in the $n$-tuple given by
\begin{equation}
    n = 2\sum_i^k u^I_i + 2\sum_i^l u^F_i + 2u.
\end{equation}
Eq.~\eqref{eq:q-factor-2} can be further rewritten in terms of the parity of the number of the minus signs in the final state
\begin{equation}
    (-1)^{\tilde{q}_j} = (-1)^{n/2 + 2\sum_i^l u_i^F} \prod_{i=1}^l (-1)^{u_i^F-m^{F}_i}.
\end{equation}
Note however, that the factor $(-1)^{n/2 + 2\sum_i^l u_i^F}$ is the same for all the amplitudes. Thus we can introduce a complementary definition
\begin{equation}\label{eq:q-factor-def-final}
    (-1)^{q_j} \equiv \prod_{i=1}^l (-1)^{u_i^F-m^{F}_i}\,,
\end{equation}
which is simply the parity of the number of minuses in the final state. We can use the $(-1)^{q_j}$ factors to introduce the following redefinition of amplitudes instead of Eq.~\eqref{eq:AIj-redef-v1}
\begin{equation}\label{eq:AIj-redef-v2}
    A^{(\text{I})}_j \rightarrow (-1)^{q_j} A^{(\text{I})}_j.
\end{equation}
One can choose either of the two definitions $(-1)^{\tilde{q}_j}$ or $(-1)^{q_j}$. Even though they are not equivalent they preserve the relative signs between amplitudes and thus will result in exactly the same sum rules. Moreover, one can also use the parity of plus signs in the final state or parity of the minus signs in the initial state and the Hamiltonian. Any of these definitions will result in the same set of sum rules. For consistency, everywhere in this work we use the definitions in Eqs.~\eqref{eq:q-factor-def-final} and~\eqref{eq:AIj-redef-v2}. The resulting definitions of $a$- and $s$-type amplitudes are given in Eq.~\eqref{eq:as-comb-def-app}.

Summing up, in this appendix we have shown that
\begin{enumerate}
\item there is a one-to-one correspondence between amplitudes and RMEs of System~I and System~II,
\item the coefficients of the group-theoretical decompositions for the two systems are related by Eq.~\eqref{eq:CI-CII-relation},
\item if one redefines the amplitudes of System~II by multiplying them by the factors $(-1)^{q_j}$ defined in Eq.~\eqref{eq:q-factor-def-final}, the sum rules for the two systems take exactly the same form.
\end{enumerate}
The last statement is what we call the ``universality of sum rules'', the fact that the sum rules for any two systems from the same universality class are the same.

\section{Counting the number of amplitude sum rules}\label{app:SR_counting_doublets}

In this appendix we present a formula for the number of sum rules for a given $U$-spin set described by $n$ doublets. As we argue in Section~\ref{sec:universality} and show explicitly in Appendix~\ref{app:signs} the counting is the same no matter if the doublets belong to the initial state, final state or the Hamiltonian. Without loss of generality we consider a $U$-spin set whose $U$-spin structure is described by $n$ doublets in the final state:
\begin{equation}\label{eq:system-0-to-nd}
    0  \xrightarrow{u = 0} \left(\frac{1}{2}\right)^{\otimes n}.
\end{equation}
Using the notation of Appendix~\ref{app:RMEdecomposition} we describe the $U$-spin structure of the system via the sets $\bar u^{I,F}$ and $u$:
\begin{align}
    \bar{u}^I &= \emptyset, \qquad u = 0,\\
    u^F_j &= \frac{1}{2}, \qquad 1 \le j \le n.
\end{align}
The amplitudes of the system in the physical basis are described by the sets of $m$-QNs
\begin{equation}
    \bar{m}^I = \emptyset, \qquad \bar{m}^F = \{m_1^F, m_2^F, \dots, m_n^F\}.
\end{equation}
The total $m$-QNs of the initial, final state and the Hamiltonian are equal to zero
\begin{equation}
    M^I = M^F = m = 0.
\end{equation}

Generally, to count the number of sum rules one needs to consider the decomposition of the CKM-free amplitudes 
in terms of RMEs. This decomposition takes the form, see  Eq.~\eqref{eq:CKMfree_decomposition},
\begin{equation}\label{eq:CKMfree_decomposition-copy}
    A_j = \sum_\alpha C_{j \alpha} X_\alpha.
\end{equation}
Using Eqs.~\eqref{eq:Aj=AmImF}-\eqref{eq:def-alpha} we find that for the system under consideration the CKM-free amplitudes are
\begin{equation}
    A_j \equiv A_{(\emptyset,\bar{m}^F)},
\end{equation}
the RMEs are
\begin{equation}
    X_\alpha \equiv \mel{\bar{U}^F}{H(\bar{U}, b^\prime)}{0},
\end{equation}
and the multi-index $\alpha$ is given by
\begin{equation}
    \alpha \equiv \{\bar{U}^I, \bar{U}^F,\bar{U},b^\prime\} \equiv \{ \bar{U}^F,\bar{U},b^\prime\},
\end{equation}
where we used $\bar{U}^I = \emptyset$. The order of breaking, $b^\prime$, takes values from $0$ to $b$, where $b$ is the chosen order in breaking we consider. Note that in order to ensure non-zero RMEs, the last elements of the sets $\bar{U}^F$ and $\bar{U}$ must be the same, that is $U^F_{n-1} = U_{b^\prime -1}$. Also, note that everywhere in this Appendix we use $b^\prime$ as a generic index, and $b$ as a chosen order of breaking, up to which we want to write the sum rules.

Once we have the decomposition in Eq.~\eqref{eq:CKMfree_decomposition-copy}, the number of sum rules, $n_{SR}$, that are valid up to $b$ is given by
\begin{equation}\label{eq:nSR_gen}
    n_{SR}^{(b)} = n_A - \text{rank} \left[C_{j \alpha}\right],
\end{equation}
where $n_A$ is the number of amplitudes in the $U$-spin set, and the multi-index $\alpha$ carries information about the order of breaking $b$.

In what follows we first discuss the details of the decomposition in Eq.~\eqref{eq:CKMfree_decomposition-copy} for the special case that we consider in this appendix, next we find $n_A$ and $\text{rank} \left[C_{j \alpha}\right]$. These allow us to find the number of sum rules in terms of $n$ and $b$.

\subsection{Decomposition in terms of RMEs}

In this subsection we discuss some properties of Eq.~\eqref{eq:CKMfree_decomposition-copy} that are specific for the processes with $U$-spin structure that is described by $n$ doublets in the final state. First, we note that applying the Wigner-Eckart  theorem, Eq.~\eqref{eq:WE-gen}, to the matrix elements of the system under consideration we find
\begin{equation}\label{eq:X_doublets}
    \mel{\bar{U}^F,0}{H\left(\bar{U},0,b^\prime\right)}{0,0} \overset{\text{WE}}{=} \mel{\bar{U}^F}{H(\bar{U}, b^\prime)}{0} \equiv X_{\alpha},
\end{equation}
where we used $u = m = 0$, $U^F_{n-1} = U_{b^\prime-1}$ and $M^I = M^F = 0$. Thus in order to find sum rules for a system of $n$ doublets in the final state one does not need to use the Wigner-Eckart theorem and invoke the concept of RMEs at all. This is because in this case states as well as operators at all orders of $U$-spin breaking have their total $m$-QN equal to zero. That is, in this case the application of the Wigner-Eckart theorem does not lead to additional complexity reduction.

Using Eq.~\eqref{eq:Cjalpha} we find
\begin{equation}\label{eq:Cjalpha-special-def}
    C_{j\alpha} = C^*(\bar{u}^F; \bar{m}^F, \bar{U}^F)\times C^*(\bar{u}^H; \bar m^H, \bar{U}),
\end{equation}
where the sets $\bar u^H$ and $\bar m^H$ are such that
\begin{align}
    u^H_1 = u = 0, \qquad u^H_j = 1, \qquad 2 \le j \le b^\prime+1, \nonumber \\
    m^H_j = 0, \qquad 1 \le j \le b^\prime+1.
\end{align}

\subsection{The number of amplitudes in the $U$-spin set, $n_A$}

As we discuss in Section~\ref{sec:An-tuples-doublets}, the amplitudes $A_j$ of a $U$-spin set can be mapped one-to-one onto $n$-tuples made of ${n/2}$ pluses and ${n/2}$ minuses. The number of different $n$-tuples that satisfy this rule is the binomial coefficient and is given by
\begin{equation}\label{eq:nA}
    n_A = \binom{n}{{n/2}} = \frac{n!}{(n/2)! (n/2)!}.
\end{equation}
This is the number of possible final states of the $U$-spin set in the physical basis.

Alternatively, one can perform the counting in the $U$-spin basis. The relation between states in the physical basis and in the $U$-spin basis for the system we consider is given by
\begin{equation}
    \ket{\bar{u}^F; \bar{m}^F} = \sum_{\bar{U}^F} C^*(\bar{u}^F; \bar{m}^F, \bar{U}^F) \ket{\bar{U}^F, 0}. \label{eq:basis-rotation}
\end{equation}
Since all we do here is just a basis rotation, the number of states $\ket{\bar{u}^F; \bar{m}^F}$ in the physical basis is the same as the number of states $\ket{\bar{U}^F, 0}$ in the $U$-spin basis. The amplitudes are fully defined by the sets $\bar{m}^F$ and thus the number of amplitudes in the physical basis is the same as the number of amplitudes in the $U$-spin basis and is equal to $n_A$.

The multiplicity of irreps $U$ in the decomposition of the tensor product of $n$ doublets is given by \cite{Zachos:1992xp, Curtright:2016eni}
\begin{equation}\label{eq:Nnu}
        N_{U}^{n} =
   \frac{n! (2U + 1)}{\left({n/2} - U\right)! \left({n/2} + U + 1\right)!},
\end{equation}
where we assume that $U$ takes values that are consistent with $n$, this is the case when $(n + 2U)$ is even.
The number of different sets $\bar{U}^F$ such that $U^F_{n-1} = U$ is then equal to $N^n_{U}$. Note that $N_U^n$ can be written in terms of the entries $c(n,k)$ of Catalan's triangle~\cite{weissteinCatalansTriangle}
\begin{align}
c(n,k) \equiv \frac{(n+k)! (n-k+1)}{k! (n+1)!} 
\end{align}
as
\begin{align}
N_U^n &= c\left(\frac{n}{2}+U,\frac{n}{2}-U\right)\,.
\end{align}
By summing $N_U^n$ over all possible $U$ we find the number of basis elements in the $U$-spin basis and thus the number of amplitudes in the $U$-spin set
\begin{equation}\label{eq:nA_basis}
    n_A = \sum_{U=0}^{{n}/{2}} N^{n}_{U}.
\end{equation}
One can check that Eq.~\eqref{eq:nA_basis} is equal to Eq.~\eqref{eq:nA} for all $n$ as it should be.

\subsection{Finding the rank of the matrix $\left[C_{j\alpha}\right]$}

The elements of the matrix $C_{j\alpha}$ are defined in Eq.~\eqref{eq:Cjalpha-special-def}. First, we note that the number of sets $\bar{U}^F$ present in the decomposition in Eq.~\eqref{eq:CKMfree_decomposition-copy} is determined by $b$, since $U^F_{n-1}$ must satisfy $U^F_{n-1} \le b$. Second, studying Eq.~\eqref{eq:Cjalpha-special-def}, we note that the factors $C^*(\bar{u}^H; \bar{m}^H,\bar{U})$ are independent of $\bar{m}^F$. 
Note that $C^*(\bar{u}^H; \bar{m}^H,\bar{U})$ depends on $b^\prime$ and the specific set $\bar U$. Thus, for a fixed matrix element this factor can be absorbed by a redefinition of the matrix elements in the $U$-spin basis. 
This is equivalent to multiplying columns of the corresponding Clebsch-Gordan coefficient matrix by arbitrary numbers $\neq 0$. This operation does not change the rank.
As a result, the rank of the matrix $\left[C_{j\alpha}\right]$ is fully determined by the rank of the matrix that we denote as $\left[C_{\bar{m}^F\bar{U}^F}\right]$, with elements given by the coefficients $C^*(\bar{u}^F;\bar{m}^F,\bar{U}^F)$. 

The matrix $C^*(\bar{u}^F;\bar{m}^F,\bar{U}^F)$ is the rotation matrix between the physical and the $U$-spin basis of states, see Eq.~(\ref{eq:basis-rotation}). This means that the number of states on the left is equal to the number of states on the right. All states $\ket{\bar U^F,0}$ are linearly independent. Therefore, the rank of the matrix $\left[C_{\bar{m}^F\bar{U}^F}\right]$ is found by counting the number of different sets $\bar{U}^F$ allowed at the chosen order of breaking $b$.

Consequently, we can use Eq.~\eqref{eq:Nnu} to find
\begin{equation}\label{eq:rank}
    \text{rank } \left[C_{j\alpha}\right] = \sum_{U=0}^{b} N_U^n\,.
\end{equation}
Consider, for example, $b = 0$. In this case the only non-vanishing matrix elements have $U^F_{n-1} = 0$ and there are $N^n_0$ of them. For $b = 1$ the non-vanishing matrix elements can have $U^F_{n-1} = 0$ and $U^F_{n-1}=1$. Thus, in this case there are $N^n_0 + N^n_1$ matrix elements.

The fact that the rank of $\left[C_{j\alpha}\right]$ is determined by the rank of $\left[C_{\bar{m}^F\bar{U}^F}\right]$ can easily be understood. Let's consider two non-zero matrix elements $X_{\alpha_1}$ and $X_{\alpha_2}$ with fixed $\bar U^F$, such that 
\begin{equation}
    \alpha_1 = \{\bar{U}^F, \bar{U}_1, b_1^\prime\}, \qquad \alpha_2 = \{\bar{U}^F, \bar{U}_2, b_2^\prime\},
\end{equation}
where $b_1^\prime < b_2^\prime$. Note that in order for both matrix elements to exist, $b_1^\prime$ and $b_2^\prime$ must have the same parity because of Eq.~\eqref{eq:b-parity}. Now, according to Eq.~\eqref{eq:Cjalpha-special-def} the two matrix elements will always enter decompositions of amplitudes in the same linear combinations (since $C^*(\bar{u}^H;\bar{m}^H,\bar{U})$ is independent on $\bar{m}^F$) and thus the addition of the matrix element $X_{\alpha_2}$ at the higher order $b_2^\prime$ will not change the rank of $\left[C_{j\alpha}\right]$.

The matrix elements that change the rank come from the highest irrep in the tensor product $H_\varepsilon^{\otimes b_2^\prime}$, that is the terms for which $U_{b_2^\prime-1} = b_2^\prime$. Such matrix elements will be also described by new sets $\bar{U}^F$ with $U^F_{b_2^\prime-1} = b_2^\prime$ that were not generated at lower orders in breaking.

Equivalently, one can say that the terms for which $U_{b^\prime-1} = U^F_{n-1} < b^\prime$ can be always absorbed into the terms at lower orders of breaking and when performing the counting only the terms with $U_{b^\prime-1} = U^F_{n-1} = b^\prime$ can provide new $U$-spin structure.

\subsection{The number of sum rules}

Combining the results Eqs.~\eqref{eq:nSR_gen},~\eqref{eq:nA_basis} and~\eqref{eq:rank} we arrive at the following expression for the number of sum rules up to the considered order $b$ for a system of $n$ doublets
\begin{equation}\label{eq:nSR_doublets}
    n_{SR}^{(b)} = \sum_{U = b + 1}^{{n}/{2}} N^n_U = \frac{n!}{\left({n/2} + b + 1\right)!\left({n/2} - b - 1\right)!},
\end{equation}
where $N^n_U$ is defined in Eq.~\eqref{eq:Nnu}. This expression gives the number of sum rules for any system of $n$ doublets, no matter if they are in the initial state, final state or represent operators. From Eq.~\eqref{eq:nSR_doublets}, the maximum order of breaking when there are still sum rules between amplitudes of a $U$-spin system is $b_{\rm max} = {n}/{2} - 1$. There is exactly one sum rule at this order.

\section{Proof of the sum rule theorem for a system of $n$ doublets}\label{app:ThII_doublets}

In what follows we prove a Theorem and two Corollaries that together establish the algorithm of finding the sum rules for systems of $n$ doublets without writing explicitly the group-theoretical decomposition of amplitudes.

In this Appendix we consider a $U$-spin set of processes that can be described by $n$ doublets. We assume that all doublets belong to the final state and only consider CKM-free amplitudes. 

\subsection{Notations and definitions}

We start by listing definitions and notations that are used in the formulation and the proof of the theorem.

\begin{itemize}
\item $b$ is the order of breaking. The highest order at which there are still sum rules in the system is $b_{\text{max}} = {n/2} - 1$, see Eq.~\eqref{eq:nSR_doublets}. As in Appendix~\ref{app:SR_counting_doublets} we use $b^\prime$ as a generic index that may take values between $0$ and $b$.
\item 
$a_i$ and $s_i$ are $a$-type and $s$-type amplitudes respectively.
\item
We recall the definition of a set of subsets of $U$-spin pairs of the system $S^{(k)} = \{S_j^{(k)}\}$. Each $U$-spin pair is defined by the corresponding $n$-tuple, which in our convention starts with a minus sign. A subset $S_j^{(k)}$ contains all the $U$-spin pairs whose $n$-tuples share $k$ minuses at the same positions including the first minus. The index $j$ is used to enumerate all such subsets. The number of subsets $S_j^{(k)}$ is denoted by $n_S^{(k)}$ and the number of amplitude pairs in each subset by $n_A^{(k)}$. In Sec.~\ref{sec:n_doublet_system} we show that
\begin{equation}\label{eq:nSk-nAk-def}
    n_S^{(k)} = \binom{n-1}{k-1}, \hspace{25pt} n_A^{(k)} = \binom{n-k}{{n/2}-k}.
\end{equation}
\item
We use the notation $a_i \in S^{(k)}_j$ and $s_i \in S^{(k)}_j$ to denote the $a$- and $s$-type amplitudes corresponding to $U$-spin pairs that are in $S^{(k)}_j$. For shortness, instead of saying ``the amplitudes that correspond to $U$-spin pairs from the subset $S_j^{(k)}$'' we simply say ``amplitudes from $S_j^{(k)}$''.
\item
We define $S_1^{(k)}$ to be a specific subset of the amplitude pairs which can be schematically written as follows    \begin{equation}\label{eq:amp}
S_1^{(k)}  = (\underbrace{-, -, ..., -}_{k}; \{\underbrace{-, ..., -}_{{n/2}-k}\underbrace{+, ..., +}_{{n/2}}\}).
\end{equation}
That is, the first $k$ signs before the semicolon are fixed to be minuses and the curly brackets indicate all possible orderings of minuses and pluses after the semicolon.
For example, with $n = 6$ and $k = 2$, the subset of interest is $S_1^{(2)}$ and is given by
\begin{equation} \label{eq:S1-examp}
S_1^{(2)} = \left\{\left(-, -; -, +, +, + \right), \left(-, -; +, -, +, + \right),
        \left(-, -; +, +, -, + \right), \left(-, -; +, +, +, - \right)\right\}.
\end{equation}
\item 
$N_{U}^{n}$ is the number of different irreps $U$ in the decomposition of the tensor product of $n$ doublets which is given in Eq.~\eqref{eq:Nnu}. 
\item $n_{X-a}^{(b)}$ and $n_{X-s}^{(b)}$ are the numbers of linearly independent combinations of matrix elements that enter the decompositions of the $a$- and $s$-type amplitudes of subset $S_1^{(k)}$, respectively, up to an order of breaking $b$. As we show in step 4 of the proof below, these numbers are the same for all subsets from the set $S^{(k)}$.
\item The sets $\bar u^{I, F, H}$, $\bar U^{I, F}$, $\bar U$, $\bar{m}^{I,F,H}$ and $u$, $m$, $M^{I, F}$ are defined in Appendix~\ref{app:RMEdecomposition}. For a system described by $n$ doublets in the final state we have \begin{align}
\bar{u}^I &= \emptyset, \qquad u = m = 0, \nonumber  \\
\bar{u}^F &= \{u^F_1, \dots, u^F_n\}, \qquad u^F_j = \frac{1}{2}, \qquad 1 \le j\le n,\nonumber \\
\bar{u}^H &= \{u^H_1, u^H_2, \dots, u^H_{b+1}\}, \qquad u^H_1 = 0,\qquad u^H_j = 1,\qquad 2 \le j \le b+1, \nonumber\\
\bar m^H &= \{m^H_1, m^H_2, \dots, m^H_{b+1}\}, \qquad m^H_j = 0, \qquad 1 \le j \le b+1. 
\end{align}
\item
When performing the tensor product of the representations $\bar u^F$ in the final state we choose to divide the set into two subsets
\begin{equation}
\bar u^F = \bar u^{(1)} \cup \bar u^{(2)},
\end{equation}
such that set $\bar u^{(1)}$ contains $k$ irreps and set $\bar u^{(2)}$ contains the remaining $n-k$ irreps. Since for the system under consideration all irreps are doublets, we have
\begin{equation}
u^{(1)}_j = \frac{1}{2} \qquad \text{for } 1\le j \le k, \qquad u^{(2)}_j = \frac{1}{2}\qquad \text{for } 1\le j \le n-k.
\end{equation}
The corresponding sets of $m$-QNs are denoted as $\bar m^{(1)}$ and $\bar m^{(2)}$ and we have
\begin{equation}
\bar m^F = \bar m^{(1)} \cup \bar m^{(2)}.
\end{equation}
For the subset of $n$-tuples $S_1^{(k)}$ defined in Eq.~\eqref{eq:amp} we can choose the set $\bar u^{(1)}$ such that
\begin{equation}
m^{(1)}_j = -\frac{1}{2}\qquad \text{for } 1 \le j \le k.
\end{equation}
In this case we have
\begin{equation}
M^{(1)} = \sum_{j=1}^k m^{(1)}_j = -\frac{k}{2}, \qquad M^{(2)} = \sum_{j=1}^{n-k} m^{(2)}_j = \frac{k}{2}.
\end{equation}

We use $\bar U^{(1)}$ and $\bar{U}^{(2)}$ to denote the sets of $U$-type QNs. For the final state we choose to perform the tensor product in the following order:
\begin{equation}\label{eq:tensor_prod}
\mathop{\otimes}_i^n d_i = \underbrace{\left( d_1 \otimes d_2 \otimes ... d_{k} \right)}_{k} \otimes \underbrace{\left( d_{k+1} \otimes d_{k + 2} \otimes ... d_{n} \right)}_{n - k},
\end{equation}
where $d_i$ are doublets. We use $\bar{U}^{(1)}$ to denote the $U$-type QNs for the tensor product of the first $k$ doublets, and $\bar{U}^{(2)}$ to denote the tensor product of the last $n-k$ doublets. Thus we write the set of $U$-type QNs in the final state as follows:
    \begin{equation}
        \bar{U}^F = \{U^{(1)}_1, U^{(1)}_2, \dots, U^{(1)}_{k-1}, U^{(2)}_1, U^{(2)}_2, \dots, U^{(2)}_{n-k-1}, U^F_{n-1}\} = \bar{U}^{(1)} \cup \bar{U}^{(2)} \cup U^F_{n-1}.
    \end{equation}

\end{itemize}

\subsection{Formulation of the sum rule theorem}

Consider a $U$-spin set of $n$ doublets in the final stat such that $n$ is even.

\textbf{Theorem:} 
\begin{enumerate}
\item 
For any even $n$, an even (odd) order of breaking $b \le {n}/{2} - 1$, and for any $k = {n}/{2}-b$, there exist exactly one $a$($s$)-type sum rule among the amplitudes from each  $S_j^{(k)}$. As a result, there are $n_S^{(k)}$ $a$($s$)-type sum rules among the $n_A^{(k)}$ amplitudes from every subset $S_j^{(k)}$.

\item These sum rules are valid up to order of breaking $b$ and are broken by corrections of order $b+1$.  
\item The sum rules are given by
\begin{equation}\label{eq:SR1}
\sum_{a_i \in S^{(k)}_j} a_i = 0 \hspace{25pt} \text{and} \hspace{25pt} \sum_{s_i \in S^{(k)}_j} s_i = 0,
\end{equation}
where for even (odd) $b$ the sums are taken over all $a$($s$)-type amplitudes from subsets $S^{(k)}_j$. Note that for even $b$ there are $a$-type sum rules but no $s$-type sum rules. For odd $b$ there are $s$-type sum rules but no $a$-type sum rules.
\end{enumerate}

\textbf{Corollary I.} The sum rules described by the Theorem provide the full set of sum rules for the system. For a system described by $n$ doublets the number of sum rules at order of breaking $b \le {n/2} - 1$ is 
\begin{equation}
    n_{SR}^{(b)} = n_S^{({n}/{2}-b)} + n_S^{({n}/{2}-b-1)},
\end{equation}
where $n_{SR}^{(b)}$ is given in Eq.~\eqref{eq:nSR_doublets}.

\textbf{Corollary II.} All the sum rules for a system of $n$ doublets at any order $b \le {n}/{2}-1$ can be found using Table~\ref{tab:sum-rules-rule}.

\begin{table}[t]
\centering
\begin{tabular}{|c|c|}
\hline
$a$-type & $s$-type \\
\hline
$n_{SR-a}^{(b)} = n_S^{(k)}$ & $n_{SR-s}^{(b)} = n_S^{(k)}$\\
$j$th sum rule: \(\displaystyle \sum_{a_i \in S^{(k)}_j} a_i \), &
$j$th sum rule: \(\displaystyle \sum_{s_i \in S^{(k)}_j} s_i \),
\\ ~~with $i = 1, 2, ..., n_A^{(k)}$, $j = 1, 2, ..., n_S^{(k)}$~~   & ~~ with $i = 1, 2, ..., n_A^{(k)}$, $j = 1, 2, ..., n_S^{(k)}$~~\\
\hline
\multicolumn{2}{|c|}{even $b$}\\
\hline
$k = \frac{n}{2}-b$ & $k = \frac{n}{2} - b - 1$\\
\hline
\multicolumn{2}{|c|}{odd $b$}\\
\hline
$k = \frac{n}{2}-b-1$ & $k = \frac{n}{2} - b$\\
\hline
\end{tabular}
\caption{Summary of the $a$- and $s$-type sum rules at even and odd orders of $U$-spin breaking~$b$. Note that $k$ takes different values for $a$- and $s$-type sum rules. This means specifically, that the respective sums of the $a_i$ and $s_i$ go over different sets $S_j^{(k)}$. \label{tab:sum-rules-rule}}
\end{table}

\textbf{Outline of the proof:}
\begin{itemize}[leftmargin=2.2cm]
\item[\textbf{Step 1:}] For an arbitrary $1 \le k \le {n/2}$, we consider one specific subset of amplitude pairs $S^{(k)}_1 \in S^{(k)}$ that we define below. For the subset $S_1^{(k)}$, we count the numbers of linearly independent combinations of matrix elements that enter the group-theoretical decompositions of the $a$- and $s$-type amplitudes up to  order of breaking $b$. We denote these numbers as $n_{X-a}^{(b)}$ and $n_{X-s}^{(b)}$ respectively.
\item[\textbf{Step 2:}] We consider $S_1^{(k)}$ for $k = {n/2}-b$ and show that in this case the following relations hold for even $b$
\begin{equation}
n_A^{(k)} - n_{X-a}^{(b)} = 1, \hspace{25pt} n_A^{(k)} - n_{X-s}^{(b)} = 0,  \hspace{25pt} \forall n, b,
    \end{equation}
    and for odd $b$
    \begin{equation}
        n_A^{(k)} - n_{X-a}^{(b)} = 0, \hspace{25pt} n_A^{(k)} - n_{X-s}^{(b)} = 1,  \hspace{25pt} \forall n, b.
\end{equation}
This means that at $b$ even (odd) there is exactly one $a$($s$)-type sum rule and no $s$($a$)-type sum rules among the amplitudes from the subset $S_1^{(k)}$, where $k = {n/2}-b$.
\item[\textbf{Step 3:}] We show that the $a$($s$)-type sum rule among the amplitudes from $S_1^{(k)}$, with $k={n/2}-b$, has the symmetric form given in Eq. \eqref{eq:SR1}.
\item[\textbf{Step 4:}] We show that the results of steps 2 and 3 hold for all subsets from the set $S^{(k)}$. This proves the statement of the theorem.
\end{itemize}

\subsection{Step 1}
Step 1 follows very closely the counting performed in Appendix~\ref{app:SR_counting_doublets} with minor differences due to the specific basis choice and the fact that we focus on the  $S_1^{(k)}$ subset.

\subsubsection{Linearly independent combinations of matrix elements}

We consider the subset $S_1^{(k)}$ defined in Eq.~\eqref{eq:amp}. This subset contains $n_A^{(k)}$ amplitude pairs. Our aim is to perform the counting of $n_{X-a}^{(b)}$ and $n_{X-s}^{(b)}$. That is, the numbers of linearly independent combinations of matrix elements that enter the decompositions of the $a$- and $s$-type amplitudes from the subset under consideration at order $b$.

As we already mentioned, without loss of generality we choose the initial state and the Hamiltonian to be singlets. All the matrix elements of interest, for both $a$- and $s$-type amplitudes, have the form
\begin{equation}\label{eq:me_gen}
    \mel{\bar U^F}{H(\bar{U},b^\prime)}{0}, \qquad \text{where} \qquad 0 \le U_{b^\prime} \le b^\prime, \qquad 0 \le b^\prime \le b.
\end{equation}
In order for a matrix element to be non-vanishing the relation $U^F_{n-1} = U_{b^\prime}$ must be satisfied. Matrix elements with even $b^\prime$ contribute only to $s$-type amplitudes  and matrix elements with odd $b'$ contribute only to $a$-type amplitudes, see the discussion of the decoupling of $a$- and $s$-type sum rules in Section~\ref{sec:U-spin-amp-pairs}.

We consider an $a$($s$)-type amplitude from the subset $S_1^{(k)}$ defined in Eq.~\eqref{eq:amp}. This is equivalent to considering a certain ordering of signs in the $n$-tuple. Since the subset is such that all $n$-tuples have their first $k$ signs fixed, to define an amplitude from the subset it is enough to indicate the ordering of the remaining $n-k$ signs. This ordering is given by the sets $\bar m^{(2)}$. Below we write an expression for the coefficients $C_{j \alpha}$ that enter the decompositions of the amplitudes from the subset $S^{(k)}_1$: 
\begin{equation}\label{eq:me_coeff}
C_{j \alpha} = C^*(\bar u^{(1)}; \bar{m}^{(1)}, \bar U^{(1)}) \times C^*(\bar u^{(2)}; \bar{m}^{(2)}, \bar U^{(2)}) \times \mathop{C_{U^{(1)}_{k-1}, -U^{(1)}_{k-1}}}_{\hspace{12pt} U^{(2)}_{n-k-1}, U^{(1)}_{k-1}}^{\hspace{-10pt} U^F_{n-1}, 0} \times C^*(\bar{u}^H; \bar{m}^H, \bar U).
\end{equation}

Similarly to the case considered in Appendix~\ref{app:SR_counting_doublets} we conclude that the rank of the matrix that connects the amplitudes from the subset $S_1^{(k)}$ in the physical basis with RMEs is determined by $C^*(\bar u^{(2)}; \bar{m}^{(2)}, \bar{U}^{(2)})$. This is the case since $C^*(\bar u^{(2)}; \bar{m}^{(2)}, \bar{U}^{(2)})$ is the only part of $C_{j \alpha}$ that depends on $\bar m^{(2)}$ and thus can not be absorbed via re-definitions of RMEs. Note that, by construction, $\bar{m}^{(1)}$ is the same for all RMEs.

The rank of $C_{j \alpha}$ is equal to the number of linearly independent combinations of RMEs entering the group-theoretical decomposition of the $a$($s$)-type amplitudes. Thus to find $n_{X-a}^{(b)}$($n_{X-s}^{(b)}$) all we need to do is to count the number of different sets $\bar{U}^{(2)}$ that result in a non-zero value of Eq.~\eqref{eq:me_coeff}.

Note, that due to the decoupling of the $a$-type and $s$-type amplitudes, in order to find both $n_{X-a}^{(b)}$ and $n_{X-s}^{(b)}$ the counting needs to be done separately.

\subsubsection{Counting the number of sets $\bar{U}^{(2)}$}

To count the number of different sets $\bar{U}^{(2)}$ that could appear at order  $b$ for the subset under consideration we use Eq.~\eqref{eq:Nnu}. We need to be careful when imposing the limits in which the elements of $\bar U^{(2)}$ can vary for specific $k$, $n$, and $b$.

In order to count the number of different sets $\bar{U}^{(2)}$ we consider the tensor product of $n$ doublets in the order defined in Eq.~\eqref{eq:tensor_prod}, that is:
\begin{equation}\label{eq:basis_proof}
    \mathop{\otimes}_i^n d_i = \underbrace{\left( d_1 \otimes d_2 \otimes ... d_{k} \right)}_{k} \otimes \underbrace{\left( d_{k+1} \otimes d_{k + 2} \otimes ... d_{n} \right)}_{n-k}.
\end{equation}

$\bar{U}^{(1)}$ is fixed for all amplitudes in the subset $S^{(k)}_{1}$:
\begin{equation}
    U^{(1)}_1 = 1, \qquad U^{(1)}_2 = \frac{3}{2}, \qquad \dots \quad, \quad U^{(1)}_{k-1} = \frac{k}{2}.
\end{equation}
This is because the absolute value of the third component is equal to the highest irrep for all intermediate tensor products of the first $k$ doublets. There, however, could be several different sets $\bar{U}^{(2)}$. All of them are such that the last element, $U^{(2)}_{n-k-1}$, satisfies the following:
\begin{equation}\label{eq:J2_limits}
    \frac{k}{2}\le U^{(2)}_{n-k-1} \le \frac{k}{2} + b.
\end{equation}

\begin{itemize}
\item $U^{(2)}_{n-k-1} \ge {k/2}$ since the total $m$-QN of the doublets after the semicolon is $M^{(2)} = {k/2}$.
\item $U^{(2)}_{n-k-1} \le \frac{k}{2} + b$ since at order $b$ we consider up to $b$ insertions of the spurion operator, meaning that the maximum value of $U^F_{n-1}$ is equal to $b$. Now, as $U^F_{n-1}$ is constructed from adding the angular momenta $U^{(1)}_{k-1}$ and $U^{(2)}_{n-k-1}$, we have 
\beq
\left|U^{(1)}_{k-1}-U^{(2)}_{n-k-1}\right| \leq U^F_{n-1} \leq U^{(1)}_{k-1}+U^{(2)}_{n-k-1}.
\eeq
Since $U^{(2)}_{n-k-1} \geq U^{(1)}_{k-1}$, it follows that $U^{(2)}_{n-k-1}-U^{(1)}_{k-1}\leq U^F_{n-1}$. Therefore, the values of $U^{(2)}_{n-k-1}$ such that $U^F_{n-1} \le b$ can be found from
    \begin{equation}
        U^{(2)}_{n-k-1} - U^{(1)}_{k-1} \le b \hspace{10pt} \Rightarrow \hspace{10pt} U^{(2)}_{n-k-1} \le U^{(1)}_{k-1} + b 
    \end{equation}
    Greater values of $U^{(2)}_{n-k-1}$ give rise only to $U^F_{n-1} > b$.     
\end{itemize}
Using Eq.~\eqref{eq:Nnu} we can find the number of different sets $\bar{U}^{(2)}$ for every fixed value of $U^{(2)}_{n-k-1}$ from the interval above, the sum of these numbers gives the number of all sets $\bar{U}^{(2)}$ that one can have for the chosen $k$, $n$, and $b$.

In the next subsection we use this fact in order to obtain the explicit expressions for the numbers of matrix elements that enter the decompositions of the $a$- and $s$-type amplitudes.

\subsubsection{Counting $n_{X-a}^{(b)}$ and $n_{X-s}^{(b)}$}
The counting of $n_{X-a}^{(b)}$ and $n_{X-s}^{(b)}$ is different for even and odd $b$ because matrix elements with even $b$ enter the decompositions of the $s$-type amplitudes, while matrix elements with odd $b$ enter the decompositions of the $a$-type amplitudes. 

We start the discussion for the case of even $b$. In that case,
$n_{X-s}^{(b)}$ can be found as the number of all the sets $\bar{U}^{(2)}$ that satisfy the condition in Eq.~\eqref{eq:J2_limits}. This is the case, since all $U^{(2)}_{n-k-1}$ from the interval given in Eq.~(\ref{eq:J2_limits}) can appear in the decompositions of the $s$-type amplitudes. Thus, using Eq.~\eqref{eq:Nnu}, for even $b$ we find
\begin{equation}\label{eq:n_X-s^b_even}
    n_{X-s}^{(b)} = \sum_{U = {k/2}}^{{k/2}+b} N^{n-k}_{U}.
\end{equation}
Here, $U$ takes the values $k/2,\, k/2+1,\, k/2+2\,\dots, k/2+b$. 
The matrix elements with even $b$, however, do not contribute to $a$-type amplitudes, thus only $U^{(2)}_{n-k-1}$ that satisfy $k/2 \le U^{(2)}_{n-k-1} \le k/2 + b - 1$ enter the decompositions of $a$-type sum rules:
\begin{equation}\label{eq:n_X-a^b_even}
    n_{X-a}^{(b)} = \sum_{U = {k/2}}^{{k/2}+b - 1} N^{n-k}_{U}.
\end{equation}

For odd $b$ the situation is reversed and we have
\begin{equation}\label{eq:n_X-s-a^b_odd}
n_{X-a}^{(b)} = \sum_{U = {k/2}}^{{k/2}+b} N^{n-k}_{U}, \hspace{25pt} n_{X-s}^{(b)} = \sum_{U = {k/2}}^{{k/2}+b - 1} N^{n-k}_{U}.
\end{equation}

\subsection{Step 2}

Using Eqs.~\eqref{eq:nSk-nAk-def},~\eqref{eq:Nnu},~\eqref{eq:n_X-a^b_even} and \eqref{eq:n_X-s^b_even} and setting $k = {n}/{2} - b$ we see that for even $b$ 
\begin{equation}
\begin{gathered}
    n_A^{({n/2} - b)} - n_{X-a}^{(b)} = 1, \hspace{15pt} \forall n, b,\\
    n_A^{({n/2} - b)} - n_{X-s}^{(b)} = 0, \hspace{15pt} \forall n, b.\\
\end{gathered}
\end{equation}
This means that at order $b$ there is exactly one sum rule among the $a$-type amplitudes of the subset $S_1^{(k)}$. There are no $s$-type sum rules among the amplitudes from $S_1^{(k)}$ at this order. 

Similarly for odd $b$ we use Eqs.~\eqref{eq:nSk-nAk-def},~\eqref{eq:Nnu} and~\eqref{eq:n_X-s-a^b_odd} to find
\begin{equation}
\begin{gathered}
    n_A^{({n/2} - b)} - n_{X-a}^{(b)} = 0, \hspace{15pt} \forall n, b,\\
    n_A^{({n/2} - b)} - n_{X-s}^{(b)} = 1, \hspace{15pt} \forall n, b.\\
\end{gathered}
\end{equation}

Summing up, we have obtained the following result. For any $n$ and order of breaking $b\leq {n}/{2}-1$ there exists a subset $S_1^{(k)}$ with $k = {n/2} - b$, such that there is exactly one sum rule among the amplitudes of the subset. It is an $a$-type sum rule for even $b$ and an $s$-type sum rule for odd $b$. 

\subsection{Step 3}

We found that there exists exactly one sum rule between amplitudes in the subset under consideration. A change of basis does not affect the number of sum rules between amplitudes from the subset nor the form of the sum rules. By a change of basis for the subset, particularly the change of the order in which one takes the tensor product of the $n-k$ doublets after the semicolon in Eq.~\eqref{eq:amp}, any amplitude in the subset can be exchanged with any other amplitude in the subset. This means that the sum rules must be symmetric under the exchange of any two amplitudes. This implies the following symmetric form for sum rules
\begin{equation}\label{eq:SR}
\sum_{a_i \in S_1^{(k)}} a_i = 0 \hspace{15pt} \text{and} \hspace{15pt} \sum_{s_i \in S_1^{(k)}} s_i = 0
\end{equation}
for even and odd $b$ respectively, and $k = {n/2} - b$.  The sums are taken over all the amplitudes in $S_1^{(k)}$.

\subsection{Step 4}

Now we are ready to discuss the entire set of subsets of the $U$-spin pairs $S^{(k)}$ with $k = {n/2} - b$. Eq.~\eqref{eq:nSk-nAk-def} defines $n_S^{(k)}$ and $n_A^{(k)}$ which are the number of subsets in $S^{(k)}$ and the number of $U$-spin pairs in each subset, respectively.

As in step 3 we use the fact that the choice of the specific $U$-spin basis, that is, the choice of the order in which the tensor product is taken, does not affect the number of sum rules between amplitudes nor the form of the sum rules. For any subset there is a basis choice such that the group theoretical decomposition of amplitudes takes the same form as the decomposition for the subset $S_1^{(k)}$ in the basis of Eq.~\eqref{eq:basis_proof}. Thus each of the $n_S^{(k)}$ subsets has exactly one sum rule among its amplitudes. The sum rule has the symmetric form given in Eq.~\eqref{eq:SR}.

In other words, we have proven that for any $n$ and even(odd) $b\leq {n/2}-1$ there are $n_S^{\left({n/2}-b\right)}$ $a$($s$)-type sum rules, that are valid up to order $b$ and broken at order $b+1$, and that are given in Eq.~\eqref{eq:SR1}.

\subsection{Corollary I}

Let us start by considering even $b$. As we have shown above, at even~$b$ the system has at least $n_S^{(\frac{n}{2}-b)}$ $a$-type sum rules.  $b+1$ is odd and there exist at least $n_S^{({n/2} - b - 1)}$ $s$-type sum rules. The $s$-type sum rules also hold at order $b$ since only matrix elements with even $b$ contribute to the $s$-type amplitudes. Together this implies that, according to the theorem, the number of $a$- and $s$-type sum rules for the case of even $b$ is at least $n_S^{(\frac{n}{2}-b)} + n_S^{(\frac{n}{2} -b - 1)}$. Consideration of odd $b$ leads to the same result.

Now, we can compare this number with the total number of sum rules $n_{SR}^{(b)}$ given in Eq.~(\ref{eq:nSR_doublets}). Using Eq.~\eqref{eq:nSk-nAk-def} for $n_S^{(k)}$, we find
\begin{equation}
n_{SR}^{(b)} = n_S^{({n}/{2} - b)} + n_S^{({n}/{2}-b-1)} \hspace{25pt} \forall n,b\,.
\end{equation}
We see that the counting of sum rules predicted by the theorem at order $b$ is the same as the counting of all sum rules of the system. This means that all sum rules of the $U$-spin system at any order of breaking can be found as symmetric sums of amplitudes from the subsets of sets $S^{(k)}$.

\subsection{Corollary II}

Table~\ref{tab:sum-rules-rule} summarizes the statements of the Theorem and Corollary I. At any order of breaking $b \le {n/2} - 1$ the $U$-spin system has $a$- and $s$-type sum rules which are given as sums over the amplitudes of subsets $S^{(k)}$, where $k$ takes different values for $a$- and $s$-type sum rules, depending on the parity of $b$.
This difference in the definition of $k$ is due to the decoupling of the $a$- and $s$-type amplitudes. That is, the fact that the $a$-type amplitudes contain only contributions from matrix elements that appear with odd $b$ and $s$-type amplitudes that have matrix elements that appear at even $b$.

\section{Generalization to the case of arbitrary irreps \label{app:mu-factor}}

As we outline in Section~\ref{sec:sym}, the key ideas that are used in obtaining the sum rules for arbitrary systems from the sum rules of systems of doublets are basis rotation and symmetrization.
In this appendix we provide extra details and proofs for that process.

\subsection{Basis rotation for amplitudes}
Consider a system of $r$ arbitrary $U$-spin irreps $u_0, u_1, \dots, u_{r-1}$. We would like to obtain the sum rules for this system from the sum rules of the underlying system of $n$ would-be doublets, where $n$ is given in Eq.~\eqref{eq:would-be-n-def}. As in Section~\ref{sec:sym} we denote the system of $n$ would-be doublets as ``the original system'' and the system of $r$ arbitrary representations as ``the new system.''

Each irrep $u_j$ of the new system is constructed as the highest irrep in the tensor product of $n_j = 2u_j$ doublets of the original system. Being the highest irrep implies that the irrep is totally symmetric
with respect to the interchange of the representations in the original system.  
We denote the amplitudes of the original system as $A^{(d)}(\bar{m}_0, \bar{m}_1, \dots, \bar{m}_{r-1})$, where the label $(d)$ indicates that this is an amplitude of a system of doublets. 
Note that in the main text we only considered one symmetrization, while here we consider the more general case of $r$ symmetrizations.
Each $\bar m_j$ is a set of $2u_j$ $m$-QNs of the doublets of the original system that we use in order to build the irrep $u_j$ of the new system. For each of the tensor products of the $n_j$ doublets of the original system we perform a basis rotation according to Eq.~\eqref{eq:basis_rot_def}. We arrive at the following result
\begin{equation}\label{eq:doublet_to_u_basis_rot_r_irreps}
A^{(d)}(\bar{m}_0, \dots, \bar{m}_{r-1}) = \sum_{\bar{U}_0, \dots, \bar{U}_{r-1}} \left(\prod_{j=0}^{r-1}  C^*\left(\bar{m}_j,\bar{U}_j\right) \right) A\left(\bar{U}_0, \dots, \bar{U}_{r-1}, M_0, \dots, M_{r-1}\right),
\end{equation}
where $M_j$ is the sum of the $m$-QNs of the set $\bar m_j$. 
Each set $\bar U_j$ contains $n_j-1$ elements. Note that as in Eq.~\eqref{eq:basis_rot_def}, the sum in Eq.~\eqref{eq:doublet_to_u_basis_rot_r_irreps} goes over all the possible sets $\bar{U}_0, \dots, \bar{U}_{r-1}$, and not over the particular elements in these sets.

Now, consider one specific representation $u_j$ of the new system. Among all the possible sets of $U$-type QNs $\bar U_j$ only one has the total $U$-spin $u_j$. We denote this set of $U$-type QNs as $\bar{U}_j^{(h)}$, where the label $(h)$ highlights that the set corresponds to the highest possible representation in the tensor product. Thus, out of all the terms that enter the RHS of Eq.~\eqref{eq:doublet_to_u_basis_rot_r_irreps} only one amplitude belongs to the new system that we are interested in. We denote this amplitude as $A(M_0, M_1, \dots, M_{r-1})$.

The procedure of obtaining the sum rules for the new system can be performed in two steps. First, we perform the basis rotation as in Eq.~\eqref{eq:doublet_to_u_basis_rot_r_irreps} and rewrite the sum rules of the original system in terms of the amplitudes in the RHS of Eq.~\eqref{eq:doublet_to_u_basis_rot_r_irreps}. Second, we manipulate the sum rules to obtain relations that only contain the amplitudes with the highest irreps, that is, $A(M_0, M_1, \dots, M_{r-1})$. We show below that the latter is guaranteed to be realizable due to the decoupling of sum rules corresponding to different combinations of sets $\bar U_j$.

As a consequence of the decoupling, in order to obtain the sum rules of the new system from the sum rules of the original system it is enough to just perform the following substitution inside the sum rules of the original system
\begin{equation}\label{eq:doublet_to_u_basis_subst_r_irreps}
A^{(d)}(\bar{m}_0, \dots, \bar{m}_{r-1}) \, \longrightarrow \,\, \left(\prod_{j=0}^{r-1}  C^*\left(\bar{m}_j,\bar{U}_j^{(h)}\right) \right) A\left(M_0, \dots, M_{r-1}\right).
\end{equation}
Note that the mapping in Eq.~(\ref{eq:doublet_to_u_basis_subst_r_irreps}) is not injective, \emph{i.e.}~in general several $A^{(d)}(\bar{m}_0, \dots, \bar{m}_{r-1})$ are mapped onto the same $A\left(M_0, \dots, M_{r-1}\right)$.

In the next subsection we explain why the decoupling takes place.

\subsection{Decoupling}

Below we expand on why the sum rules in the symmetrized basis decouple. In short the decoupling takes place due to the fact that different RMEs cannot cancel each other,~\emph{i.e.}~the respective contributions have to cancel separately in each sum rule.

Note first that in order to show that a system of sum rules decouples, it is enough to show that this is manifestly the case in one specific basis. In any other basis the decoupling might not be manifest, however, the decoupling still will be the underlying feature of the system. Therefore, for the argument of this subsection, we construct a specific basis of RMEs in which the decoupling is manifest.
Recall that different bases are generated by changing the order of the tensor products.

When we are given a system of $n$ doublets and perform the decomposition of the amplitudes in terms of RMEs, as described in detail in Appendix~\ref{app:RMEdecomposition}, we can always choose a basis of RMEs  such that it mimics the new system. To construct such a basis we perform the tensor product of the doublets of the original system in a specific order as follows. We first separately multiply the doublets of each subset of $n_j$ doublets that are eventually used to build the irrep $u_j$. For each $u_j$ we use the same order of the tensor product that is encoded in the corresponding set $\bar U_j$ that is used to define the amplitudes in the RHS of Eq.~\eqref{eq:doublet_to_u_basis_rot_r_irreps}. The remaining tensor products can be arbitrary.

For the purpose of the argument, we introduce the following notations 
\begin{align}
\bar U &= \{\bar{U}_0, \dots, \bar{U}_{r-1}\}, \\  \bar M & = \{M_0, \dots, M_{r-1}\}, \\ A\left(\bar U, \bar{M}\right) &\equiv A\left(\bar{U}_0, \dots, \bar{U}_{r-1}, M_0, \dots, M_{r-1}\right).
\end{align}
In the basis constructed above, the amplitudes $A\left(\bar U, \bar M\right)$ can be written only in terms of RMEs that have $\bar U$ as a part of their multi-index $\alpha$ and no other RME could be present in the decomposition. That is, we see that sets of amplitudes with different $\bar{U}$ are decomposed in terms of different non-overlapping sets of RMEs. This shows the decoupling: Any sum rule that holds between the amplitudes of the original system must also hold when the amplitudes of the original system are replaced by the amplitudes $A\left(\bar U, \bar M\right)$ with the factors as in Eq.~\eqref{eq:doublet_to_u_basis_rot_r_irreps}. Thus the substitution in Eq.~\eqref{eq:doublet_to_u_basis_subst_r_irreps} is justified.

\subsection{The symmetry factor \label{eq:deriving-the-symmetry-factor}}

To perform the substitution in Eq.~\eqref{eq:doublet_to_u_basis_subst_r_irreps} we need to to know the coefficients $C^*(\bar m_j, \bar U^{(h)}_j)$. A straightforward way to find these coefficients would be to use the definition of coefficients $C^*$ given in Eq.~\eqref{eq:basis_rot_def}. This approach requires us to specify a $U$-spin basis and evaluate the coefficients of interest in this chosen basis. However, since the coefficients $C^*(\bar m_j, \bar U^{(h)}_j)$ that appear in the RHS of Eq.~\eqref{eq:doublet_to_u_basis_subst_r_irreps} correspond to the highest representation in the tensor product of $n_j$ doublets, one can find them using simple combinatorics. In what follows we use the symmetry of the highest representation in the tensor product of doublets to derive a basis independent expression for the coefficients $C^*(\bar m_j, \bar U_j^{(h)})$.

Consider a state $\ket{u_j,M}$. As above we build this state from the product of $n_j$ doublets such that $n_j = 2u_j$, and thus $u_j$ is the highest irrep in the tensor product. The highest representation is totally symmetric, which means that the coefficients in the basis rotation analogous to Eq.~(\ref{eq:basis_rot_def}) 
should be the same for all sets $\bar m_j$ that appear in the decomposition of $\ket{u_j, M}$. We define this universal prefactor $C_\text{sym}(u_j,M)$ such that
\begin{align}
\label{eq:C*_Csym_relation}
\ket{u_j, M} = \frac{1}{\sqrt{C_\text{sym}(u_j, M)}}\sum_{\bar m_j} \delta_{M_j, M}\ket{\bar m_j}\,.
\end{align}
Note that $\ket{\bar m_j}$ represents a tensor product of $n_j$ doublets with $m$-QNs from the set $\bar m_j$.
Note further that all $\bar m_j$ we sum over in Eq.~(\ref{eq:C*_Csym_relation}) have the same number of elements, namely $n_j=2u_j$ and that the Kronecker delta ensures that the sum goes only over the sets $\bar m_j$ with  total $m$-QN equal to $M$. The normalization condition implies that the square of the coefficient $C_\text{sym}(u_j,M)$ is equal to the number of different sets $\bar{m}_j$ with the fixed value $M_j = M$, so we arrive at
\begin{equation}\label{eq:counting-mj}
   C_\text{sym}(u_j, M_j) = C(2u_j, u_j-M_j) = C(2u_j, y_j),
\end{equation}
where $C(*,*)$ is a binomial coefficient. The first binomial coefficient is written in terms of the $m$-QN $M_j$ of the representation $u_j$, while the latter uses the number of minus signs $y_j$ from the $y_j$-notation introduced in Section~\ref{sec:gen_1d}.

Eqn.~(\ref{eq:counting-mj}) can be seen as follows. All $\ket{\bar{m}_j}$ that contribute to Eq.~\eqref{eq:C*_Csym_relation} can be represented using $2u_j$ signs. We count the number of ways to arrange minus signs such that we have a total $m$-QN $M_j$. There are two possibilities:
\begin{itemize}
\item $M_j\geq 0$:  Then in order to contribute, $\ket{\bar{m}_j}$ must contain $2M_j$ \lq\lq{}$+$\rq\rq{}-signs, and on top of that an equal number of $\frac{2 u_j-2 M_j}{2}$ \lq\lq{}$+$\rq\rq{}- and \lq\lq{}$-$\rq\rq{}-signs. Thus the total number of \lq\lq{}$-$\rq\rq{}-signs is $y_j = u_j-M_j$, which can be chosen from a total of $2u_j$ signs, it follows Eq.~(\ref{eq:counting-mj}).
\item $M_j<0$: Then in order to contribute, $\ket{\bar{m}_j}$ must contain $2\vert M_j\vert$ \lq\lq{}$-$\rq\rq{}-signs, and on top of that an equal number of $\frac{2 u_j-2 \vert M_j\vert}{2}$ \lq\lq{}$+$\rq\rq{}- and \lq\lq{}$-$\rq\rq{}-signs. This makes a total of $y_j = u_j+\vert M_j\vert = u_j-M_j$ minus signs, it follows again Eq.~(\ref{eq:counting-mj}).
\end{itemize}

Finally, we relate our results in Eqs.~\eqref{eq:C*_Csym_relation} and~\eqref{eq:counting-mj} to the coefficient $C^*(\bar{m}_j, \bar U_j^{(h)})$ in Eq.~(\ref{eq:doublet_to_u_basis_subst_r_irreps}). To do this we take Eq.~\eqref{eq:C*_Csym_relation} and multiply it by $\bra{u_j, M}$ on both sides
\begin{align}
    1 &= \frac{1}{\sqrt{C_{\text{sym}}(u_j,M)}} \sum_{\bar{m}_j} \delta_{M_j,M} \braket{u_j, M}{\bar{m}_j}\\
    &=\frac{C^*(\bar{m}_j, \bar U_j^{(h)})}{\sqrt{C_\text{sym}(u_j, M)}}C_\text{sym}(u_j, M) \\
    &= C^*(\bar{m}_j, \bar U_j^{(h)}) \sqrt{C_\text{sym}(u_j, M)},
\end{align}
and where we also used, following from Eq.~\eqref{eq:basis_rot_def},
\begin{align}
\braket{u_j, M}{\bar{m}_j} &= C^*(\bar{m}_j, \bar U_j^{(h)})\,.
\end{align}
It follows:
\begin{equation}
    C^*(\bar{m}_j, \bar U_j^{(h)}) = \frac{1}{\sqrt{C_\text{sym}(u_j, M)}}.
\end{equation}

\subsection{The $\mu$-factor \label{sec:mu-Factor-Appx}}

In Section~\ref{sec:halves-lattice} we study the geometrical approach to writing sum rules for doublets-only systems. The method can be straightforwardly generalized to the case of systems with at least one doublet, while the rest of the representations are arbitrary.

For the lattice in the general case each node is assigned a $\mu$-factor. Note that the $\mu$-factor used in the lattice formalism is different from, but related to the symmetry factor derived in Sec.~\ref{eq:deriving-the-symmetry-factor} above.
The reason for the difference is as follows. When performing the replacement Eq.~(\ref{eq:doublet_to_u_basis_subst_r_irreps}), several different amplitudes from the LHS are mapped onto the same amplitude in the RHS. While in the purely algebraic algorithm, these are then automatically added together, in the lattice algorithm we have to count explicitly how many amplitudes contribute in this way to the same symmetrized amplitude.

There are therefore two sources that contribute to the $\mu$-factors of lattice nodes. One comes from the symmetry factor that we derive above. As we showed, when obtaining the sum rules for the new system from the sum rules for the original system of doublets all we need to do is to perform the substitution given in Eq.~\eqref{eq:doublet_to_u_basis_subst_r_irreps}. Thus each amplitude/node of the lattice for the new system gains a factor of $1/\sqrt{C(2u_j, y_j)}$ for each higher representation, see Eqs.~(\ref{eq:doublet_to_u_basis_subst_r_irreps}, \ref{eq:counting-mj}).

The other contribution to the $\mu$-factor is due to the fact that when transitioning from the system of doublets to the system of arbitrary representations, different $n$-tuples of the doublets-only system can be mapped onto a single generalized $n$-tuple of the system of arbitrary representations. For each representation $u_j$ and for a fixed $y_j$ (which corresponds to a fixed $M_j$) the contribution to the $\mu$-factor is given by the number of different amplitudes of the system of doublets that are mapped into a single amplitude of the general system. 

What we actually need to count in the geometrical method is how many nodes from the lattice for doublets-only system are mapped onto a given node of the lattice that corresponds to the system of arbitrary representations. Consider the example of the lattice point
\begin{equation}\label{eq:example-lattice-point}
(\underset{0}{-},\underset{1}{---++}, \underset{2}{+ +}) = (1, 1, 1)\,,
\end{equation}
where we consider the case $2 u_1=5$, and $M_1=-1/2$, i.e. $y_1=3$.
How many lattice points, \emph{i.e.}~amplitudes, correspond to this node in the lattice for the doublets? To start the counting, the number of ways three minus signs can be assigned to the five available positions is given by the binomial coefficient $C(5,3)=10$. However, in order to get the total number of doublet lattice points that correspond to Eq.~(\ref{eq:example-lattice-point}), we also have to account for the fact that in the lattice there are points that correspond to identical amplitudes when the labels of the three minus signs (in the doublet lattice) are interchanged, see Eq.~(\ref{eq:example-permutations-lattice-points}). This gives another factor $3!$ on top of the binomial coefficient. 
Therefore, the total number of lattice points in the doublet lattice that are mapped onto the same point in the new lattice Eq.~(\ref{eq:example-lattice-point}) is $C(5,3)\times 3! = 60$. 

Alternatively, we can perform the counting also by directly counting the number of arrangements of three distinguishable minus-signs into five numbered positions, no matter what is the permutation of the plus-signs. This is equal to the total number of permutations 5! divided by the permutations of the plus-signs 2!, leading of course to the same result $5!/2!= 60$.

In the general case this translates to the number of ordered arrangements of $y_j$ minus signs into the $2 u_j$ positions, given by
\begin{align}
P(2u_j,y_j) = C(2u_j,y_j) \times y_j! = \frac{(2u_j)!}{(2u_j-y_j)!}\,.
\end{align}
In words: the number of lattice points of the doublet lattice that correspond to one lattice point in the lattice of the higher representation $u_j$ (and number of minus signs $y_j$, determined by the value of $M_j$) is given by the number of unordered ways $C(2u_j, y_j) $ to put the $y_j$ minus signs into the $2u_j$ positions, times the number of ways one can order the minus signs, which is $y_j!$.

Thus, we conclude that each irrep $u_j$ contributes a factor of
\beq
\mu_j= 
\sqrt{C(2u_j,y_j)} \times y_j!\,. \label{eq:complete-mu-factor}
\eeq
The total $\mu$-factor of a node is then given by
\beq
\mu[y_1,y_2,...,y_{r-1}]=\prod_{j=1}^{r-1} \mu_j\,.
\eeq

\section{Decomposition of the $C_b \to L_b P^- P^+$ system in terms of RMEs}\label{app:CbtoLbPP}

In this Appendix we perform the decomposition of the CKM-free amplitudes of the $C_b\to L_b P^- P^+$ set in terms of RMEs. We perform the decomposition up to $b_{\text{max}} = 2$. Table~\ref{tab:CbtoLPP-b01} shows the decompositions of the CKM-free amplitudes in terms of RMEs with $b =0$ and $b = 1$, which we list below:
\begin{align}\label{eq:RMEb01}
    X_1 &= \mel{\frac{3}{2}}{_{1}1}{\frac{1}{2}}, & X_2 &= \mel{\frac{1}{2}}{_{1}1}{\frac{1}{2}},  & X_3 &= \mel{\frac{1}{2}}{_{0}1}{\frac{1}{2}} \nonumber \\
    X_4 &= \mel{\frac{3}{2}}{_1 \left(1 \times 1_\varepsilon\right)_2}{\frac{1}{2}}, & X_5 &= \mel{\frac{3}{2}}{_1 \left(1 \times 1_\varepsilon \right)_1}{\frac{1}{2}}, & X_6 &= \mel{\frac{1}{2}}{_1 \left(1\times 1_\varepsilon\right)_1}{\frac{1}{2}} \nonumber \\
     X_7 &= \mel{\frac{1}{2}}{_0 \left(1 \times 1_\varepsilon \right)_1}{\frac{1}{2}}, & X_8 &= \mel{\frac{1}{2}}{_1 \left(1 \times 1_\varepsilon\right)_0}{\frac{1}{2}}, &  X_9 &= \mel{\frac{1}{2}}{_0 \left(1 \times 1_\varepsilon\right)_0}{\frac{1}{2}}.
\end{align}
For the states we use the notation as in Ref.~\cite{Grossman:2018ptn}, where the subindex for the final state is the intermediate representation in the tensor product of three doublets. The Hamilton operator $H^1$ is denoted by ``$1$'', the spurion operator is denoted by ``$1_\varepsilon$'', and expressions of the form $\left(u_1 \times u_2\right)_{u_3}$ denote the $u_3$ irrep in the tensor product of $u_1$ and $u_2$.

Table~\ref{tab:CbtoLPP-b2} shows the contributions to the decompositions from RMEs with $b = 2$. All $b = 2$ RMEs are listed below:
\begin{align}\label{eq:RMEb2}
X_{10} & = \mel{\frac{3}{2}}{_1 \left(\left(1 \times 1_\varepsilon\right)_2 \times 1_\varepsilon\right)_2}{\frac{1}{2}}, & X_{11} & = \mel{\frac{3}{2}}{_1 \left(\left(1\times1_\varepsilon\right)_2\times 1_\varepsilon\right)_1}{\frac{1}{2}}, \nonumber\\
X_{12} &= \mel{\frac{1}{2}}{_1 \left(\left(1\times1_\varepsilon\right)_2\times 1_\varepsilon\right)_1}{\frac{1}{2}}, & X_{13} & = \mel{\frac{1}{2}}{_0 \left(\left(1 \times 1_\varepsilon\right)_2 \times 1_\varepsilon\right)_1}{\frac{1}{2}}, \nonumber\\
X_{14} & = \mel{\frac{3}{2}}{_1 \left(\left(1\times1_\varepsilon\right)_1\times1_\varepsilon\right)_2}{\frac{1}{2}}, & X_{15} & = \mel{\frac{3}{2}}{_1 \left(\left(1\times 1_\varepsilon\right)_1 \times 1_\varepsilon\right)_1}{\frac{1}{2}}, \nonumber\\
X_{16} & = \mel{\frac{1}{2}}{_1 \left(\left(1\times1_\varepsilon\right)_1 \times 1_\varepsilon\right)_1}{\frac{1}{2}}, & X_{17} & = \mel{\frac{1}{2}}{_0 \left(\left(1 \times 1_\varepsilon\right)_1 \times 1_\varepsilon\right)_1}{\frac{1}{2}}, \nonumber \\
X_{18} &= \mel{\frac{3}{2}}{_1 \left(\left(1\times 1_\varepsilon\right)_0\times 1_\varepsilon\right)_1}{\frac{1}{2}}, & X_{19} &= \mel{\frac{1}{2}}{_1 \left(\left(1\times1_\varepsilon\right)_0\times1_\varepsilon\right)_1}{\frac{1}{2}},\nonumber\\
X_{20}& = \mel{\frac{1}{2}}{_0 \left(\left(1 \times 1_\varepsilon\right)_0\times 1_\varepsilon\right)_1}{\frac{1}{2}}. & &
\end{align}
When writing the decompositions of the amplitudes of the $C_b \to L_b P^- P^+$ set in Tables~\ref{tab:CbtoLPP-b01} and~\ref{tab:CbtoLPP-b2}, we perform the tensor products in the final state in the order $\left(\left(L_b \otimes P^-\right)\otimes P^+\right)$. The tensor product for the Hamilton operators are taken in the order shown in Eqs.~\eqref{eq:RMEb01} and~\eqref{eq:RMEb2}.

To find the $b=0$ sum rules one needs to consider only three RMEs $X_{1}$, $X_2$, $X_3$. For $b=1$ sum rules one needs to also include the RMEs from $X_4$ through $X_9$. In order to write the sum rules that hold up to $b = 2$ all the RMEs $X_1\,-\,X_{20}$ must be considered. Note that not all the RMEs are linearly independent: the rank of the combined matrix that includes all 20 RMEs is equal to 13
thus resulting in one sum rule that holds at order $b = 2$. One can check explicitly that all the sum rules listed in Section~\ref{sec:CbtoLbPP} are indeed in agreement with Tables~\ref{tab:CbtoLPP-b01} and~\ref{tab:CbtoLPP-b2}. We also checked that for $b=3$ the rank of the matrix saturates, i.e.~there are no higher order sum rules. We do not write the explicit table here.

\begingroup
\squeezetable
\begin{table}
\centering
\begin{tabular}{c|c|c|c|c|c|c|c|c|c}
\hline\hline
 \text{Decay amplitude} & ~~$X_1$~~  & ~~$X_2$~~ & ~~$X_3$~~ & ~~$X_4$~~ & ~~$X_5$~~ & ~~$X_6$~~ & ~~$X_7$~~ & ~~$X_8$~~ & ~~$X_9$~~ \\
 \hline\hline
 $A\left(\Lambda_c^+ \to \Sigma^+ K^- K^+\right)$ & $\frac{1}{3}$ & $-\frac{2}{3}$ & $0$ & $\frac{1}{\sqrt{10}}$ & $-\frac{1}{3 \sqrt{2}}$ & $\frac{\sqrt{2}}{3}$ & $0$ & $0$ & $0$\\
 $A\left(\Xi_c^+\to p\pi^- \pi^+\right)$ & $\frac{1}{3}$ & $-\frac{2}{3}$ & $0$ & $-\frac{1}{\sqrt{10}}$ & $\frac{1}{3 \sqrt{2}}$ & $-\frac{\sqrt{2}}{3}$ & $0$ & $0$ & $0$ \\
 $A\left(\Lambda_c^+\to \Sigma^+ \pi^- \pi^+\right)$ & $\frac{1}{3}$ & $\frac{1}{3}$ & $-\frac{1}{\sqrt{3}}$ & $\frac{1}{\sqrt{10}}$ & $-\frac{1}{3 \sqrt{2}}$ & $-\frac{1}{3 \sqrt{2}}$ & $\frac{1}{\sqrt{6}}$ & $0$ & $0$ \\
 $A\left(\Xi_c^+\to p K^- K^+\right)$ & $\frac{1}{3}$ & $\frac{1}{3}$ & $-\frac{1}{\sqrt{3}}$ & $-\frac{1}{\sqrt{10}}$ & $\frac{1}{3 \sqrt{2}}$ & $\frac{1}{3 \sqrt{2}}$ & $-\frac{1}{\sqrt{6}}$ & $0$ & $0$ \\
 $A\left(\Lambda_c^+\to \Sigma^+ \pi^- K^+\right)$ & $\frac{\sqrt{2}}{3}$ & $-\frac{1}{3 \sqrt{2}}$ & $-\frac{1}{\sqrt{6}}$ & $\frac{2}{3 \sqrt{5}}$ & $0$ & $0$ & $0$ & $\frac{1}{3 \sqrt{2}}$ & $\frac{1}{\sqrt{6}}$ \\
 $A\left(\Xi_c^+\to p K^- \pi^+\right)$ & $\frac{\sqrt{2}}{3}$ & $-\frac{1}{3 \sqrt{2}}$ & $-\frac{1}{\sqrt{6}}$ & $-\frac{2}{3 \sqrt{5}}$ & $0$ & $0$ & $0$ & $-\frac{1}{3 \sqrt{2}}$ & $-\frac{1}{\sqrt{6}}$ \\
 $A\left(\Lambda_c^+\to p K^- \pi^+ \right)$ & $\frac{1}{3}$ & $\frac{1}{3}$ & $\frac{1}{\sqrt{3}}$ & $\frac{1}{\sqrt{10}}$ & $-\frac{1}{3 \sqrt{2}}$ & $-\frac{1}{3 \sqrt{2}}$ & $-\frac{1}{\sqrt{6}}$ & $0$ & $0$ \\
 $A\left(\Xi_c^+\to \Sigma^+ \pi^- K^+\right)$ & $\frac{1}{3}$ & $\frac{1}{3}$ & $\frac{1}{\sqrt{3}}$ & $-\frac{1}{\sqrt{10}}$ & $\frac{1}{3 \sqrt{2}}$ & $\frac{1}{3 \sqrt{2}}$ & $\frac{1}{\sqrt{6}}$ & $0$ & $0$ \\
 $A\left(\Lambda_c^+\to p K^- K^+\right)$ & $\frac{\sqrt{2}}{3}$ & $-\frac{1}{3 \sqrt{2}}$ & $\frac{1}{\sqrt{6}}$ & $\frac{2}{3 \sqrt{5}}$ & $0$ & $0$ & $0$ & $\frac{1}{3 \sqrt{2}}$ & $-\frac{1}{\sqrt{6}}$ \\
 $A\left(\Xi_c^+\to \Sigma^+ \pi^- \pi^+\right)$ & $\frac{\sqrt{2}}{3}$ & $-\frac{1}{3 \sqrt{2}}$ & $\frac{1}{\sqrt{6}}$ & $-\frac{2}{3 \sqrt{5}}$ & $0$ & $0$ & $0$ & $-\frac{1}{3 \sqrt{2}}$ & $\frac{1}{\sqrt{6}}$ \\
 $A\left(\Lambda_c^+\to p \pi^- \pi^+\right)$ & $\frac{\sqrt{2}}{3}$ & $\frac{\sqrt{2}}{3}$ & $0$ & $\frac{2}{3 \sqrt{5}}$ & $0$ & $0$ & $0$ & $-\frac{\sqrt{2}}{3}$ & $0$ \\
 $A\left(\Xi_c^+\to \Sigma^+ K^- K^+\right)$ & $\frac{\sqrt{2}}{3}$ & $\frac{\sqrt{2}}{3}$ & $0$ & $-\frac{2}{3 \sqrt{5}}$ & $0$ & $0$ & $0$ & $\frac{\sqrt{2}}{3}$ & $0$ \\
 $A\left(\Lambda_c^+\to p \pi^- K^+\right)$ & $1$ & $0$ & $0$ & $\frac{1}{\sqrt{10}}$ & $\frac{1}{\sqrt{2}}$ & $0$ & $0$ & $0$ & $0$ \\
 $A\left(\Xi_c^+\to \Sigma^+ K^- \pi^+\right)$ & $1$ & $0$ & $0$ & $-\frac{1}{\sqrt{10}}$ & $-\frac{1}{\sqrt{2}}$ & $0$ & $0$ & $0$ & $0$ \\
 \hline\hline
\end{tabular}
\caption{RME decomposition of $C_b \to L_bP^-P^+$ amplitudes up to first order $U$-spin breaking ($b=0$ and $b=1$). The corresponding RMEs $X_\alpha$ are defined in Eq.~\eqref{eq:RMEb01}.\label{tab:CbtoLPP-b01}}
\end{table}
\endgroup

\begingroup
\squeezetable
\begin{table}
\centering
\begin{tabular}{c|c|c|c|c|c|c|c|c|c|c|c}
\hline\hline
 \text{Decay amplitude} & ~~$X_{10}$~~  & ~~$X_{11}$~~ & ~~$X_{12}$~~ & ~~$X_{13}$~~ & ~~$X_{14}$~~ & ~~$X_{15}$~~ & ~~$X_{16}$~~ & ~~$X_{17}$~~ & ~~$X_{18}$~~&~~$X_{19}$~~&~~$X_{20}$~~\\
\hline\hline
$A\left(\Lambda_c^+\to \Sigma^+ K^- K^+\right)$ & $-\frac{1}{2 \sqrt{15}}$ & $-\frac{1}{2 \sqrt{15}}$ & $\frac{1}{\sqrt{15}}$ & $0$ & $-\frac{1}{2 \sqrt{5}}$ & $\frac{1}{6}$ & $-\frac{1}{3}$ & $0$ & $0$ & $0$ & $0$ \\
$A\left(\Xi_c^+\to p\pi^- \pi^+\right)$ & $-\frac{1}{2 \sqrt{15}}$ & $-\frac{1}{2 \sqrt{15}}$ & $\frac{1}{\sqrt{15}}$ & $0$ & $-\frac{1}{2 \sqrt{5}}$ & $\frac{1}{6}$ & $-\frac{1}{3}$ & $0$ & $0$ & $0$ & $0$ \\
$A\left(\Lambda_c^+\to \Sigma^+ \pi^- \pi^+\right)$ & $-\frac{1}{2 \sqrt{15}}$ & $-\frac{1}{2 \sqrt{15}}$ & $-\frac{1}{2 \sqrt{15}}$ & $\frac{1}{2 \sqrt{5}}$ & $-\frac{1}{2 \sqrt{5}}$ & $\frac{1}{6}$ & $\frac{1}{6}$ & $-\frac{1}{2 \sqrt{3}}$ & $0$ & $0$ & $0$ \\
$A\left(\Xi_c^+\to p K^- K^+\right)$ & $-\frac{1}{2 \sqrt{15}}$ & $-\frac{1}{2 \sqrt{15}}$ & $-\frac{1}{2 \sqrt{15}}$ & $\frac{1}{2 \sqrt{5}}$ & $-\frac{1}{2 \sqrt{5}}$ & $\frac{1}{6}$ & $\frac{1}{6}$ & $-\frac{1}{2 \sqrt{3}}$ & $0$ & $0$ & $0$ \\
$A\left(\Lambda_c^+\to \Sigma^+ \pi^- K^+\right)$ & $0$ & $-\frac{2}{3} \sqrt{\frac{2}{15}}$ & $\frac{1}{3}\sqrt{\frac{2}{15}}$ & $\frac{1}{3}\sqrt{\frac{2}{5}}$ & $0$ & $0$ & $0$ & $0$ & $-\frac{1}{3}\sqrt{\frac{2}{3}}$ & $\frac{1}{3 \sqrt{6}}$ & $\frac{1}{3 \sqrt{2}}$ \\
$A\left(\Xi_c^+\to p K^- \pi^+\right)$ & $0$ & $-\frac{2}{3} \sqrt{\frac{2}{15}}$ & $\frac{1}{3}\sqrt{\frac{2}{15}}$ & $\frac{1}{3}\sqrt{\frac{2}{5}}$ & $0$ & $0$ & $0$ & $0$ & $-\frac{1}{3}\sqrt{\frac{2}{3}}$ & $\frac{1}{3 \sqrt{6}}$ & $\frac{1}{3 \sqrt{2}}$ \\
$A\left(\Lambda_c^+\to p K^- \pi^+\right)$ & $-\frac{1}{2 \sqrt{15}}$ & $-\frac{1}{2 \sqrt{15}}$ & $-\frac{1}{2 \sqrt{15}}$ & $-\frac{1}{2 \sqrt{5}}$ & $-\frac{1}{2 \sqrt{5}}$ & $\frac{1}{6}$ & $\frac{1}{6}$ & $\frac{1}{2 \sqrt{3}}$ & $0$ & $0$ & $0$ \\
$A\left(\Xi_c^+\to \Sigma^+ \pi^- K^+\right)$ & $-\frac{1}{2 \sqrt{15}}$ & $-\frac{1}{2 \sqrt{15}}$ & $-\frac{1}{2 \sqrt{15}}$ & $-\frac{1}{2 \sqrt{5}}$ & $-\frac{1}{2 \sqrt{5}}$ & $\frac{1}{6}$ & $\frac{1}{6}$ & $\frac{1}{2 \sqrt{3}}$ & $0$ & $0$ & $0$ \\
$A\left(\Lambda_c^+\to p K^- K^+\right)$ & $0$ & $-\frac{2}{3} \sqrt{\frac{2}{15}}$ & $\frac{1}{3}\sqrt{\frac{2}{15}}$ & $-\frac{1}{3}\sqrt{\frac{2}{5}}$ & $0$ & $0$ & $0$ & $0$ & $-\frac{1}{3}\sqrt{\frac{2}{3}}$ & $\frac{1}{3 \sqrt{6}}$ & $-\frac{1}{3 \sqrt{2}}$ \\
$A\left(\Xi_c^+\to \Sigma^+ \pi^- \pi^+\right)$ & $0$ & $-\frac{2}{3} \sqrt{\frac{2}{15}}$ & $\frac{1}{3}\sqrt{\frac{2}{15}}$ & $-\frac{1}{3}\sqrt{\frac{2}{5}}$ & $0$ & $0$ & $0$ & $0$ & $-\frac{1}{3}\sqrt{\frac{2}{3}}$ & $\frac{1}{3 \sqrt{6}}$ & $-\frac{1}{3 \sqrt{2}}$\\
$A\left(\Lambda_c^+\to p \pi^- \pi^+\right)$ & $0$ & $-\frac{2}{3} \sqrt{\frac{2}{15}}$ & $-\frac{2}{3} \sqrt{\frac{2}{15}}$ & $0$ & $0$ & $0$ & $0$ & $0$ & $-\frac{1}{3}\sqrt{\frac{2}{3}}$ & $-\frac{1}{3}\sqrt{\frac{2}{3}}$ & $0$ \\
$A\left(\Xi_c^+\to \Sigma^+ K^- K^+\right)$ & $0$ & $-\frac{2}{3} \sqrt{\frac{2}{15}}$ & $-\frac{2}{3} \sqrt{\frac{2}{15}}$ & $0$ & $0$ & $0$ & $0$ & $0$ & $-\frac{1}{3}\sqrt{\frac{2}{3}}$ & $-\frac{1}{3}\sqrt{\frac{2}{3}}$ & $0$ \\
$A\left(\Lambda_c^+\to p \pi^- K^+\right)$ & $\frac{1}{2 \sqrt{15}}$ & $-\frac{1}{2}\sqrt{\frac{3}{5}}$ & $0$ & $0$ & $\frac{1}{2 \sqrt{5}}$ & $\frac{1}{2}$ & $0$ & $0$ & $0$ & $0$ & $0$ \\
$A\left(\Xi_c^+\to \Sigma^+ K^- \pi^+\right)$ & $\frac{1}{2 \sqrt{15}}$ & $-\frac{1}{2}\sqrt{\frac{3}{5}}$ & $0$ & $0$ & $\frac{1}{2 \sqrt{5}}$ & $\frac{1}{2}$ & $0$ & $0$ & $0$ & $0$ & $0$ \\
\hline\hline
\end{tabular}
\caption{Contributions to the RME decomposition $C_b \to L_b P^- P^+$ from second order $U$-spin breaking ($b=2$). The corresponding RMEs $X_\alpha$ are defined in Eq.(\ref{eq:RMEb2}).\label{tab:CbtoLPP-b2}}
\end{table}
\endgroup

\section{The $n=6$, three triplets system}\label{app:3t}

In this appendix we derive in detail the sum rules for the system of three triplets. The sum rules for this system can be obtained from the sum rules for the system of two doublets and two triplets. Thus we introduce the following definitions:
\begin{itemize}
\item System I: a $U$-spin set described by two doublets and two triplets, $u_0 = u_1 = 1/2$ and $u_2 = u_3 = 1$. The amplitudes and the corresponding nodes of the lattice for this system are as follows:
\begin{align}
(1,2): \qquad  A_7^\text{(I)} &= \left(-,-,-+,++\right), & A_{52}^\text{(I)} &= \left(+,+, -+,--\right), \nonumber\\
(1,3): \qquad  A_{13}^\text{(I)} &= \left(-,-, ++,-+\right), & A_{49}^\text{(I)} &= \left(+,+, --,-+\right),\nonumber\\
(2,2): \qquad A_{19}^\text{(I)} &= \left(-,+,--,++\right), & A_{44}^\text{(I)}& = \left(+,-,++,-- \right), \nonumber\\
(2,3): \qquad A_{21}^\text{(I)} &= \left(-,+,-+,-+\right), & A_{37}^\text{(I)} &=\left(+,-,-+,-+\right),\nonumber\\
(3,3): \qquad A_{28}^\text{(I)}&= \left(-,+,++,--\right), & A_{35}^\text{(I)} & = \left(+,-,--,++\right).
\end{align}
\item System II: a $U$-spin set described by three triplets, $u_0 = u_1 = u_2 = 1$. The amplitudes for this system are: 
    \begin{align}
     A_7^\text{(II)} & = \left(--,-+,++\right), & A_{52}^\text{(II)} & = \left(++, -+,--\right), \nonumber\\
     A_{13}^\text{(II)} & = \left(--, ++,-+\right), & A_{49}^\text{(II)} & = \left(++, --,-+\right),\nonumber\\
     A_{19}^\text{(II)} &= \left(-+,--,++ \right), & A_{28}^\text{(II)} & = \left(-+,++,--\right),\nonumber\\
     A_{21}^\text{(II)} & = \left(-+,-+,-+\right). & &
    \end{align}
\end{itemize}

The sum rules for System I can be read off Fig.~\ref{fig:n6-2d-2t}. We obtain the following trivial $a$-type sum rules valid at $b=0$:
\begin{equation}\label{eq:3t-a-b0}
    a^\text{(I)}_{(1,2)} = a^\text{(I)}_{(1,3)} = a^\text{(I)}_{(2,2)} = a^\text{(I)}_{(2,3)} = a^\text{(I)}_{(3,3)} = 0.
\end{equation}
The $s$-type sum rules up to $b = 1$ are given by
\begin{equation}\label{eq:3t-s-b1}
    s_{(1,2)}^\text{(I)} + s_{(1,3)}^\text{(I)} = 0,  \qquad s_{(1,2)}^\text{(I)} + \sqrt{2}s_{(2,2)}^\text{(I)} + \sqrt{2} s_{(2,3)}^\text{(I)} = 0, \qquad s_{(1,3)}^\text{(I)} + \sqrt{2}s_{(2,3)}^\text{(I)} + \sqrt{2}s_{(3,3)}^\text{(I)} = 0.
\end{equation}
Finally, the $b=2$ $a$-type sum rule is
\begin{equation}\label{eq:3t-a-b2}
    \sqrt{2}a_{(1,2)}^\text{(I)} + \sqrt{2}a_{(1,3)}^\text{(I)} + a_{(2,2)}^\text{(I)} + 2 a_{(2,3)}^\text{(I)} + a_{(3,3)}^\text{(I)} = 0.
\end{equation}

To obtain the sum rules for System~II we perform the substitutions as described in Section~\ref{sec:sym}. We have
\begin{align}
    &(1,2): &
    A_{7}^{\text{(I)}} &\rightarrow A_7^{\text{(II)}}, & 
    A_{52}^{\text{(I)}} &\rightarrow A_{52}^{\text{(II)}}, \nonumber\\
    &(1,3): &
    A_{13}^{\text{(I)}} &\rightarrow A_{13}^{\text{(II)}}, & 
    A_{49}^{\text{(I)}} &\rightarrow A_{49}^{\text{(II)}}, \nonumber\\
    &(2, 2): &
    A_{19}^{\text{(I)}} &\rightarrow \frac{1}{\sqrt{2}}A^\text{(II)}_{19}, & 
    A^\text{(I)}_{44} &\rightarrow \frac{1}{\sqrt{2}}A^{\text{(II)}}_{28} \nonumber \\
    &(2,3): &
    A_{21}^\text{(I)} &\rightarrow \frac{1}{\sqrt{2}} A_{21}^{\text{(II)}}, & 
    A^\text{(I)}_{37} &\rightarrow \frac{1}{\sqrt{2}} A^{\text{(II)}}_{21}, \nonumber \\
    &(3,3): &
    A_{28}^{\text{(I)}} &\rightarrow \frac{1}{\sqrt{2}} A^{\text{(II)}}_{28}, & 
    A_{35}^\text{(I)} &\rightarrow \frac{1}{\sqrt{2}} A_{19}^{\text{(II)}}.
\end{align}
As System~II is an all-integer system, we have $p=n/2$, i.e.~for $n=6$ it follows $p=3$.
As $p$ is odd, for the mapping on the self-conjugate amplitude of System~II we have
\begin{align}
    a_{(2,3)}^{\text{(I)}} &\rightarrow 2\times \frac{1}{\sqrt{2}} A_{21}^{\text{(II)}} = \sqrt{2}A_{21}^{\text{(II)}},\\
    s_{(2,3)}^{\text{(I)}} &\rightarrow 0.
\end{align}
Furthermore, no matter if the representations belong to the initial state, final state or the Hamiltonian, we can rewrite the $a$-type sum rules of Eq.~\eqref{eq:3t-a-b0} in terms of the amplitudes of the system of three triplets as
\begin{equation}\label{eq:3t-b0}
    A_7^\text{(II)} + A_{52}^\text{(II)} = A_{13}^{\text{(II)}} + A_{49}^\text{(II)} = A_{19}^\text{(II)} + A_{28}^\text{(II)} = A_{21}^\text{(II)} = 0\,.
\end{equation}
Note that the $a$-type sum rules $a_{(2,2)}^{\text{(I)}} = 0$ and $a_{(3,3)}^{\text{(I)}} = 0$ in Eq.~\eqref{eq:3t-a-b0} result in an identical sum rule for the system of three triplets. Thus the total number of $b=0$ $a$-type sum rules for the system of three triplets is one less than the corresponding number of $a$-type sum rules for the system of two doublets and two triplets given in Eq.~\eqref{eq:3t-a-b0}.

For the $s$-type sum rules in Eq.~\eqref{eq:3t-s-b1} we obtain
\begin{align}\label{eq:3t-b1}
    (-1)^{q_7} \left( A_7^{\text{(II)}} - A_{52}^\text{(II)}\right) + (-1)^{q_{13}} \left(A_{13}^\text{(II)} - A_{49}^{\text{(II)}}\right) = 0, \nonumber\\
    (-1)^{q_7} \left( A_7^{\text{(II)}} - A_{52}^\text{(II)}\right) + (-1)^{q_{19}} \left(A^{\text{(II)}}_{19} - A_{28}^\text{(II)}\right) = 0,
\end{align}
where we list only two sum rules since only two out of the three are linearly independent. 

The $b = 2$ $a$-type sum rule in Eq.~\eqref{eq:3t-a-b2} takes the following form in terms of the amplitudes of the system of three triplets
\begin{equation}\label{eq:3t-b2}
    (-1)^{q_7}\left(A_7^\text{(II)} + A_{52}^\text{(II)}\right) + (-1)^{q_{13}}\left( A_{13}^{\text{(II)}} + A_{49}^\text{(II)} \right) + (-1)^{q_{19}} \left(A_{19}^\text{(II)} + A_{28}^\text{(II)}\right) + (-1)^{q_{21}}2A_{21}^\text{(II)} = 0.
\end{equation}
The $(-1)^{q_i}$ factors in Eqs.~\eqref{eq:3t-b1} and~\eqref{eq:3t-b2} are to be found for each specific system using Eq.~\eqref{eq:qi-gen}.


\end{appendix}

\bibliography{GGS.bib}

\bibliographystyle{apsrev4-1}

\end{document}